\documentclass{aa}  

\usepackage{graphicx}
\usepackage{xcolor}
\usepackage{txfonts}
\usepackage{booktabs}
\usepackage{multirow}

\usepackage{lipsum}
\usepackage{caption}

\usepackage{tikz}
\usepackage{mdframed}

\usepackage{makecell}

\usepackage[colorlinks=true,linkcolor=blue,allcolors=blue]{hyperref}

\newcommand{\OIII}{[\ion{O}{III}]}
\newcommand{\Hbeta}{H$\beta$\xspace}
\newcommand{\lya}{Ly$\alpha$\xspace}

\definecolor{mycolor}{rgb}{0.122, 0.435, 0.698}
\definecolor{jmcolor}{rgb}{0.7, 0.3, 0.0}

\newmdenv[innerlinewidth=1.0pt, roundcorner=6pt,linecolor=mycolor,innerleftmargin=6pt,innerrightmargin=6pt,innertopmargin=4pt,innerbottommargin=2pt]{mybox}

\newmdenv[innerlinewidth=1.0pt, roundcorner=6pt,linecolor=jmcolor,innerleftmargin=6pt,innerrightmargin=6pt,innertopmargin=4pt,innerbottommargin=2pt]{jmbox}

\begin{document}

   \title{Anatomy of an ionized bubble: NIRCam grism spectroscopy of the $z=6.6$ double-peaked Lyman-$\alpha$ emitter COLA1 and its environment}
    \titlerunning{The JWST Slitless Ionized Bubble Survey}

   \author{
    Alberto Torralba-Torregrosa\inst{\ref{inst1}, \ref{inst2}}\thanks{E-mail: alberto.torralba@uv.es}
    \and Jorryt Matthee\inst{\ref{inst3}}
    \and Rohan P. Naidu\inst{\ref{inst4}}\thanks{NASA Hubble Fellow}
    \and Ruari Mackenzie\inst{\ref{inst_RM}}
    \and Gabriele Pezzulli\inst{\ref{inst_GP}}
    \and Anne Hutter\inst{\ref{inst_AH1}, \ref{inst_AH2}}
    \and Pablo Arnalte-Mur\inst{\ref{inst1}, \ref{inst2}}
    \and Siddhartha Gurung-López\inst{\ref{inst1}, \ref{inst2}}
    \and Sandro Tacchella\inst{\ref{inst_ST1}, \ref{inst_ST2}}
    \and Pascal Oesch\inst{\ref{inst_PO}, \ref{inst_AH1}, \ref{inst_AH2}}
    \and Daichi Kashino\inst{\ref{inst_DK}}
    \and Charlie Conroy\inst{\ref{inst_CC}}
    \and David Sobral\inst{\ref{inst_DS}}
    }

   \institute{
    Observatori Astron\`omic de la Universitat de Val\`encia, Ed. Instituts d’Investigaci\'o, Parc Cient\'ific. C/ Catedr\'atico Jos\'e Beltr\'an, n2, 46980 Paterna, Valencia, Spain\label{inst1}
    \and Departament d’Astronomia i Astrof\'isica, Universitat de Val\`encia, 46100 Burjassot, Spain\label{inst2}
    \and Institute of Science and Technology Austria (ISTA), Am Campus 1, 3400 Klosterneuburg, Austria \label{inst3}
    \and MIT Kavli Institute for Astrophysics and Space Research, 77 Massachusetts Ave., Cambridge, MA 02139, USA \label{inst4}
    \and Department of Physics, ETH Z{\"u}rich, Wolfgang-Pauli-Strasse 27, Z{\"u}rich, 8093, Switzerland \label{inst_RM}
    \and Kapteyn Astronomical Institute, University of Groningen, Landleven 12, NL-9747 AD Groningen, the Netherlands\label{inst_GP}
    \and Cosmic Dawn Center (DAWN), Copenhagen, Denmark\label{inst_AH1}
    \and Niels Bohr Institute, University of Copenhagen, Jagtvej 128, DK-2200, Copenhagen N, Denmark. \label{inst_AH2}
    \and Kavli Institute for Cosmology, University of Cambridge, Madingley Road, Cambridge, CB3 0HA, UK \label{inst_ST1}
    \and Cavendish Laboratory, University of Cambridge, 19 JJ Thomson Avenue, Cambridge CB3 0HE, UK \label{inst_ST2}
    \and Department of Astronomy, University of Geneva, Chemin Pegasi 51, 1290 Versoix, Switzerland \label{inst_PO}
    \and National Astronomical Observatory of Japan, 2-21-1 Osawa, Mitaka, Tokyo 181-8588, Japan  \label{inst_DK}
    \and Center for Astrophysics $\mid$ Harvard \& Smithsonian, 60 Garden St, Cambridge, MA 02138, USA\label{inst_CC}
    \and BNP Paribas Corporate \& Institutional Banking, Torre Ocidente Rua Galileu Galilei, 1500-392 Lisbon, Portugal \label{inst_DS}
    }

   \date{Accepted XXX. Received YYY; in original form ZZZ}

  \abstract{
The increasingly neutral intergalactic gas at $z>6$ impacts the Lyman-$\alpha$ (Ly$\alpha$) flux observed from galaxies. One luminous galaxy, COLA1, stands out because of its unique double-peaked Ly$\alpha$ line at $z=6.6$, unseen in any simulation of reionization. Here, we present JWST/NIRCam wide-field slitless spectroscopy in a 21 arcmin$^{2}$ field centered on COLA1. We find 141 galaxies spectroscopically selected through the \OIII\ doublet at $5.35<z<6.95$, with 40 of these sources showing H$\beta$. For COLA1, we additionally detect \OIII$_{4363}$ as well as H$\gamma$. We measure a systemic redshift of $z=6.5917$ for COLA1, confirming the classical double-peak nature of the Ly$\alpha$ profile. This implies that it resides in a highly ionized bubble and that it is leaking ionizing photons with a high escape fraction of $f_{\rm esc}{\rm (LyC)}=20$--$50$\%, making it a prime laboratory to study Lyman continuum escape in the Epoch of Reionization. COLA1 shows all the signs of a prolific ionizer with a Ly$\alpha$ escape fraction of $81\pm5\%$, Balmer decrement indicating no dust, a steep UV slope ($\beta_{\rm UV}=-3.2\pm 0.4$), and a star-formation surface density $\gtrsim 10\times$ that of typical galaxies at similar redshift. We detect five galaxies in COLA1's close environment ($\Delta z<0.02$). Exploiting the high spectroscopic completeness inherent to grism surveys, and using mock simulations that fully mimic the selection function, we show that the number of detected companions is very typical for a normal similarly UV-bright ($M_{\rm{UV}}\sim-21.3$) galaxy -- that is, the ionized bubble around COLA1 is unlikely to be due to an excessively large over-density. Instead, the measured ionizing properties suggest that COLA1 by itself might be powering the bubble required to explain its double-peaked Ly$\alpha$ profile ($R_{\rm ion}\approx 0.7$ pMpc), with only minor contributions from detected neighbors ($-19.5 \lesssim M_{\rm UV} \lesssim -17.5$). 
  }


   \maketitle

\section{Introduction}

The Epoch of Reionization (EoR) is the last major phase transition in the Universe. During this period, the bulk of hydrogen gas in the intergalactic medium (IGM) transitioned from neutral to fully ionized. The most recent observations of the Lyman-$\alpha$ (Ly$\alpha$; $\lambda=1215.67$ \AA) forest in distant quasar spectra indicate that this process was complete at $z\approx 5.5$ \citep[e.g.,][]{Greig17, Bosman22}, whereas the optical depth to the cosmic microwave background as measured by the Planck satellite places the average reionization redshift around $z\approx 7.8$--$8.8$ \citep{Planck16}. The start and duration of the reionization process are still uncertain \citep[see e.g.,][]{Robertson22, Gnedin22}, and they are linked to the sources of ionization \citep[e.g.,][]{Sharma18,Finkelstein19,Naidu20}. Star formation in galaxies is the leading candidate to have provided the majority of ionizing photons \citep[e.g.,][]{Bouwens15,Dayal20}, in particular because ultraviolet (UV) luminous active galactic nuclei (AGN) are very rare at $z>5$ \citep[e.g.,][]{Kulkarni19}, while UV-faint AGN are more common but appear heavily obscured \cite[e.g.,][]{Kocevski23,Matthee23b,Greene23,Kokorev24}.

For star formation, the main uncertainty is whether rare bright or numerous faint galaxies dominated the budget. This hinges upon the luminosity dependence of the escape fraction of ionizing photons (LyC, $\lambda<912$ \AA). Detecting LyC photons directly from galaxies becomes increasingly challenging at higher redshifts, as they are attenuated by the increasingly denser and more neutral IGM \citep{Inoue14,Steidel18}. Some semiempirical and hydrodynamical models suggest that faint ($M_{\rm UV}\gtrsim -18$) galaxies dominate \citep[e.g.,][]{Ocvirk21,Trebitsch21}; others point toward a more important contribution from late-emerging luminous galaxies \citep[in particular supported by the seemingly rapid and late timeline of reionziation;][]{Naidu20,Nakane23}, or toward a more complex picture in which the mass of the dominant agent of reionization evolves during the process \citep{Dayal23} and for most of the time is of intermediate mass ($M_*\sim 10^7\ M_\sun$; \citealt{Rosdahl22}, see also e.g., \citealt{Ma18}).

As the neutral gas in the IGM absorbs the LyC photons that escaped from galaxies in the EoR, our knowledge of the amount of LyC leakage and its dependence on the properties of galaxies relies on the local and intermediate redshift Universe at $z\lesssim 3$. The main physical processes that prevent ionizing photon escape are absorption by dust and neutral hydrogen in the interstellar medium (ISM). Likewise, in the $z=0$--$3$ Universe, the dust-sensitive observed UV slope has been recognized as a predictor of LyC escape \citep[e.g.,][]{Chisholm22}, albeit with significant scatter, whereas line ratios that trace the ionization parameter (e.g., [OIII]/[OII]) have yielded inconclusive results \citep[e.g.,][]{Izotov18,Naidu18,Flury22}. The LyC escape fraction further appears to correlate with the star formation rate surface density \citep[e.g.,][]{Flury22}, as this quantity is linked to the presence of large-scale galactic outflows \citep[e.g.,][]{Heckman16}. Recent studies indicate that high LyC escape is associated with strong UV emission lines as CIV and HeII \citep[e.g.,][]{Naidu22,Mainali22,Mascia22,Saxena22,Schaerer22,Kramarenko24}. This suggests a link between the stellar populations that lead to these high ionization lines and the processes that enable LyC escape.

The radiative transfer of Ly$\alpha$ photons is sensitive to the dust attenuation and the neutral gas column density \citep[e.g.,][]{Verhamme15,Kakiichi&Gronke21}. As these are the same barriers to LyC escape, the strength and shape, in particular the peak separation, of the Ly$\alpha$ emission line from galaxies is a powerful indirect tracer of ionizing photon leakage \citep[e.g.,][]{Izotov18,Steidel18,Flury22,Pahl23}. However, the main downside is that measurements of the Ly$\alpha$ line-profile as it emerges from the ISM can generally not be obtained for sources in the EoR. As neutral intergalactic gas in the EoR impacts the \lya line \citep[e.g.,][]{Haiman02,Pentericci14,Mason18,Gurung-Lopez20}, we expect to observe lower EW \lya emission, with redder velocity offsets as they reside in increasingly smaller ionized regions beyond $z>6$ \citep[e.g.,][]{Hayes23}. Indeed, JWST spectroscopy has already revealed such systems with relatively faint and redshifted Ly$\alpha$ emission at $z>6$ \citep[e.g.,][]{Bunker23,Jung23}. This imprint from the neutral gas in the IGM, which particularly impacts the blue side of the Ly$\alpha$ line, tends to hide the information on the ISM properties (and hence, LyC escape) encoded in galaxies' Ly$\alpha$ profiles.

An exception to this are galaxies that reside in such large ionized regions where \lya photons could already have redshifted enough before encountering neutral gas in the IGM such that their emergent profile is visible. Currently, a handful of such galaxies are known at $z>5$ \citep{Hu16,Songaila18,Bosman20,Meyer21}. COLA1, located in the COSMOS field at $z=6.6$, is the most luminous galaxy among those with double-peaked \lya line known at $z>5$. COLA1 has a very narrow peak separation of 220 km s$^{-1}$ (suggestive of a high LyC escape fraction of $\approx30$ \%; \citealt{Verhamme17,Izotov18}) and the only double-peak that has been confirmed by multiple instruments (Keck/DEIMOS, \citealt{Hu16}; VLT/X-shooter, \citealt{Matthee18}). However, with only the \lya line detected, the systemic redshift was uncertain, allowing alternative explanations for the particular shape of the \lya profile, such as a merger \citep[see][for a discussion]{Matthee18}.

Besides using these systems to infer their LyC escape fraction from the Ly$\alpha$ peak separation, these lines are also suited to probe the sizes of the ionized regions around these galaxies \citep[e.g.,][]{Hu16,Matthee18,Mason&Gronke20}. However, sight-lines that allow the detection of double peaks such as the one observed in COLA1 appear extremely rare in hydrodynamical radiative transfer reionization simulations, virtually non existent for luminous galaxies at $z\sim7$ \citep{Gronke21}. Therefore, an immediate question that arises is whether such a galaxy resides in a peculiarly large over-density of galaxies, similar to other high-redshift Ly$\alpha$ lines \citep[e.g.,][]{Wistok23}.

To measure the systemic redshift and ionizing output of the COLA1 galaxy, together with directly probing galaxies in its environment, we will present results from a survey of COLA1 and its field with {\it JWST}/NIRCam imaging and grism Wide Field Slitless Spectroscopy (WFSS) based on a Cycle 1 program (PID 1933, PIs Matthee \& Naidu). We observe with the F356W filter, which simultaneously probes the key diagnostic lines H$\gamma$, H$\beta$ and the \OIII\ doublet for the bright COLA1 galaxy, and (primarily) the strong \OIII\ doublet in the $\sim10$ cMpc environment from $z=5.3-6.9$. The sensitive imaging in the short-wavelength at 1-2 micron probes the rest-frame UV light of the identified galaxies. The rationale behind our survey design is that the typical strong rest-frame optical emission lines, indicated from {\it Spitzer}/IRAC photometry \citep[e.g.,][]{Raiter10,Labbe13, deBarros14}, would facilitate the direct identification of the galaxies in the environment using the NIRCam grism data. Indeed, spectroscopic observations of large samples of distant galaxies confirm that the strong rest-frame optical lines are common \citep[e.g.,][]{Atek23, Cameron23, Fujimoto23, Matthee23a, Oesch23,Windhorst23}. The relatively high resolution of the NIRCam grism, combined with the relatively featureless observed continuum of the major source of light contamination from galaxies at $z\sim1-2$, can indeed be exploited to obtain large samples of spectroscopically selected galaxies with redshifts as accurate as $\sim50$ km s$^{-1}$ \citep[e.g.,][]{Kashino23}.

In Sect.~\ref{sec:data_description} we describe the observational data used in this work. In Sect.~\ref{sec:methods} we detail the \OIII\ emitter selection method, the procedures used to extract the line fluxes of the sample, the accuracy of the redshift measurement of COLA1 and we introduce mock catalogs of the field. In Sect.~\ref{sec:results_1} we describe the measured UV properties and optical emission lines of COLA1 and we investigate COLA1 in the context of the full \OIII\ sample, making use of mock observations. In Sect.~\ref{sec:discussion} we discuss the results, and compare COLA1 to other LyC leakers in the literature. Finally, in Sect.~\ref{sec:summary} we summarize the contents of this work.

Throughout this work we use a $\Lambda$CDM cosmology as described by \texttt{Planck18}\ \citep{Planck18}, with $\Omega_\Lambda=0.69$, $\Omega_\text{M}=0.31$, and $H_0=67.7$ km\,s$^{-1}$\,Mpc$^{-1}$. All photometric magnitudes are given in the AB system \citep{Oke83}.

\section{Data}\label{sec:data_description}

\subsection{JWST/NIRCam}

\subsubsection{Observations}

\begin{figure*}
    \centering
    \includegraphics[width=\linewidth]{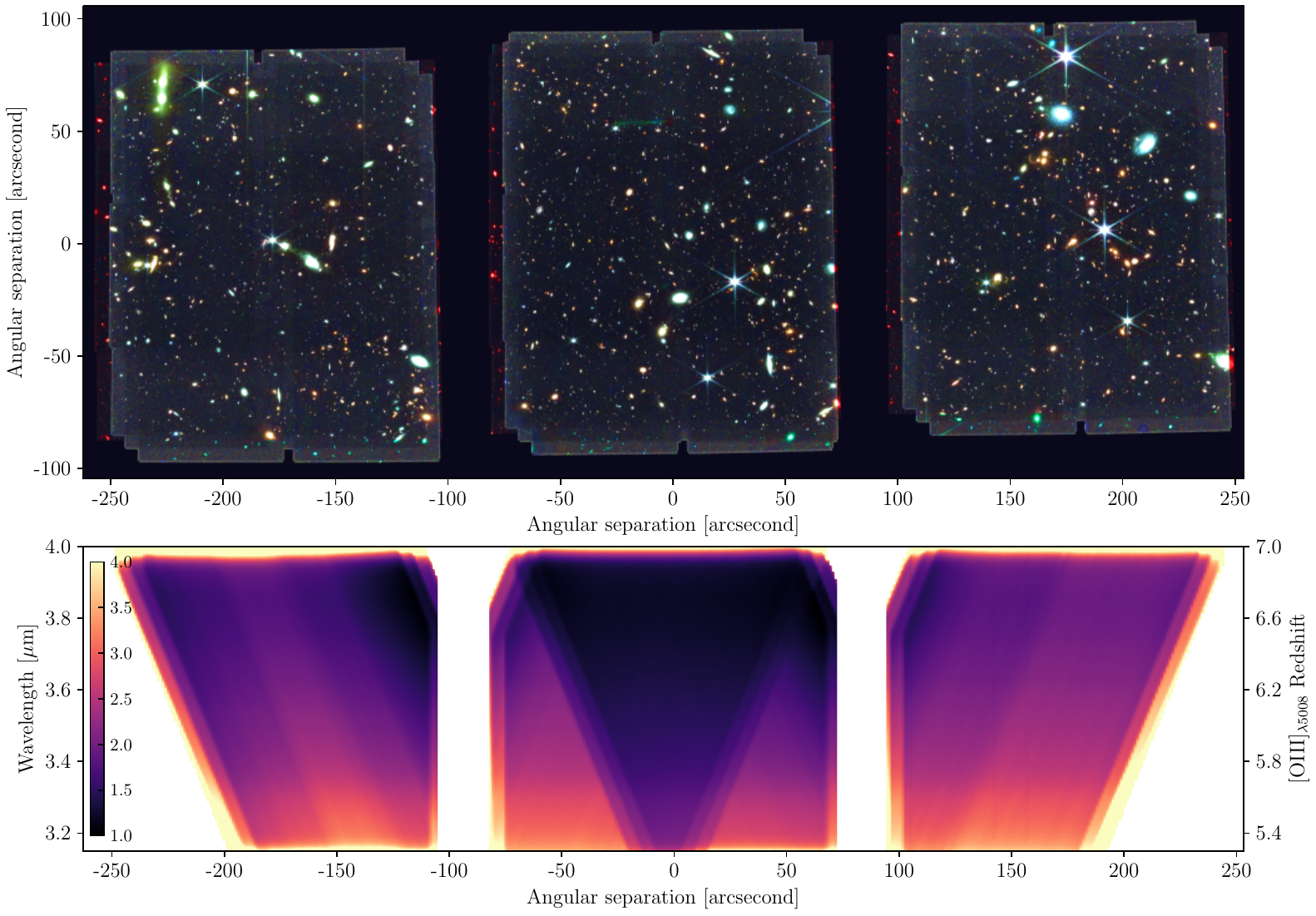}
    \caption{{\bf Top:} False-color {\it JWST} NIRCam image of the COLA1 field (C1F) coverage based on the F115W/F200W/F356W images. The vertical and horizontal axes represent the angular separation relative to COLA1, which is located in the center of the middle tile. {\bf Bottom:} A slice of the error cube of the C1F in the horizontal direction of the top panel. This illustrates how the field of view depends on wavelength, and that it is maximal around 3.8 $\mu$m, which is the undispersed wavelength of the NIRCam grism. We note that the \OIII\ doublet at the COLA1 redshift falls exactly at this wavelength. The units are relative to the minimum flux error across the field.}
    \label{fig:field_coverage}
\end{figure*}

We use imaging and Wide Field Slitless Spectroscopy (WFSS) with NIRCam targeting a mosaic centered around COLA1 through JWST program 1933 (PIs Matthee \& Naidu). Observations were performed on January 6 and 7, 2023. The coverage consists of four visits in a $2\times 2$ mosaic with a position angle of 295 deg, which yields three slightly separated regions of $2.5\arcmin\times 2.8\arcmin$. The total area of the COLA1 field (hereafter C1F) is 21 arcmin$^2$, shown in top panel of Fig.~\ref{fig:field_coverage}.

We use GrismR, which disperses spectra in the horizontal direction of the detectors, in opposite directions for modules A and B. This is similar to the strategy employed by the EIGER survey \citep{Kashino23}, albeit here maximizing the area with cross-dispersed coverage at the expense of gaps. The central $1.5\arcmin\times2.6\arcmin$ are covered by four visits. Another area of 11.4 arcmin$^2$ is covered by two visits and one of 6.2 arcmin$^2$ by a single visit. Grism spectroscopy inherently obtains a 2D spectrum for every source in the field of view, without target preselection. However, due to this extensive coverage, the spectra are prone to be contaminated due to blending of spectra from different sources, and the background is also dispersed, adding noise.

The F356W filter ($\lambda\sim 3.1$--$4.0$ $\mu$m) is used in combination with the grism to mainly cover redshifted \Hbeta + \OIII\ lines at $z=5.5$--$6.9$ and to cover H$\gamma$ at $z>6.3$ (including the redshift of COLA1). The spatial limits of the detector in the spectral dispersion directions of grism modules A and B define the regions of the field where it is possible to observe the \Hbeta + \OIII\ at a given redshift (see bottom panel of Fig.~\ref{fig:field_coverage}). The central region uniquely has grism spectra in both grism modules, facilitating contamination and confusion removal, in addition to allowing for deeper WFSS data. In a very similar manner to \cite{Kashino23}, we use the \OIII$_{\lambda4960, \lambda5008}$ doublet for source identification in the grism data. The fixed peak separation of the \OIII\ doublet, in combination with its intrinsic flux ratio of \OIII$_{\lambda5008}$/\OIII$_{\lambda4960}=2.98$, makes this doublet ideal to perform this task. The \OIII\ doublet often appears as a triplet with \Hbeta ($\lambda 4863$), allowing for an even more secure source identification in those cases.

Simultaneously with the deep grism exposures, we obtain imaging data in the F115W and F200W filters in the same part of the sky. Direct and out of field images are taken using the F150W and F356W filters to facilitate the identification of all spectral traces on the grism data.
The total spectroscopic on sky integration time ranges from 8.5 to 34 ks for the F356W/GrismR data, depending on the position in the observed field and according to the number of visits as described above. The F115W and F200W imaging was equal to 42 and 58 \% of the F356W/GrismR exposure time, respectively. Direct and out of field images amount to 1.5--6 ks of exposure time in the F150W and F356W imaging. 

For data rate considerations, we chose 3/4 groups/Int Medium8 for grism and 3 groups/int Deep8 for direct imaging. This led to a significant number of cosmic ray hits that are especially challenging to filter out in regions of the imaging with few exposures.

\subsubsection{Data reduction and quality}

\begin{figure*}
    \centering
    \includegraphics[width=\linewidth]{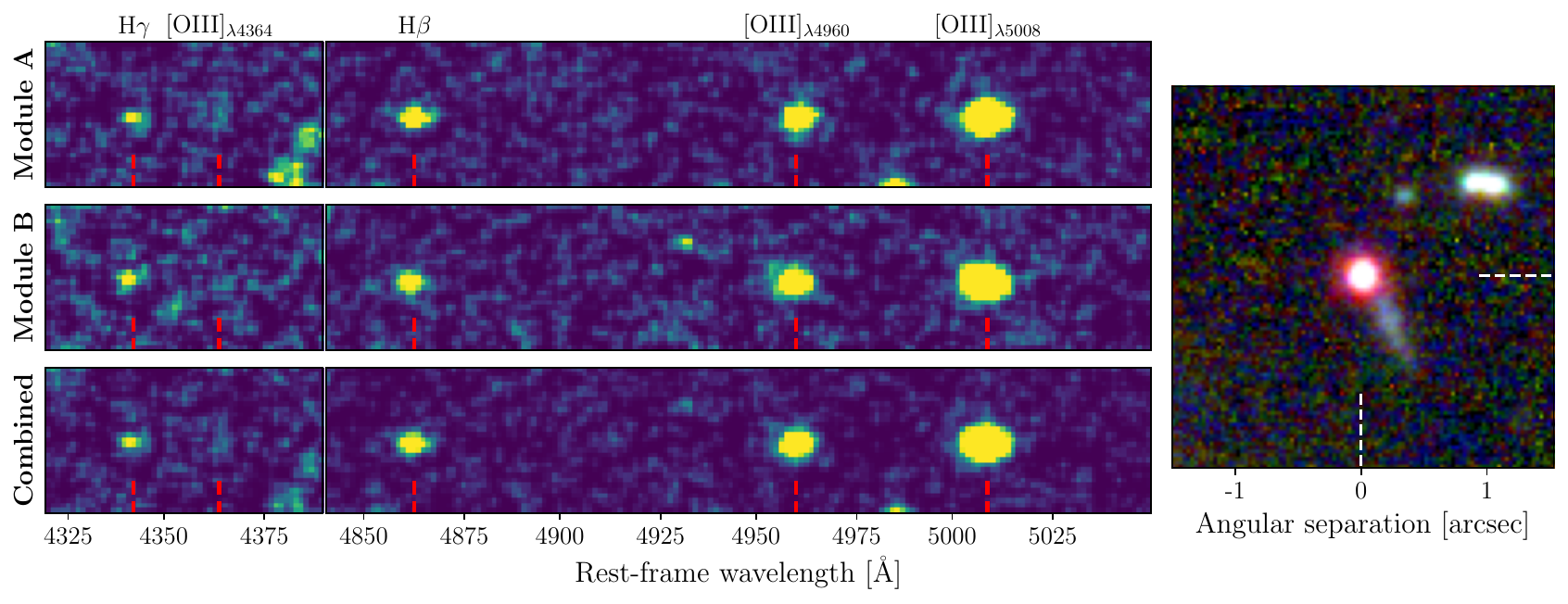}
    \caption{2D grism spectrum of COLA1 in the regions where the most luminous lines are visible. The three rows of panels on the left side of the figure contain the EMLINE fluxes of modules A and B, as well as the average flux of both modules. The x axis shows the rest-frame wavelength assuming a redshift of $z=6.59165$. The right panel displays the false-color NIRCam image of COLA1, oriented with our position angle of 295 deg. We note that the blue extended `tail' is a $B$-band detected foreground contaminant at $z\leq2.5$ (see Appendix~\ref{sec:tail_removal}; see also \citealt{Matthee18}).}
    \label{fig:COLA1_2D_spec}
\end{figure*}

The NIRCam imaging and WFSS data were reduced using the \texttt{jwst} pipeline v-1.11.0\footnote{https://github.com/spacetelescope/jwst}, using CRDS version 11.16.20 and the \texttt{jwst\_1097.pmap} context, with modifications developed for the EIGER survey \citep{Kashino23}.

For the imaging data, we first use the standard \texttt{Detector1} and \texttt{Image2} steps and then align the images to the astrometry from stars and spherically symmetric objects (i.e., whose centroid is unambiguously defined in our high resolution images) in the COSMOS2020 catalog \citep{Weaver22}, which has been aligned to GAIA. We use this indirect method as only few GAIA-detected stars are present in our NIRCam images. By creating source-masked pixel-based median stacks of the imaging in a specific filter/module/camera combination, we subtract the sky and stray-light features (so-called wisps). Cosmic-ray related artefacts (snowballs) are masked following \cite{Merlin22}. The $1/f$ read noise in our imaging data is subtracted using the median sky in quarter rows, columns and in the four amplifiers, respectively. Finally, images are combined with \texttt{Image3} to a common grid with pixel scale of 0.03\arcsec.

For our full source catalog, we use the F356W image as detection image. The SW images are convolved (assuming the point spread functions, PSFs, of webbpsf) to match the F356W PSF. Aperture-matched photometry is measured with Kron apertures, see \cite{Kashino23} for details. The method from \cite{Finkelstein22} is used to derive errors on the photometry. We derive the relation between blank sky variation and local variance for various aperture sizes, and for each object apply the derived scaling. We measure a typical (deepest) 5 sigma point-source sensitivity of 27.7 (28.5), 27.4 (28.1), 28.4 (29.0), 28.2 (28.9) in F115W, F150W, F200W and F356W, respectively. The deepest sensitivity by design is in the central region in Fig.~\ref{fig:field_coverage}, where COLA1 is located.

For the WFSS data, we also use a combination of the \texttt{jwst} pipeline (version 1.11.0) and specific modifications developed in the context of the EIGER project. As detailed in \cite{Kashino23}, we first process each exposure with the \texttt{Detector1} step and assign an initial WCS using \texttt{Spec2}. The images are flat-fielded using \texttt{Image2} and we remove the sky background by subtracting the median value in each column. We then split this so-called `SCI' image in an emission-line (`EMLINE') and continuum component. The continuum emission in each row is obtained by calculating a running median in the dispersion direction. The median is calculated within a kernel that has a hole in the center not to over-subtract lines themselves. The continuum image is then used to create the emission-line map. The process is run in two iterations. After the first iteration, emission-lines are detected in the emission-line map with source-extractor, and then masked in the second iteration to create the continuum image. 

Finally, we obtain relative astrometry offsets between the astrometry of each grism exposure and the final image stack by using the F200W images that are taken simultaneously with each of the grism images. We carefully selected a combination of stars and galaxies with low apparent ellipticity, with steep light profiles and derived small astrometric corrections for each exposure. The good alignment of the spectrum of COLA1 (Fig.~\ref{fig:COLA1_2D_spec}) in the two modules, each a combination of observations spanning two independent visits, validates the astrometric corrections. The 5$\sigma$ line-flux sensitivity of the grism data ranges from $0.7$--$1.5\times 10^{-18}$ erg\,s$^{-1}$\,cm$^{-2}$, but varies with wavelength and position as illustrated in the bottom panel of Fig.~\ref{fig:field_coverage}.

\subsection{VLT/X-shooter}\label{sec:vlt_xshooter}

We also present an updated VLT/X-shooter spectrum of COLA1 covering the Lyman-$\alpha$ line in the VIS arm. The observations (from ESO programs 100.A-0213 and 102.A-0652) and data reduction are described in detail in \cite{Matthee21}. The total exposure time constitutes 13.2 ks, which is a factor 1.8 longer than the data used to obtain the spectrum presented in \cite{Matthee18}. The spectral resolution is R=4100.

\section{Methods}\label{sec:methods}

In this section, first we describe the methods used to identify the \OIII\ emitter sample in the field (Sect.~\ref{sec:doublet_identification}), and measure line fluxes and systemic redshifts from the grism data (Sects.~\ref{sec:z_and_f_measurement} and \ref{sec:redshift_calibration}). Secondly, we describe the SED modeling code used to infer properties of the sample from the photometric and spectroscopic data (Sect.~\ref{sec:SED_fitting}). Finally we introduce mock catalogs of the field, employed to characterize the environment of COLA1 (Sect.~\ref{sec:mocks_description}).

\subsection{{\normalfont\OIII\ }doublet identification}\label{sec:doublet_identification}

The WFSS technique yields a spectrum for all the objects in the spectroscopic field of view that is shown in Fig. $\ref{fig:field_coverage}$. We extracted 2D spectra from the GrismR (SCI, EMLINE and CONT) images for the sample of all F356W detected sources (following the so-called ``forward'' method outlined in \citealt{Kashino23}). This sample typically contains sources down to a F356W magnitude of 28 but extends to magnitude of 29.5 in the deepest areas. We ran SExtractor \citep{Bertin&Arnouts96} on each coadded EMLINE 2D spectrum to identify pairs of emission lines, regardless of any photometric preselection apart from detection. Then, we selected an initial \OIII\ doublet candidate sample by identifying pairs of lines with S/N $>3$, with wavelengths that could be compatible with \OIII$_{\lambda4960, \lambda5008}$ at $z=5.33$--$6.93$, with a maximum offset of 3 pixels (0.09\arcsec) in the spatial centroid (i.e., perpendicular to the spectral direction). We required a peak separation for the \OIII\ doublet according to the initial redshift solution, within a 15 \AA\ tolerance. The theoretical line flux ratio of the \OIII\ doublet is $f_{\lambda5008}/f_{\lambda4960}=2.98$ \citep{Storey00}, and therefore we conservatively imposed a line flux ratio of $1.2<f_{\lambda5008}/f_{\lambda4960}<6$. This search produced an initial sample of 3221 candidates.

The parameters used for the search of this preliminary catalog are very tolerant to enable the identification of sources with complex morphologies, for which the centroiding of the multiple lines may be quite uncertain. This leads to a complete yet highly contaminated sample.
The major sources of contamination are lines that appear in the spectra as false detections (cosmic rays, due to the low data  rate settings, and diffraction spikes in the imaging data), pairs of lines from different galaxies that lay in the same grism spectral axis as the real source of the emission-lines, line doublets other than \OIII\ (i.e., \Hbeta + \OIII$_{\lambda 4960}$) that assigned to the wrong object, and residuals from our continuum filtering in the spectra of stars and very bright galaxies. We perform visual inspection of all 3221 doublet candidates in the catalog to obtain a clean sample of 144 \OIII\ doublets.

Some of our candidates are multiple-component systems that are sometimes blended into a single catalog entry, sometimes not (see Fig. $\ref{fig:example_spec}$ for some examples). Following \cite{Matthee23a}, we define a `system' as a group of two or more galaxies with a projected separation smaller than 2\arcsec\ and an absolute redshift difference of $\Delta z<1000$ km\,s$^{-1}$. In such systems, the measured line flux of each single component is possibly contaminated by their companions. Moreover, whether the components of systems can disentangled or not is an effect of our arbitrary line of sight to the system a the spatial resolution of the images. We find three such systems within our list of galaxies. We define the main component of each system as the brightest galaxy in F356W, and add the photometry and line fluxes of their companion to the main component in the catalog. The companions are then removed from our galaxy catalog, and the systems are regarded as single objects, leaving a sample of 141 \OIII\ emitters. Among this sample, 40 objects ($\sim 30\%$) have a measurement of \Hbeta flux with S/N $>3$ (13 with S/N $>5$).

A complete list of all the selected sources is found in Table~\ref{tab:O3_cat}, and some examples of their grism spectra are shown in Appendix~\ref{sec:example_spec}. In order to carry out this task, we created inspection images similar to Fig.~\ref{fig:COLA1_2D_spec}. Two of the authors (ATT, JM) agreed on the final sample. In order to visually select our candidates, both peaks of the \OIII\ doublet must be present in the available grism modules. The shape of these emission lines must not substantially differ from the morphology seen in the detection image, nor between modules. It has to be taken into account that, since modules A and B disperse photons in opposite directions, the transverse spatial component is mirrored between modules in our extraction method. Also, both peaks of the doublet must not be significantly misaligned in the spatial direction, nor with the detection image. Additionally, color is another useful indicator for identifying contaminants. In the color representation we chose, \OIII\ emitters present red-purple colors, as a result of typically blue colors in ${\rm F115W} - {\rm F200W}$ (i.e., their UV continuum), and \OIII -boosted red colors in ${\rm F200W} - {\rm F356W}$. Sometimes an \OIII\ doublet is assigned to more than one object, due to the intrinsic degeneracy of the grism wavelength solution -- galaxies that line up along the dispersion direction end up with emission lines in the same row. In those cases we assign the correct object matching the morphologies of the grism emission lines and photometric image.

\subsection{Redshift and line fluxes measurements}\label{sec:z_and_f_measurement}

In order to measure the line fluxes and redshifts, we extract the 1D spectra of the candidates. We use the \OIII$_{\lambda 5008}$ line to trace the morphology of each object, as it is the most luminous emission line in our range. First, we select regions in the grism 2D EMLINE spectra containing the \OIII$_{\lambda 5008}$ line, and integrate them along the wavelength direction, for both modules A and B, whenever they are available. Then, we fit the spatial profile of the \OIII$_{\lambda 5008}$ emission with a single Gaussian for most cases, and double or triple Gaussian components for a few objects that present multiple clumps compatible with \OIII\ emission at the given redshift. Finally, we optimally extract the 2D spectra with the shape derived from the Gaussian fit to obtain the 1D spectrum of each candidate \citep{Horne86}. 

The next step is to obtain precise redshifts for each candidate from the \OIII\ doublets. We use Gaussian profiles to fit the line-profiles doublets from the 1D spectra. For these fits, we use a flat prior for the redshift with $\delta z=\pm 0.02$ around the redshift solution given by the candidate selection pipeline. The fitted function is a double component Gaussian for which we impose a fixed peak separation of $(5008.24 - 4960.295)\times (1 + z)$ \AA , corresponding to the expected observed wavelengths of the \OIII\ doublet. The normalization of the $\lambda5008$ and $\lambda4960$ Gaussian components is initialized with a ratio of 2.98:1, with a fudge factor $f$ to allow uncertainties, for which we use a flat prior, $f\in[0.5, 2]$. We extract the redshift for each candidate from the best fit.

Lastly, we measure the line fluxes. We fit individual Gaussian profiles to \OIII$_{\lambda 5008}$, \OIII$_{\lambda 4960}$ and \Hbeta. We use narrow flat priors for the observed wavelengths of each line, based on the redshifts extracted from the \OIII\ doublets. In the particular case of COLA1, we also measure fluxes of the \OIII$_{\lambda 4364}$ and H$\gamma$, due to their exceptional S/N, as detailed in Sect.~\ref{sec:redshift_calibration} below. Throughout this work, we use the average flux from the measurements of modules A and B (a single measurement is used for those sources where only one module is available).

\subsection{Redshift calibration}\label{sec:redshift_calibration}

In grism 2D spectra, there is an intrinsic degeneracy between the spatial position and the dispersed wavelength of the emission lines. Due to that, the accuracy of the astrometric alignment and the grism trace model directly affect the redshift measurement. While an absolute verification of the wavelength solution is not possible without independent redshift measurements, we compared the redshifts measured in modules A and B, for all 24 objects in our \OIII-emitter catalog that are detected in both modules. This subsample presents a median offset of $z_\mathrm{modA}-z_\mathrm{modB}=0.0016 \pm 0.0018$, which is comparable to the redshift estimates difference of COLA1 between both grism modules: $z_\mathrm{modA}^\mathrm{COLA1}-z_\mathrm{modB}^\mathrm{COLA1}=0.0012$.

\begin{figure}
    \centering
    \includegraphics[width=\linewidth]{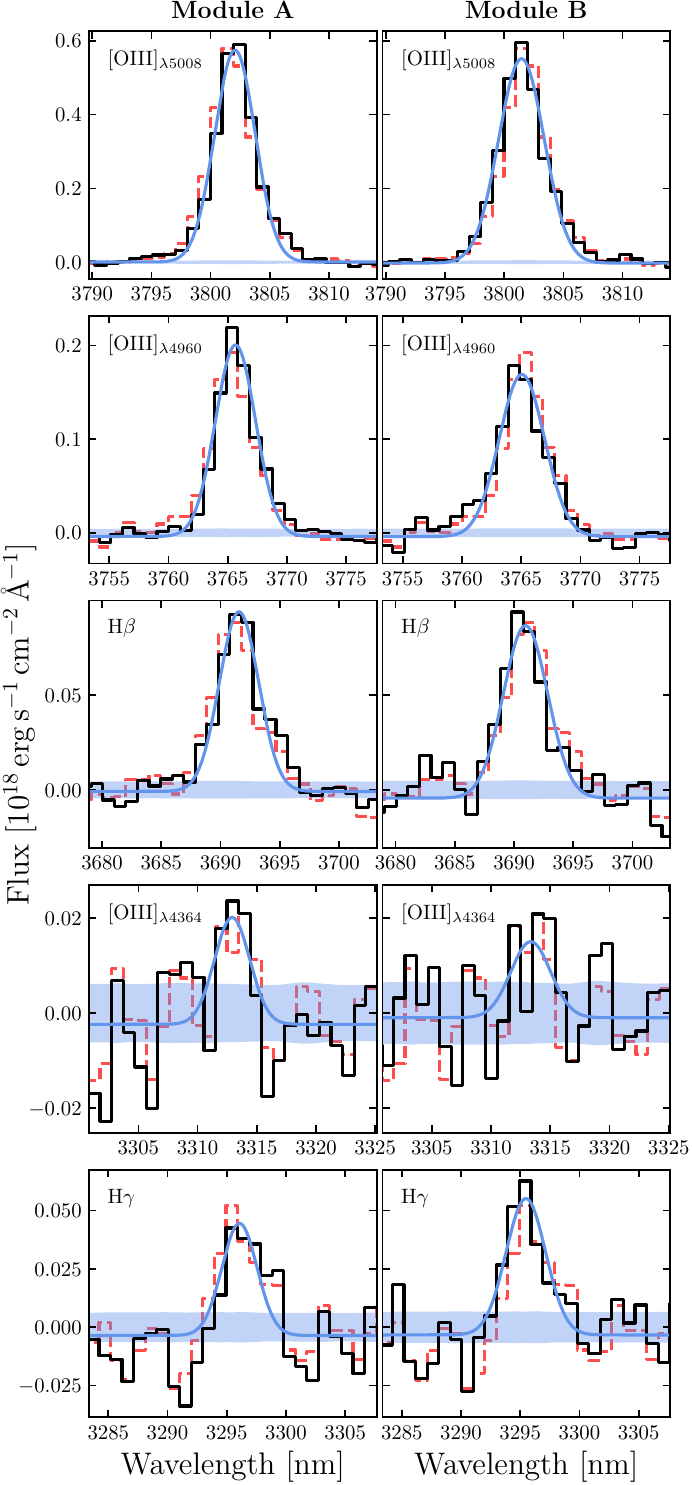}
    \caption{Gaussian line fits (blue line) to the extracted 1D COLA1 spectrum (black line) for \OIII$_{\lambda 5008}$, \OIII$_{\lambda 4960}$, \Hbeta , \OIII$_{\lambda 4364}$ and H$\gamma$. We show the results for module A and B separately. The 1$\sigma$ noise level of the 1D spectrum is represented as the blue shaded region around the zero level. The dashed red lines represent the average flux between modules A and B.}
    \label{fig:COLA1_line_fits}
\end{figure}

\begin{figure}
    \centering
    \includegraphics[width=\linewidth]{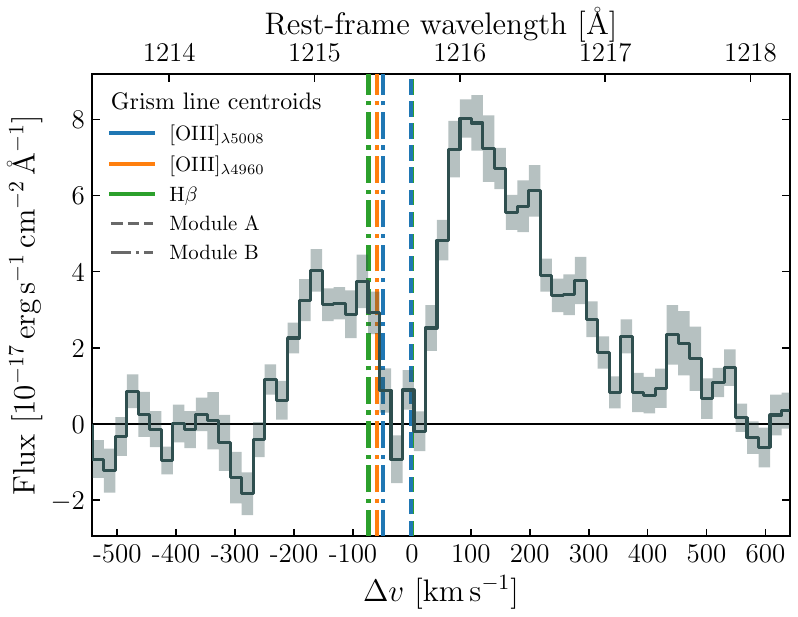}
    \caption{\lya profile of COLA1 obtained from the VLT/X-shooter UV spectrum. We assume a systemic redshift of $z=6.59165$, the average redshift measured in the NIRCam grism data in module A. The dashed (dot-dashed) color lines mark the position of the systemic \lya wavelength suggested by the redshifts obtained from the centroids of \OIII$_{\lambda 5008}$, \OIII$_{\lambda 4960}$ and \Hbeta lines in module A (module B) of the WFSS data. For module A the variation in the horizontal direction is lower than the width of the marker lines.}
    \label{fig:Lya_profile}
\end{figure}

We fit independent Gaussian profiles for all the detected lines of COLA1: \OIII$_{\lambda 5008}$, \OIII$_{\lambda 4960}$, \Hbeta , \OIII$_{\lambda 4364}$ and H$\gamma$, in the separate module A and B data. In Fig.~\ref{fig:COLA1_line_fits}, we show the fits for every detected emission line detected in COLA1 spectra. The fluxes and S/N for all these lines in COLA1's spectrum are listed in Table~\ref{tab:COLA1_line_redshifts}. Our integration times were planned to ensure a H$\gamma$ detection with a S/N $\sim5$ (based on the \lya escape fraction estimated following \citealt{sobralmatthee19}). This goal has been reached. Remarkably, we also detect \OIII$_{\lambda 4364}$ with S/N $\approx3$, combining grism modules A and B. In Table~\ref{tab:COLA1_line_redshifts} we also show the measured redshifts for the three highest S/N lines. In the case of module A, the measured redshift for both lines of the \OIII\ doublet and \Hbeta are in great agreement, with an average of $z=6.59165$ (weighted by the inverse squared error of the line flux). This redshift is in line with the systemic redshift of the \lya line estimated from the center of the two \lya peaks in the VLT/X-shooter data \citep[see Fig.~\ref{fig:Lya_profile}, $z_{{\rm Ly}\alpha}^{\rm sys}=6.591$;][]{Matthee18}. In module B we obtain an average redshift of $z=6.58997$ for the three high S/N lines. This value is slightly lower to module A, which is consistent with the above mentioned systematic redshift difference between modules. Moreover, the measured redshift for these three lines in module B decreases with the wavelength of the emission line ($z_{\lambda 5008}>z_{\lambda 4960}>z_{\lambda 4863}$). This can be explained by a small systematic calibration error in the astrometry or wavelength solution of module B in our grism data. As a consequence of this, the redshifts measured by module B might be slightly underestimated. Hereafter, we adopt the average redshift measured in module A ($z=6.59165$) as the fiducial COLA1 redshift.

\begin{table}
    \centering
    \caption{Measured redshifts, fluxes and S/N of various emission-lines from COLA1 in our NIRCam WFSS data.}
    \label{tab:COLA1_line_redshifts}
    \resizebox{\linewidth}{!}{
    \begin{tabular}{ccccccc}
         \toprule
         \multirow{2}{*}{Line} & \multicolumn{3}{c}{Module A} & \multicolumn{3}{c}{Module B} \\
          & Redshift & Flux$^\star$ & S/N &  Redshift & Flux$^\star$ & S/N \\
         \midrule
         \OIII$_{\lambda 5008}$ & 6.59160 & 24.60 & 45.2 & 6.59042 & 26.75 & 48.4\\
         \OIII$_{\lambda 4960}$ & 6.59166 & 8.68 & 38.2 & 6.59014 & 8.29 & 24.2\\
         \Hbeta & 6.59166 & 3.94 & 19.5 & 6.58979 & 4.26 & 16.4\\
         \OIII$_{\lambda 4364}$ & - & 0.84 & 2.7 & - & 0.67 & 1.8\\
         H$\gamma$ & - & 1.79 & 6.4 & - & 2.45 & 8.0\\
         \bottomrule
    \end{tabular}
    }
    \tablefoot{$^\star$Flux units are $10^{-18}$\,erg\,s$^{-1}$\,cm$^{-2}$.}
\end{table}

\subsection{SED modeling}\label{sec:SED_fitting}

\begin{figure}
    \centering
    \includegraphics[width=\linewidth]{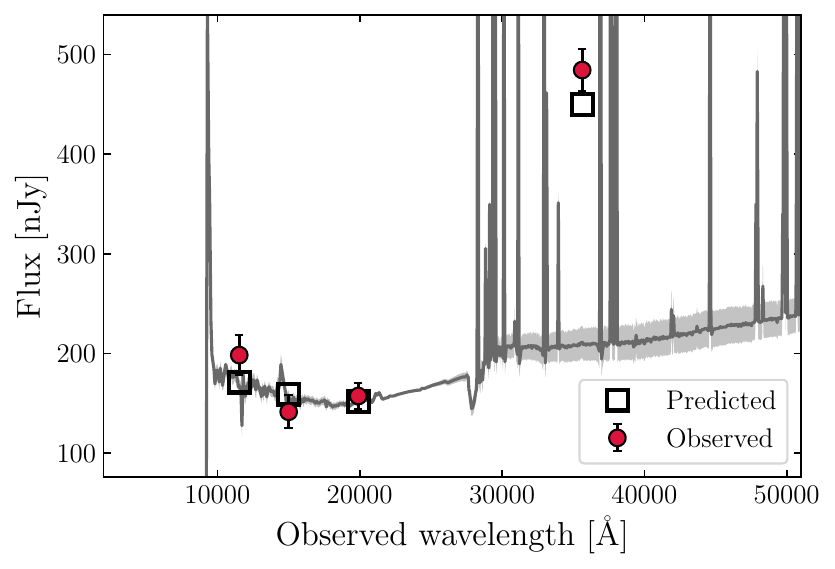}
    \caption{Best-fit \texttt{Prospector} spectral energy distribution for COLA1 (gray line) and all models allowed within 1$\sigma$ (gray shaded regions). The errors of the observed photometry are increased by a factor three for visual clarity. The code fails to reproduce the steep blue slope seen in the observed F115W-F150W color. The predicted F356W flux is also lower than observed, causing a slight underestimate the EW of the optical lines \OIII +\Hbeta.}
    \label{fig:Prospector_spectrum}
\end{figure}

We performed spectrophotometric SED fitting of all the \OIII\ emitters in our sample using the \texttt{Prospector} code \citep{Johnson21}, including nebular emission and nebular continuum implemented via \texttt{Cloudy} \citep{Ferland98, Ferland13, Byler17} and FSPS \citep{FSPS1,FSPS2,FSPS3} with the MIST stellar models \citep[][]{Choi17}. We follow the setup described in \citet{Naidu22schrodinger}, which in turn is adapted from \citet{Tacchella22}. We fit the F115W, F150W, F200W and F356W photometric fluxes, as well as \Hbeta and \OIII$_{\lambda 5008, 4960}$ line-fluxes simultaneously. The free parameters in the model include seven bins for the nonparametric star-formation history, with the first two being fixed to 0--5 Myr and 5--10 Myr (e.g., \citealt{Tacchella23}), and the remaining logarithmically spaced out to $z=20$. The other free parameters capture the total stellar mass formed, the stellar metallicity, the gas-phase metallicity, dust attenuation (a screen and an additional component for young stars), and the ionization parameter. We assume a \citet{Chabrier03} initial mass function, with a $150$ $M_\odot$ cutoff.

Figure \ref{fig:Prospector_spectrum} shows the best-fit spectrum, the predicted photometry, and draws from the posterior that are compared to the observed data. The model predicts a best-fitting UV slope of $\beta_{\rm UV}=-2.23\pm 0.05$, which is at odds with the UV photometry that shows a peculiar "U" shape where the F150W flux lies below F115W and F200W (see Sect.~\ref{sec:uv_properties}). Keeping in mind that significant LyC leakage is at play, we fit a model with additional parameters that capture the escape fraction \citep{Conroy12}. In particular, a fraction of OB stars (uniform prior between 0 to $100\%$) is allowed to shine through the birth-cloud and dust-screen entirely unimpeded, which mimics ``escape through holes;'' in other words, an ionization-bounded nebula (e.g., Fig. 1 in \citealt{Zackrisson13}). The predicted escape fraction from this fit is $f_{\rm esc}{\rm (LyC)}=33\pm 8$\%. This high of $f_{\rm esc}{\rm (LyC)}$ is in agreement with the values obtained through different observational estimators in Sect.~\ref{sec:escape_frac} below.
However, due to the sparse data, we note the quality of our fit is effectively indistinguishable from a fit assuming $f_{\rm{esc}}=0$ and we defer investigation of this issue to future work, in particular when more photometric bands are available. This exploration highlights how COLA1 provides a unique test case, directly in the EoR, to hone estimators of LyC $f_{\rm{esc}}$.

COLA1 shows an average EW(\OIII +\Hbeta) when compared with \OIII-selected galaxies in the EIGER sample \citep{Matthee23a}. Using the grism fluxes of the \OIII +\Hbeta lines and subtracting them from the F356W photometry to get a direct estimation of the continuum we get a value of EW$_0$(\OIII +\Hbeta) $= 875 \pm 5$ \AA , which is very consistent, with the \texttt{Prospector} prediction of EW$_0$(\OIII +\Hbeta) $= 870^{+90}_{-80}$ \AA. The fitted stellar mass is $\log_{10}(M_* / M_\sun) = 9.93^{+0.07}_{-0.09}$, but we note that these errors may be underestimated as they are partly bounded by the range in models that are included in \texttt{Prospector}. For example, if we would allow more recent and or exotic models, such as new implementations of binary star physics \citep{Lecroq24} or a top heavy initial mass function \citep[e.g.,][]{Cameron23b}, it may well be possible that a somewhat lower stellar mass and star formation rate is required to reproduce COLA1s spectrum.

\begin{figure}
    \centering
    \includegraphics[width=\linewidth]{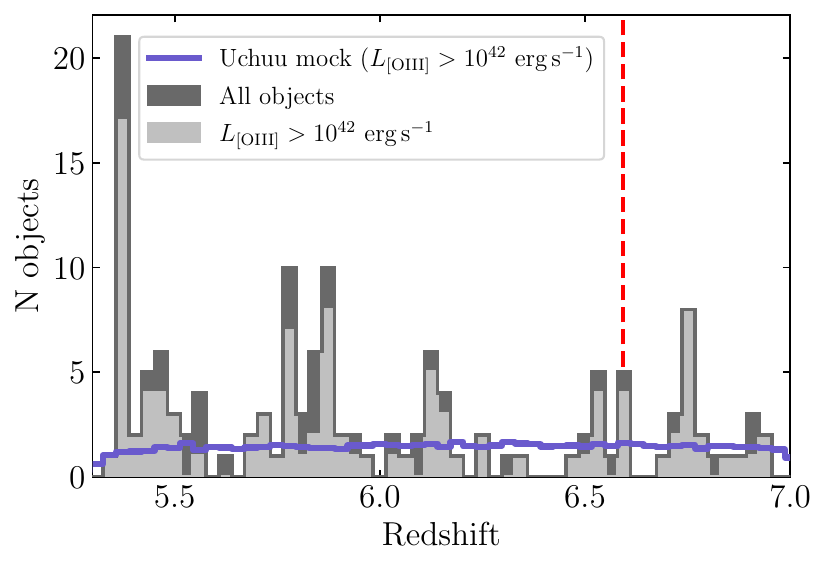}
    \caption{Redshift distribution of the objects selected via \OIII\ doublet search. The redshift of COLA1 is marked with a dashed red line. The blue solid line represents the average number of objects detected in the 1000 random Uchuu mocks, which converges to the average expectation value.}
    \label{fig:overdensity_histogram}
\end{figure}

\subsection{Galaxy mock catalogs in the COLA1 Field}\label{sec:mocks_description}

In order to identify and quantify over-densities of the detected \OIII\ emitters in our data (shown in Fig. $\ref{fig:overdensity_histogram}$), we need to account for the selection function. As illustrated in Fig.~\ref{fig:field_coverage}, the field of view and sensitivity are both nontrivially dependent on wavelength. Moreover, another obvious factor in our selection function is that we targeted JWST to center on the bright galaxy COLA1. Because of these complications, we compare our observed distribution of galaxies to detailed mock observed samples of \OIII\ emitters that are derived by forward-modeling simulations through our selection function (i.e., the error cube that is illustrated in bottom panel of Fig.~\ref{fig:field_coverage}). This methodology has been developed by Mackenzie et al. (in prep.) in the context of understanding the environments of luminous quasars in the EIGER survey (see also \citealt{Eilers24}), but is easily applicable to our survey. 

As detailed in Mackenzie et al. (in prep.), large N-body simulations are combined with the {\scshape UniverseMachine} model \citep{Behroozi19} to obtain the SFR of each simulated halo. Then, by matching the rank order of the SFR and the \OIII\ luminosity (the property on which we select galaxies), one can derive an empirical description to obtain the \OIII\ luminosity of each simulated halo that is calibrated to match the observed \OIII\ luminosity function at $z\approx6$. The observed \OIII\ luminosity function is an updated version of the one presented in \cite{Matthee23a}, with minor changes. Since the clustering strength and the \OIII\ to UV luminosity relations of the simulated galaxies match the observed counterparts, this method seems to place the right galaxies in the right halos. The method is applied to the Uchuu \citep{Ishiyama21} dark matter only simulation. The Uchuu simulation is among the largest, (2 Gpc/$h$)$^3$, and therefore contains the rarest structures and large samples of halos hosting galaxies like COLA1, at the expense of resolution (i.e., it only resolves galaxies with \OIII\ luminosity above $10^{42}$ erg s$^{-1}$, which is well above our sensitivity threshold). The median value of the halo mass for galaxies that are detected in the mocks is $\log_{10}(M_{\rm halo} / M_\sun)=10.8^{+0.3}_{-0.3}$.

Mock samples of \OIII\ emitters can be drawn from this simulation by first translating the error cube to a completeness cube, which is based on an empirically derived relation between the input S/N of the \OIII$_{4960}$ line and the completeness measured from injection experiments (Mackenzie et al. in prep). Then, the center of our observations is placed either in random locations of the simulations (so called `random mocks'), or on specifically chosen locations that center on a galaxy with a similar UV or \OIII\ luminosity as COLA1 ($M_{\rm UV}=-21.2$ to $-21.4$, or $L_{\rm \OIII}=10^{43.2\pm 0.2}$ erg\,s$^{-1}$) and similar redshift ($z=6.57 - 6.61$), so called `C1F mocks'. The random mocks are useful to understand any structure that we identified in the (independent) foreground or background of COLA1, as it incorporates realistic galaxy clustering, unlike galaxy samples that are purely randomly distributed following our completeness cube. The C1F mocks are useful to address whether the over-density around COLA1 is particularly special compared to galaxies with similar luminosities, which may be the case because of the specific Ly$\alpha$ double-peak selection.

\begin{figure}
    \centering
    \includegraphics[width=\linewidth]{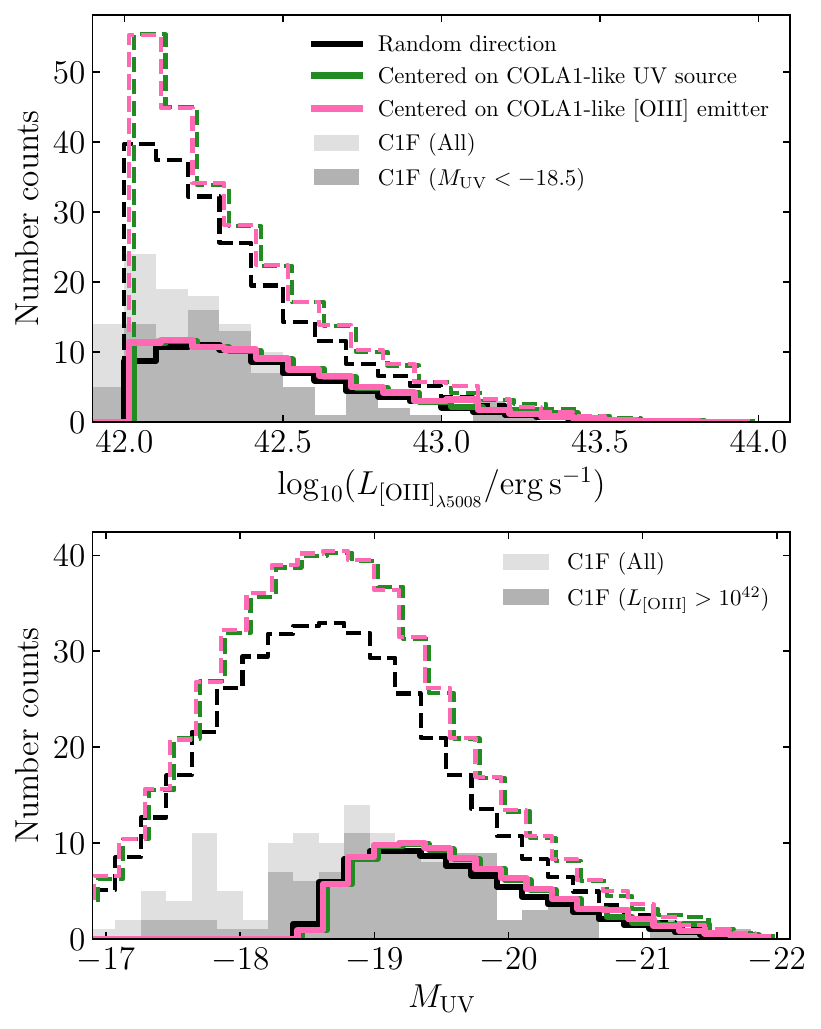}
    \caption{Distribution of \OIII$_{\lambda5008}$ luminosity and UV absolute magnitude of the Uchuu mocks and C1F. The black, pink and green histograms represent the distributions in Uchuu with random directions and centered in objects with similar $L_{\rm \OIII}$ and $M_{\rm UV}$ to COLA1, respectively. The dashed histograms represent all the objects in the field, and the solid ones only the detected objects. The bins of the three versions of the mock are slightly shifted for visual clarity. The solid gray histograms represent the distributions of the C1F \OIII\ emitters. The distributions of C1F match the ones of the Uchuu mocks, after running the detection algorithm.}
    \label{fig:mock_muv_o3_dist}
\end{figure}

In total, we draw 1000 mock galaxy samples (of each type) from the Uchuu simulations (stochastically accounting for the completeness function), which is the maximum number of fully independent mocks. As a validation of the mocks, we show that the distributions of \OIII\ luminosity and UV magnitude of the objects in the Uchuu mocks match relatively well the observed distributions in our sample (Fig.~\ref{fig:mock_muv_o3_dist}). This means that the \OIII\ LF and the \OIII\ - M$_{\rm UV}$ relation in the mocks, the EIGER and C1F data-sets are consistent with each other.  Generally, these distributions appear to be independent of the pointing direction choice (i.e., whether the mock is centering in a random position or towards a COLA1-like object).

We compare the redshift distribution of our sample of 141 \OIII\ emitters with the mocks described above. Due to resolution limitations of the Uchuu simulation, we compare the distribution of objects in our catalog with \OIII\ luminosity greater than $10^{42}$ erg\,s$^{-1}$ with the average number of detected objects in the mocks. The peaks in the light gray histogram of Fig.~\ref{fig:overdensity_histogram} that are higher than the average number of detections in the mocks are over-densities. There is a peak of galaxies around the redshift of COLA1: we find 6 objects with $z = 6.60 \pm 0.02$ in the sample, including COLA1 itself (4 objects if we limit the sample to $L_{\rm\OIII} > 10^{42}$ erg\,s$^{-1}$; $\delta + 1 = 1.96$)\footnote{$\delta + 1 = N / \langle N \rangle $, where $N$ number of detected objects in C1F in a given redshift interval, and $\langle N \rangle$ the average number of detected objects in the mocks.}. Interestingly, we find much more significant and stronger over-densities at other redshifts in the C1F (e.g., at $z = 6.74,\ 6.14,\ 5.86,\ 5.78,\ 5.46,\ 5.38\,(\pm 0.02)$, with $\delta +1 = 4.6,\ 3.2,\ 4.5,\ 4.3,\ 4.5,\ 11.0$, respectively). We analyze the over-density around COLA1 in detail in Section $\ref{sec:O3_sample_spatial_structure}$.

\section{The properties of COLA1: all signs of a luminous Lyman Continuum leaker}\label{sec:results_1} 

In this section, we focus on the properties of COLA1 based on our JWST data, and previous \lya data that can now be better interpreted. We confirm the double-peak nature of the \lya line from the systemic redshift (Sect.~\ref{sec:lya_profile}), study the ISM conditions from the optical emission line measurements (Sect.~\ref{sec:COLA1_ism_properties}) and the UV properties (Sect.~\ref{sec:uv_properties}); we estimate the star-formation rate and surface density (Sect.~\ref{sec:star_formation}), and the escape fraction of \lya and LyC (Sect.~\ref{sec:escape_frac}). We also investigate all other \OIII\ emitters in our dataset, with the intention to compare COLA1 with the galaxies in its environment, and quantify how special it is in comparison with these objects. Table~\ref{tab:COLA1_properties} summarizes the main properties of COLA1.

\subsection{Confirmation of the Ly$\alpha$ systemic redshift}\label{sec:lya_profile}

Cosmological radiation hydrodynamical reionization simulations predict a typical negligible transmission of the \lya blue peak for $z>6$ \citep[e.g.,][]{Weinberger18,Gronke21}. Hence, the presence of a strong blue peak at $z\sim 6.6$ is very unlikely. The \lya profile of COLA1 is shown in Fig.~\ref{fig:Lya_profile}. The main hypothesis described in \cite{Matthee18} suggests that COLA1 is located in a highly ionized region (see also \citealt{Hu16, Mason&Gronke20}), with a large enough size for the \lya photons to be redshifted out of resonance before reaching a relatively neutral IGM that would scatter them. Our measured systemic redshift using H$\beta$ and \OIII\ unambiguously confirm that the observed \lya profile is indeed a classic double-peaked \lya line, with the systemic redshift in between the two peaks \citep[e.g.,][]{Verhamme06,Gronke15}.

\subsection{Interstellar medium conditions from optical emission lines}\label{sec:COLA1_ism_properties}

\begin{table}
\renewcommand{\arraystretch}{1.25}
    \centering
    \caption{The physical properties of COLA1.}
    \label{tab:COLA1_properties}
    \begin{tabular}{ll}
         \toprule
         Property & Value \\
         \midrule
         $z$$^a$ & 6.59165\\
         $\log_{10}(\xi_{\rm ion, 0} / {\rm Hz\, erg}^{-1})$$^b$ & $25.45^{+0.04}_{-0.05}$\\
         $f_{\rm esc}({\mathrm{Ly}\alpha})$$^c$ & $81\pm5$\%\\
         $M_{\rm UV}$$^d$ & $-21.35^{+0.07}_{-0.08}$\\
         $\beta_{\rm UV}$$^e$ & $-3.2\pm 0.4$\\
         $E(B - V)$$^f$ & $0.00^{+0.02}_{-0.00}$ \\
         $T_{\rm e}$$^g$ & $1.7^{+0.4}_{-0.3} \times 10^{4}$ K \\
         EW$_0$(\OIII +\Hbeta)$^h$ & $870^{+90}_{-80}$\\
         $12+\log_{10}\left(\text{O/H}\right)_{T_e}$$^i$ & $7.88^{+0.33}_{-0.30}$ \\
         $R_{\rm UV}$$^j$ & $<$$0.26$ kpc\\
         SFR$_0$(UV)$^k$ & $9.6^{+1.7}_{-0.8}$ $M_\sun$\,yr$^{-1}$ \\
         SFR$_0$(\Hbeta)$^l$ & $10.1^{+0.9}_{-0.5}$ $M_\sun$\,yr$^{-1}$ \\
         $\log_{10}(\Sigma_{\rm SFR} / M_\sun\,{\rm yr}^{-1}\,{\rm kpc}^{-2})$(UV)$^m$ & >$1.31$ \\
         $\log_{10}(\Sigma_{\rm SFR} / M_\sun\,{\rm yr}^{-1}\,{\rm kpc}^{-2})$(\Hbeta )$^n$ & >$1.36$ \\
         \bottomrule
    \end{tabular}
    \tablefoot{$^a$ Redshift, $^b$ ionizing photon production efficiency assuming $f_{\rm esc}=0$, $^c$ \lya escape fraction, $^d$ 1500 \AA\ absolute magnitude, $^e$ power-law slope at 1500 \AA , $^f$ dust attenuation coefficient, $^g$ electron temperature, $^h$ \OIII + \Hbeta rest-frame equivalent width, $^i$ direct measurement metallicity, $^j$ F200W half-light radius, $^{k, l}$ star-formation rate assuming $f_{\rm esc}=0$ derived from $M_{\rm UV}$ and \Hbeta , respectively; $^{m, n}$ star formation rate surface density derived from both SFR indicators.}
\end{table}

The detection of the Balmer lines \Hbeta and H$\gamma$ permit to directly estimate the spectral attenuation due to dust.
Assuming an intrinsic ratio of H$\gamma$/\Hbeta $=0.47$ \citep{Miller74}, we measure $E(B - V)=0.00^{+0.02}_{-0.00}$ for COLA1 (H$\gamma$/H$\beta = 0.52\pm0.05$), whereas the value for the stack of the EIGER \OIII\ emitter sample at $z>6.25$ is $E(B - V)=0.14^{+0.16}_{-0.14}$, for comparison \citep{Matthee23a}. Following a standard \cite{Cardelli89} nebular attenuation curve \citep[see also][]{Reddy20} leads to negligible dust attenuation in COLA1.

Combining the \OIII$_{\lambda 4364}$ fluxes of modules A and B we obtain a >$3\sigma$ detection of $f($\OIII$_{\lambda 4364}) = (0.77 \pm 0.24) \times 10^{-18}$ erg\,s$^{-1}$\,cm$^{-2}$. The detection of the \OIII$_{\lambda 4364}$ line in COLA1 allows for a direct measurement of the electron temperature, based on the \OIII$_{\lambda 4364}$/\OIII$_{\lambda 5008}$ ratio. For this we use the \texttt{PyNeb} Python package \citep{Luridiana15}. For estimating the temperature, we assume an electron number density of $n_e=300$ cm$^{-3}$ \citep[e.g.,][]{Curti23}. We obtain $T_{\rm e} = 1.7^{+0.4}_{-0.3} \times 10^{4}$ K for COLA1. The measured $T_{\rm e}$ for COLA1 is compatible with the typical values for galaxies at similar redshifts in the literature \citep[e.g.,][]{Izotov20, Matthee23a, Katz23, Nakajima23, Hu24}.

We estimate the gas-phase metallicity of COLA1 from the \OIII /\Hbeta ratio following \cite{Pilyugin06}. Since there is no direct measurement of any [\ion{O}{II}] line, we assume a ratio of \OIII/[\ion{O}{II}] $=8\pm 3$, based on empirical results \citep[e.g.,][]{Reddy18, Katz23} and following \cite{Matthee23a}. We obtain $12+\log_{10}\left(\text{O/H}\right)=7.88^{+0.33}_{-0.30}$. The obtained metallicity is consistent with the average values observed in galaxies at similar redshifts \citep[e.g.,][]{Nakajima23}.  Under extreme ionization conditions, the \OIII/[\ion{O}{II}] ratio can be much higher; however, our metallicity estimation varies only slightly ($<0.1$ dex) if assumed ratios as high as $\sim 180$ \citep[e.g.,][]{Topping24}.

\subsection{Ultraviolet properties}\label{sec:uv_properties}

\begin{figure}
    \centering
    \includegraphics[width=\linewidth]{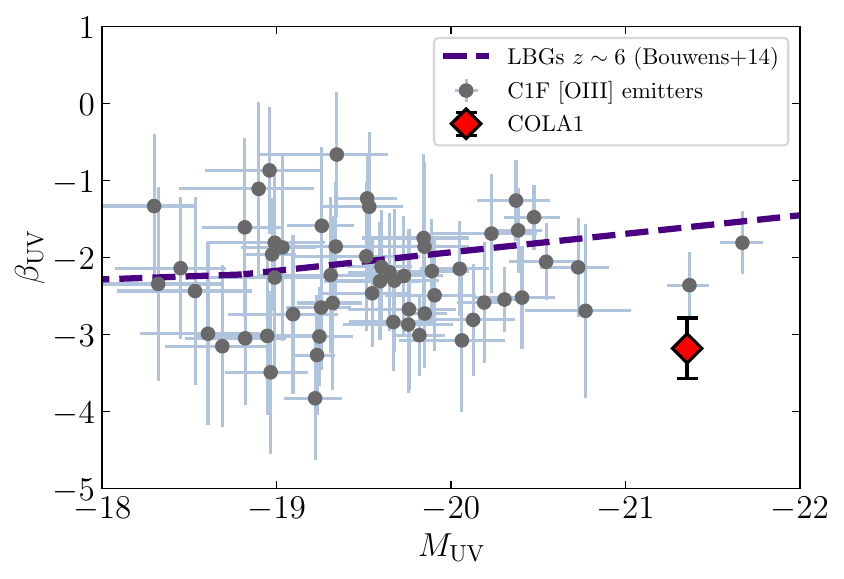}
    \caption{Observed UV slope measured by fitting a power law spectrum to the F115W and F150W photometry. The shown $M_{\rm UV}$ is computed from the best-fit used to obtain $\beta_{\rm UV}$. COLA1 presents a steep slope in comparison with other bright \OIII\ emitters in C1F. Only objects with S/N $>5$ in F115W and F150W are shown. We also compare with the estimated average $\beta_{\rm UV}$ in \protect\cite{Bouwens14} for LBGs at $z\sim 6$.}
    \label{fig:beta_powerlaw}
\end{figure}

The UV slope $\beta_{\rm UV}$, defined as the power-law slope of the UV continuum at $\lambda_{\rm rest}=1500$ \AA , is often used as a relevant diagnostic to characterize the UV properties of galaxies \citep[e.g.,][]{Wilkins12, Wilkins13}. In our data, the F115W-F150W color can be employed to probe $\beta_{\rm UV}$ using a simple power-law fit\footnote{At $z=6.6$, $\lambda_{\rm obs,  1500 \AA}=11400$ \AA.}. This method yields $\beta_{\rm UV}=-3.2\pm0.4$ for COLA1, which is steep compared to the rest of $M_{\rm UV}$-bright \OIII\ emitters in C1F (Fig.~\ref{fig:beta_powerlaw}), and with the average value for $z\sim 6$ \citep[][further discussion in Sect.~\ref{sec:COLA1_as_a_leaker}]{Bouwens14}. The UV absolute magnitudes shown in Fig.~\ref{fig:beta_powerlaw} are computed extrapolating the best power-law fit.

\begin{figure}
    \centering
    \includegraphics[width=\linewidth]{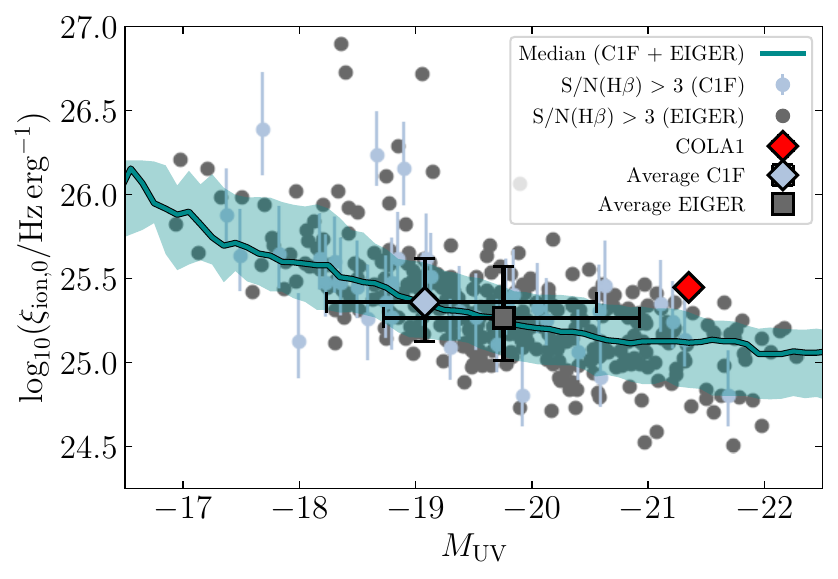}
    \caption{Ionizing photon production efficiency obtained from the \Hbeta fluxes. We compare the C1F \OIII\ emitters with the EIGER sample. The blue solid line shows the running median of the C1F+EIGER sample. COLA1 presents a significantly larger $\xi_{\rm ion, 0}$ than the C1F+EIGER median. Only objects with S/N(\Hbeta ) $> 3$ are shown. The errors of the COLA1 measurement are smaller than the marker size and the errors of the EIGER sample are omitted for visual clarity.}
    \label{fig:xiion_Hbeta}
\end{figure}

The ionizing photon production efficiency $\xi_{\rm ion, 0}$ can be inferred from the measurement of \Hbeta  as $\xi_{\rm ion, 0} = L_{{\rm H}\beta} / (c_{{\rm H}\beta} \cdot L^\lambda_{{\rm UV}})$, with the \Hbeta line coefficient $c_{{\rm H}\beta}=4.79\times 10^{-13}$ erg, assuming case B recombination, electron temperature of $\sim 10^4$ K and zero escape fraction of ionizing photons \citep{Schaerer03}. We obtain $\log_{10}(\xi_{\rm ion, 0} / {\rm Hz\, erg}^{-1})=25.45^{+0.05}_{-0.06}$ for COLA1, assuming negligible dust attenuation from the high H$\gamma$/\Hbeta ratio. In Fig.~\ref{fig:xiion_Hbeta} we show that the estimated $\xi_{\rm ion}$ for COLA1 is above the median values for similar UV magnitudes in the C1F+EIGER sample.

\subsection{Star-formation rate and $\Sigma_{\rm SFR}$}\label{sec:star_formation}

The star formation rate (SFR) can be inferred from the relation between the ionizing photon efficiency and the luminosity of H$\alpha$ (or alternatively \Hbeta), and from the UV luminosity $L_{\nu, {\rm UV}}$. From Table~2 in \cite{Theios19} we use conversion factors of $10^{41.78}$ and 
$10^{43.51}$ erg\,s$^{-1}$/$M_\sun$\,yr$^{-1}$ for $L_{{\rm H}\alpha}$ and $\nu L_{\nu, {\rm UV}}$, respectively.
which are appropriate for the ionizing photon efficiencies of the stellar populations in high-redshift galaxies. We obtain ${\rm SFR_0(H}\beta) = 10.1^{+0.9}_{-0.5}$ $M_\sun$\,yr$^{-1}$ and ${\rm SFR_0(UV) = 9.6^{+1.6}_{-0.8}}$ $M_\sun$\,yr$^{-1}$. The median ${\rm SFR}_0$(\Hbeta ) and ${\rm SFR}_0$(UV) of the complete C1F sample are $3^{+5}_{-1}$ and $3^{+3}_{-2}\ M_\sun\,{\rm yr}^{-1}$, respectively ($4^{+7}_{-1}$ and $4^{+6}_{-2}\ M_\sun\,{\rm yr}^{-1}$ if we limit the sample to S/N(\Hbeta ) $>3$).

The star-formation rate surface density is usually defined as $\Sigma_{\rm SFR} = \frac{\rm SFR / 2}{\pi R_{\rm UV}^2}$ where $R_{\rm UV}$ is the effective UV half-light radius of the galaxy \citep[e.g.,][]{Naidu20}. In order to measure the effective radius of the C1F objects, we run \texttt{SExtractor} on the F200W image to obtain \texttt{fwhm\_world}. Then, we stack 15 visually selected luminous stars, and fit a 2D Gaussian to the stack to estimate the average PSF full width at half maximum (FWHM). Finally we estimate the intrinsic FWHM as ${\rm FHWM}_{\rm int}^2 = {\rm FHWM}_{\rm obs}^2 - {\rm FHWM}_{\rm star}^2$, valid for ${\rm FHWM}_{\rm obs}^2 \gg {\rm FHWM}_{\rm star}^2$. We compute the UV radius as the physical distance corresponding to FWHM$_{\rm int}/2$\footnote{Here we assume that the light profile of COLA1 is described by a Gaussian (Sérsic profile with $n=1/2$). However, the measurement of the effective radius could vary under the assumption of different Sérsic indices.}. The median value for the complete C1F sample is $R_{\rm UV, C1F}=0.7^{+0.4}_{-0.2}$ kpc. We find that COLA1 is unresolved in the F200W image, with FWHM$_{\rm obs}=0.094$ arcsec, essentially indistinguishable from a stellar FWHM. Hence the above approximation is not valid for COLA1. This result only allows to set an upper bound to the UV size of COLA1 based on the stellar PSF, $R_{{\rm UV,COLA1}}<0.26$ kpc. This UV size is smaller than the Ly$\alpha$ size measured in \cite{Matthee18}, $R_{{\rm Ly}\alpha}=0.38$ kpc, which is common in observed Ly$\alpha$ emitters across a variety of redshifts \citep[e.g.,][]{Leclercq17}.

The upper bound in UV radius of COLA1 leads to a lower bound in the star-formation rate surface density, $\log_{10}(\Sigma_{\rm SFR} / M_\sun\,{\rm yr}^{-1}\,{\rm kpc}^{-2}) >$$1.36$ and $>$$1.31$, using SFR$_0$(\Hbeta) and SFR$_0$(UV) conversions, respectively. These lower bounds are considerably high compared to the respective median values for the C1F \OIII\ emitters with with S/N(\Hbeta) $>3$: $\log_{10}(\Sigma_{\rm SFR} / M_\sun\,{\rm yr}^{-1}\,{\rm kpc}^{-2}) = 0.3^{+0.4}_{-0.4}$ and $0.1^{+0.4}_{-0.2}$.

\begin{figure}
    \centering
    \includegraphics[width=\linewidth]{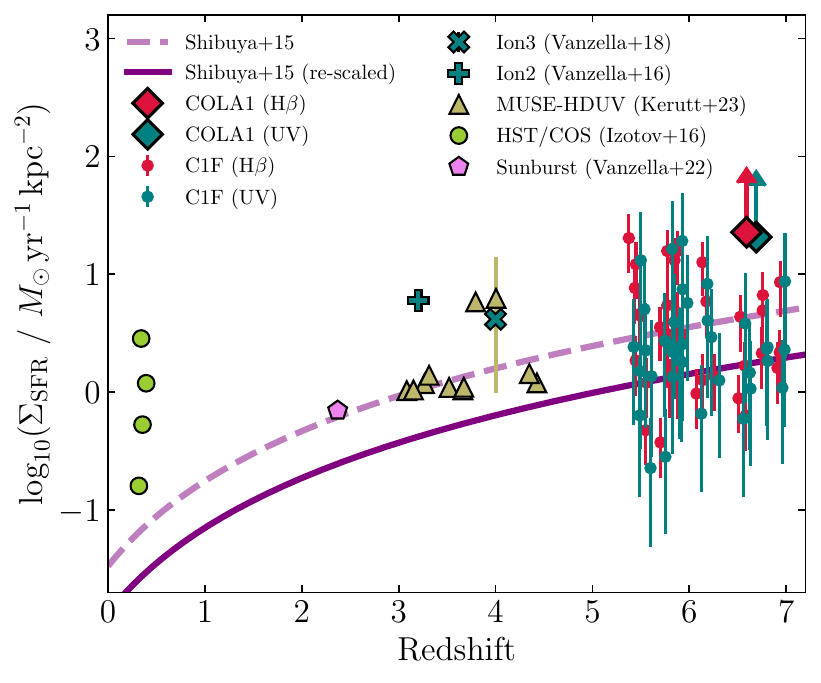}
    \caption{Star-formation rate surface density $\Sigma_{\rm SFR}$ of the C1F \OIII\ emitter sample. We show the values obtained by from SFR(UV) and SFR(\Hbeta) for every C1F object with S/N(\Hbeta) > 3. The points of C1F's SFR(\Hbeta) are slightly shifted in the horizontal axis for clarity. We compare our results with the best-fit in \protect\cite{Shibuya15}, which assumes SFR $= 10\,M_\sun$\,yr$^{-1}$ and a re-scaled version of the curve in \protect\cite{Shibuya15} that corresponds to an average of SFR $= 4\,M_\sun$\,yr$^{-1}$, as observed in the C1F+EIGER sample. We also show $\Sigma_{\rm SFR}$ for known LyC leakers in the literature \protect\citep{Izotov16, Vanzella16, Vanzella18, Vanzella22, Kerutt23}, for which we have estimated SFR(UV) with the same calibration used for COLA1. For COLA1, we show lower limits based on the upper limit of the UV half-light radius.}
    \label{fig:Sigma_SFR}
\end{figure}

In Fig.~\ref{fig:Sigma_SFR} we show the star-formation surface density lower bound measurements for COLA1 in comparison with the values for the C1F \OIII\ emitters with S/N(\Hbeta ) > 3. We compare our measurements with the observed $\Sigma_{\rm SFR}$ in \cite{Shibuya15}. For the other C1F objects we have used SFR scaling factors in \cite{Theios19} assuming a typical $\log_{10}\xi_{\rm ion}=25.3$, and dust attenuation correction corresponding to H$\gamma$/H$\beta = 0.416 \pm 0.075$, as observed in the EIGER \OIII\ sample. COLA1 presents a significantly higher $\Sigma_{\rm SFR}$ than the rest of C1F objects albeit its high $\xi_{\rm ion, 0}$ and negligible dust attenuation, both of those features implying lower $M_{\rm UV}$ or H$\beta$ to SFR conversions.

\subsection{Escape fraction of \lya and LyC photons}\label{sec:escape_frac}

The Lyman-alpha escape fraction, $f_{\rm esc}$(\lya ) is defined as the ratio between the observed \lya flux of a galaxy and the total \lya produced. We estimate the $f_{\rm esc}$(\lya) from the flux measurement of \Hbeta . Assuming case B recombination and an electron temperature of $10^4$ K, \lya /H$\alpha = 8.7$, and H$\alpha$/\Hbeta $ = 2.86$ \citep{Osterbrock89}. 
Hence, considering the measured average flux of \Hbeta in grism modules A and B, and the \lya flux from \cite{Matthee18}, we obtain $f_{\rm esc}{{\rm (Ly}\alpha )} = 81\pm5\%$. Here we have assumed negligible dust attenuation (see Sect.~\ref{sec:COLA1_ism_properties}). However, based on simulations, \cite{Choustikov24} argue that the \lya escape fractions estimated following this method are biased due to the assumptions of dust attenuation law, fixed, constant \lya /H$\alpha$ ratio and ignoring collisional \lya radiation. In particular, if considered that collisional emissivity contributes to $\sim 25\%$ of the \lya radiation, the estimate of $f_{\rm esc}$(\lya ) is biased 0.1 dex \citep[see][and references therein]{Choustikov24}. Correcting for this effect would lead to $f_{\rm esc}$(\lya ) $= 64 \pm 4\%$.


Several observational surveys and hydrodynamical simulations in the literature have investigated the correlation between the escape fraction of \lya and that of LyC photons \citep[e.g.,][]{Verhamme17}. In \cite{Begley24}, a sample of 152 star forming galaxies with $z\approx 4$--$5$ is used to obtain a linear dependence between $f_{\rm esc}({\rm LyC})$ and $f_{\rm esc}({\rm Ly}\alpha)$ inferred from H$\alpha$ measurements assuming a \lya /H$\alpha = 8.7$ ratio. Using this relation we estimate, for COLA1, $f_{\rm esc}({\rm LyC}) = 12 \pm 4 \%$. In \cite{Maji22}, hydrodynamical and radiative transfer simulations are used to calibrate the same relation, suggesting $f_{\rm esc}({\rm LyC}) = 56 \pm 7 \%$ ($39 \pm 5 \%$ if we correct for collisional \lya emission) for COLA1. Similarly, \cite{Kimm22} fitted a power-law dependence from giant molecular cloud simulations, implying $f_{\rm esc}({\rm LyC}) = 45 \pm 10 \%$ ($19 \pm 5 \%$ if we correct for collisional \lya emission) for COLA1.

Empirically, the LyC escape fraction is correlated with the peak separation of the \lya line \citep[e.g.,][]{Izotov18}. The peak separation of $\Delta v = 220 \pm 20$ km\,s$^{-1}$ observed in the VLT/X-shooter spectrum suggests $f_{\rm esc}({\rm LyC}) = 28^{+10}_{-6} \%$ \citep[see][]{Matthee18}. Furthermore, \cite{Naidu20} used a bayesian inference model to fit a power-law relation between the LyC escape fraction and the star formation rate surface density, $\Sigma_{\rm SFR}$, from observational constrains of the reionization timeline and the galaxy size evolution. Following this relation, we obtain a lower limit of $f_{\rm esc}$(LyC) $>44\%$. Lastly, \cite{Chisholm22} found a relation between the UV slope and the LyC escape fraction, for a sample of SFGs at $z\sim 0.3$ with $\beta_{\rm UV}$ up to $-2.5$. We extrapolate their best fit to get $f_{\rm esc}$(LyC) $=84^{+277}_{-65}\%$, hence a lower $1\sigma$ limit of $f_{\rm esc}$(LyC) $>20\%$.

The Ly$\alpha$ central escape fraction ($f_{\rm cen}$) is introduced in \cite{Naidu22} as a diagnostic of LyC leakage, defined as the fraction of Ly$\alpha$ flux emitted within $\pm 100$ km\,s$^{-1}$ from the systemic Ly$\alpha$ velocity. The Ly$\alpha$ central fraction traces the presence of low opacity channels that allow the direct escape of \lya . From the \lya spectrum of COLA1 (see Fig.~\ref{fig:Lya_profile}) we obtain $f_{\rm cen}\approx 25$\%. Recent radiative-transfer simulations have found that sources with $f_{\rm cen}\gtrsim10\%$ are more likely to be LyC leakers \citep{Choustikov24}. Direct comparison with the "High Escape" stack in \cite{Naidu22} suggests $f_{\rm esc}{\rm (LyC)}>20\%$.

All the $f_{\rm esc}$(LyC) estimations for COLA1 are summarized in Table~\ref{tab:fesc_LyC}. Apart from the calibration to the \lya escape fraction from \cite{Begley24}\footnote{We note, however, that almost every object in the sample of \cite{Begley24} has $f_{\rm esc}$({\lya }) $< 50\%$, so our estimated LyC escape fraction based on this calibration is subject to a significant extrapolation.}, all estimates that are tied to COLA1s \lya emission indicate that it has a very high escape fraction of ionizing photons.

\begin{table}
\renewcommand{\arraystretch}{1.25}
    \centering
    \caption{Indirect estimates of the LyC escape fraction of COLA1 derived from various empirical and theoretical methods.}
    \label{tab:fesc_LyC}
    \begin{tabular}{lcl}
         \toprule
         $f_{\rm esc}$(LyC) & Method & Reference \\
         \midrule
         $28^{+10}_{-6}\%$ & \lya peak separation & \cite{Izotov18} \\
         $12 \pm 4\%^\dag$ & $f_{\rm esc}$(\lya ) & \cite{Begley24} \\
         $56 \pm 7\%$ & $f_{\rm esc}$(\lya ) & \cite{Maji22} \\
         $45 \pm 10\%$ & $f_{\rm esc}$(\lya ) & \cite{Kimm22} \\
         $>20\%$ & $f_{\rm cen}({\rm Ly}\alpha)$ & \cite{Naidu22}\\
         $>44\%$ & $\Sigma_{\rm SFR}$ & \cite{Naidu20} \\
         $84^{+277}_{-65}\%^\dag$ & $\beta_{\rm UV}$ & \cite{Chisholm22} \\
         \bottomrule
        \end{tabular}
    \tablefoot{$^\dag$ These estimates are subject to significant extrapolation from the parent samples used to obtain the corresponding relations.}
\end{table}

 \section{The environment of COLA1: a normal over-density for its UV and \OIII\ luminosity}\label{sec:O3_sample_spatial_structure}

\begin{figure*}
    \centering
    \includegraphics[width=\linewidth]{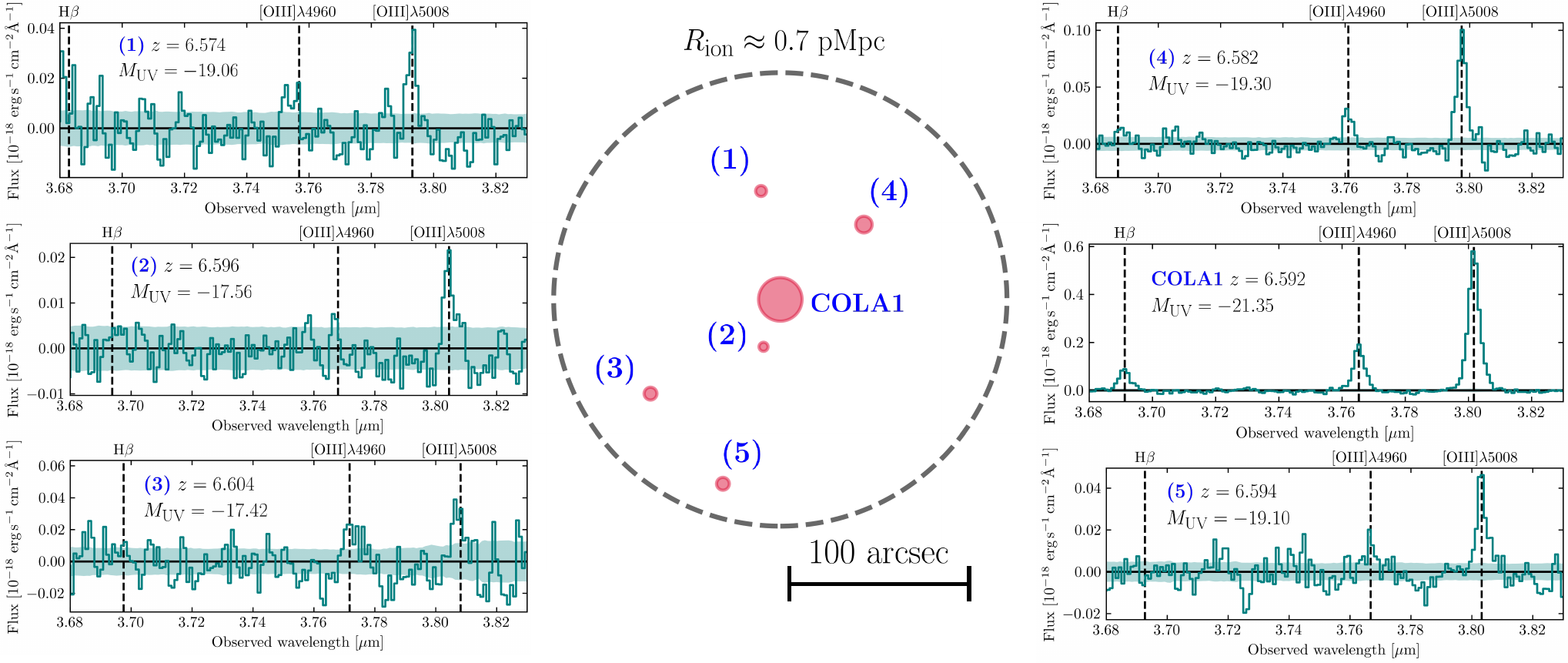}
    \caption{Projected sky positions of COLA1 and the other \OIII\ emitters with $z=z_{\rm COLA1}\pm 0.02$ in our sample (red circles). The size of the marker is proportional to the total \OIII\ flux of each source. The dashed line marks the estimated ionized bubble size \citep{Mason&Gronke20}, assuming a spherically symmetric bubble. All the sources found in the environment of COLA1 have projected distances smaller than the bubble radius, suggesting that they could be located inside the bubble. We show the 1D spectra of all the sources, in the observed wavelength range of the \OIII\ doublet and H$\beta$. The blue shaded regions mark the 1$\sigma$ confidence interval of the spectroscopic fluxes. False-color images and 2D grism spectra of these sources  can be found in Appendix~\ref{sec:example_spec}.}
    \label{fig:COLA1_overdensity}
\end{figure*}

During the EoR, galaxy over-densities are associated with an increased Ly$\alpha$ transmission \citep[e.g.,][]{Endsley22, Leonova22, Kashino23, Wistok23}, as these over-densities are more likely to form ionized bubbles, according to simulations \citep{Ocvirk20, Hutter21, Qin22, Lu23}. As sight-lines that allow for the detection of blue Ly$\alpha$ photons at $z>6$ are expected to be extremely rare \citep[e.g.,][]{Gronke21}, one could therefore expect that galaxies that show a double-peaked Ly$\alpha$ line are surrounded by an unusually high over-density.

To address exactly this question, we analyzed the over-density around COLA1 and compared it with results from mock observations of a simulation, with the methodology we described in Sect.~\ref{sec:mocks_description}. We identify five \OIII\ emitters in our sample that are close to COLA1 in redshift space ($\Delta z < 0.02$; three of which have $L_{\rm \OIII } > 10^{42}$ erg s$^{-1}$). These galaxies are all considerably fainter than COLA1, spanning UV luminosities M$_{\rm UV}=-17.4$ to M$_{\rm UV}=-19.3$. Notably, as is shown in Fig.~\ref{fig:COLA1_overdensity} the sky position of all those objects fall within COLA1's inferred ionized bubble radius \citep{Matthee18,Mason&Gronke20}, assuming a spherically symmetrical bubble, meaning that they could potentially be located within the spatial extent of the bubble. By comparing the number of identified sources with the expectation from pure random mocks (Sect.~\ref{sec:mocks_description}), we find that COLA1 is located in a mild over-density ($\delta + 1 = 1.96$). The typical halo mass for galaxies in our mocks that are surrounded by a similar number of galaxies as COLA1 is $\log_{10}(M_{\rm halo} / M_\sun)\approx11.3$.
 
\begin{figure}
    \centering
    \includegraphics[width=\linewidth]{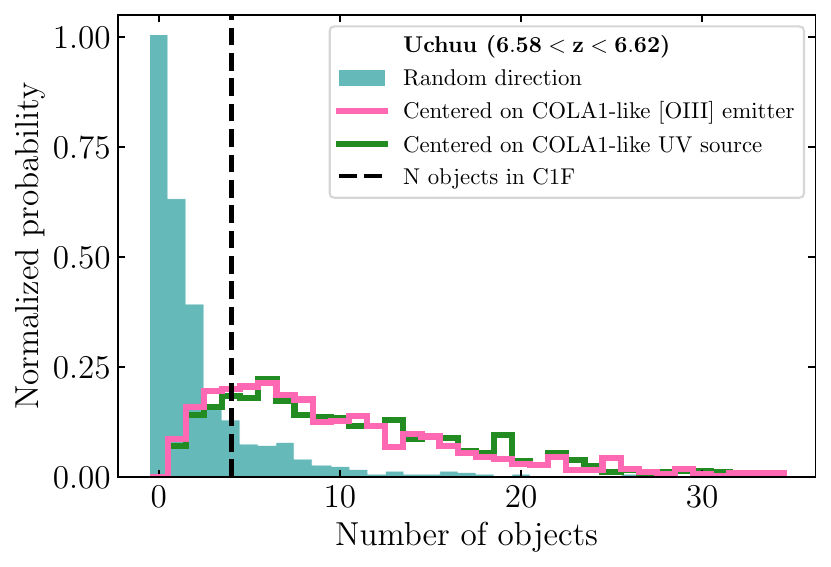}
    \caption{ Normalized relative probabilities of detecting a given number of objects with $6.58<z<6.62$ in the Uchuu mocks, compared to the number of galaxies found in the COLA1 field. We show the probability distributions for the Uchuu mocks with random pointings (solid teal histogram) and those centered in objects with similar $L_{\rm \OIII}$ and $M_{\rm UV}$ to COLA1 (pink and green, respectively).}
    \label{fig:C1F_overdensity_analysis}
\end{figure}

In Fig.~\ref{fig:C1F_overdensity_analysis} we compare the number count in the C1F \OIII\ emitter sample at $6.58<z<6.62$ ($\pm 0.02$ around the redshift of COLA1) with the number count probability distribution in the mocks specifically designed to mimic our observing strategy of centering on a UV-luminous galaxies. While the number count of the C1F sample in this narrow redshift interval is at the 84th percentile of the pure random Uchuu mocks, the number of detected galaxies in C1F around the redshift of COLA1 is very typical when comparing it to those in mocks centered around galaxies with COLA1's UV and \OIII\ luminosity.

This analysis shows that, while we do identify an over-density around COLA1, its amplitude is typical for any galaxy that is similarly UV bright at $z\sim6.5$. Thus, there is no strong indication that the detection of the double-peaked Ly$\alpha$ emission is due to an exceptionally large over-density of galaxies. Also, we remark that Fig.~\ref{fig:C1F_overdensity_analysis} shows that the possible number of neighbors that one could detected around a similarly UV bright galaxy with a similar emission-line survey on a field of $\approx21$ arcmin$^2$ has a wide distribution. This displays the wide variety of environments that could host similarly UV bright \citep[$\sim1.5 L^{\star}$; see e.g.,][]{Bouwens15} galaxies, and therefore stressing the importance of taking cosmic variance of the large scale structure into account when interpreting Ly$\alpha$ observations of galaxies in the context of reionization.

\section{Discussion}\label{sec:discussion}

In this section, we discuss the results presented in Sections~\ref{sec:results_1} and $\ref{sec:O3_sample_spatial_structure}$. We comment the measured properties of COLA1 in comparison with other SFGs and LyC leakers in the literature (Sect.~\ref{sec:COLA1_as_a_leaker}). Our results confirm the systemic redshift of COLA1, which means it must reside inside an ionized bubble (Sect.~\ref{sec:bubble_size}). We discuss the role of COLA1 and the galaxies in its proximity in ionizing their environment (Sect.~\ref{sec:COLA1_ionizing_power}) and we address the possible contribution of an AGN to COLA1's luminosity (Sect.~\ref{sec:agn}). Finally, we comment on some directions that can be followed in future works to further investigate the nature of COLA1 (Sect.~\ref{sec:future_directions}).

\subsection{COLA1 as an ionizing photon leaker}\label{sec:COLA1_as_a_leaker}

COLA1 presents an extremely blue UV slope (see Table~\ref{tab:COLA1_properties}), far bluer than the typical values of the C1F+EIGER sample and the best fit in \cite{Bouwens14} for Lyman-break galaxies at $z\sim 6$ (see Fig.~\ref{fig:beta_powerlaw}). \cite{Topping23} presented a sample of 44 galaxies with steep UV slopes ($\beta_{\rm UV} < -2.8$), and found that the SEDs of such galaxies could be explained with density-bounded models with high escape fractions ($f_{\rm esc}{\rm (LyC)} \sim 50\%$) \citep[see also][]{Topping22}. Further, \cite{Kim23} found a relation between very steep $\beta_{\rm UV}$ and LyC leaking regions of a $z=2.37$ lensed galaxy. Indeed, our estimation based on \cite{Chisholm22} using $\beta_{\rm UV}$ suggests a very high $f_{\rm esc}{\rm (LyC)}$ (see Table~\ref{tab:fesc_LyC}).

The rest-frame \OIII +\Hbeta EW of COLA1 is comparable to the typical values for galaxies with similar magnitudes in UV-selected \citep[e.g.,][]{Endsley23} and \OIII\ selected samples \citep[e.g.,][]{Matthee23a}.
The intrinsic \OIII\ and \Hbeta luminosities are remarkably high, $\log_{10} (L_{{\rm H}\beta} / {\rm erg\,s}^{-1}) = 42.33\pm 0.02$ and $\log_{10} (L_{{\rm \OIII}{\lambda5008}} / {\rm erg\,s}^{-1}) = 43.124\pm 0.007$. The high \OIII\ luminosity can be related to an elevated star formation \citep{Villa-Velez21}, as well as the high ratio of $\log_{10}({\rm \OIII}/{{\rm H}\beta}) = 0.91$ \citep{Dickey16}.

COLA1 presents significantly higher $\xi_{\rm ion, 0}$ than the average C1F+EIGER values (see Fig.~\ref{fig:xiion_Hbeta}), consistent with the measurements for high-$z$ SFGs in the literature \citep[e.g.,][]{Bouwens16, Nakajima18, Saxena23, Simmonds23, Simmonds24} and also known low-$z$ LyC leakers \citep[e.g.,][]{Izotov16}. \cite{Izotov21} showed that low-$z$ compact SFGs are likely analogs of high-$z$ SFGs. More recently, compact SFGs have been found at $z\gtrsim 6$ with similar sizes to COLA1, presenting signs of LyC leakage \citep{Mascia23b,Mascia23a}.

As stated in Sect.~\ref{sec:escape_frac}, the direct measurement of \Hbeta allows for an estimation of the \lya escape fraction, yielding a significantly high value of $f_{\rm esc}$(\lya ) $=81\pm 5\%$. This Ly$\alpha$ escape fraction is high in comparison with the typical values found in high redshift surveys (e.g., $f_{\rm esc}$(\lya ) $=11$\%, for Ly$\alpha$ selected sample at $z\sim 4.5$, \citealt{Roy23}; $f_{\rm esc}$(\lya ) $=7.3$\%, $z\sim 7.5$, mean $M_{\rm UV}=-20.25$, \citealt{Tang23}), although other objects with similarly high \lya escape fractions have recently been found for fainter systems ($f_{\rm esc}$(\lya ) $>50$\%, $M_{\rm UV}\sim-19.5$, $z\sim5$--$7$, \citealt{Chen23}; $f_{\rm esc}$(\lya ) $>70$\%, $M_{\rm UV}\sim-17$, $z=7.3$ \citealt{Saxena23b}). Moreover, as already discussed in Sect.~\ref{sec:escape_frac}, COLA1 shows a relatively high $f_{\rm cen}({\rm Ly}\alpha)\approx 25\%$, also suggestive of high LyC escape fraction.

Almost every observed LyC leaker in the literature presents higher $\Sigma_{\rm SFR}$ than the average at similar redshifts \citep[e.g.,][]{Naidu20}. This is also the case of COLA1. Comparably, the galaxies \textit{Ion2} and \textit{Ion3} show fairly higher $\Sigma_{\rm SFR}$ than the average at their respective redshifts, also presenting complex, multi-peaked \lya profiles \citep{Vanzella16, Vanzella18}.

All the measured properties convey to COLA1 being a prolific LyC leaking source, despite its very high UV luminosity. All the considered indirect methods suggest high LyC escape fraction ($f_{\rm esc}{\rm (LyC)}\sim 20$--$50$\%; see Sect.~\ref{sec:escape_frac}).

\subsection{Size of the ionized bubble}\label{sec:bubble_size}

The size of the implied ionized region in the line of sight to COLA1 can be derived by the cutoff in the blue wing of the \lya line at $\Delta v \sim -250$ km\,s$^{-1}$ (see Fig.~\ref{fig:Lya_profile}). This cutoff is particularly abrupt compared to the red peak, and defines the the line-of-sight extent in which the IGM is optically thin to Ly$\alpha$ photons \citep[so-called proximity zone;][]{Mason&Gronke20}. The blue-peak cutoff seen in the Ly$\alpha$ profile of COLA1 yields a proximity zone of $\sim 0.3$ pMpc \citep{Matthee18}. \cite{Mason&Gronke20} argue that the total size of the ionized region extends beyond the proximity zone as there may be significant damping wing absorption. They find that a bubble with a radius of at least $\sim 0.7$ pMpc is needed in order to explain the observed velocity offset of COLA1's blue \lya peak, assuming a residual neutral fraction of $x_{\rm \ion{H}{I}}\sim10^{-6}$ and defining the minimum observable velocity offset where the IGM transmission of Ly$\alpha$ photons drops to 10\%.

As is discussed in \cite{Mason&Gronke20}, for a central source to be able to ionize a bubble with the required size $\sim 0.7$ pMpc, it needs a steep UV slope ($\beta<-1.79$), large escape fraction of ionizing photons ($f_{\rm esc}^{\rm LyC}>50$\%) and an under-dense line-of-sight gas. Our results show that COLA1 aligns with the first two conditions, but we do not find a strong indication for a specifically under-dense line-of-sight (see Sect.~\ref{sec:O3_sample_spatial_structure}). However, the estimated UV slope of COLA1 is steeper than the values considered in \cite{Mason&Gronke20}, and its extreme value could explain this discrepancy as it implies a higher ionizing photon production efficiency. A more detailed analysis of the ionization conditions of the bubble is needed in order to address this topic, for example using Ly$\alpha$ line measurements from multiple neighboring galaxies and by detailed comparisons to radiative hydrodynamical simulations \citep[e.g.,][]{Gronke21}, which is beyond the scope of this work. In Sect.~\ref{sec:COLA1_ionizing_power} below we discuss the role of COLA1 in ionizing such bubble.

\subsection{COLA1 ionizing its surroundings}\label{sec:COLA1_ionizing_power}

We investigate the contribution from COLA1 and the surrounding galaxies in the $z\approx 6.6$ C1F group to the ionization of a $\sim 0.7$ pMpc bubble needed to explain COLA1's \lya profile (see Sect.~\ref{sec:bubble_size}). The capability of galaxies of ionizing their medium directly depends on their LyC escape fraction. We estimate $f_{\rm esc}$(LyC) for our $z\approx 6.6$ sample with F200W detection (4 out of those 5 objects are detected in this band) using the relation with $\Sigma_{\rm SFR}$(UV) in \cite{Naidu20}, as is described in Sect.~\ref{sec:escape_frac}, for which we get $f_{\rm esc}$(LyC) $=12^{+12}_{-6},\ 9^{+10}_{-5},\ 17^{+14}_{-8}$ and $11^{+12}_{-6}$\%. For these estimates we have assumed the average dust correction introduced in Sect.~\ref{sec:COLA1_as_a_leaker}.

\begin{figure}
    \centering
    \includegraphics[width=\linewidth]{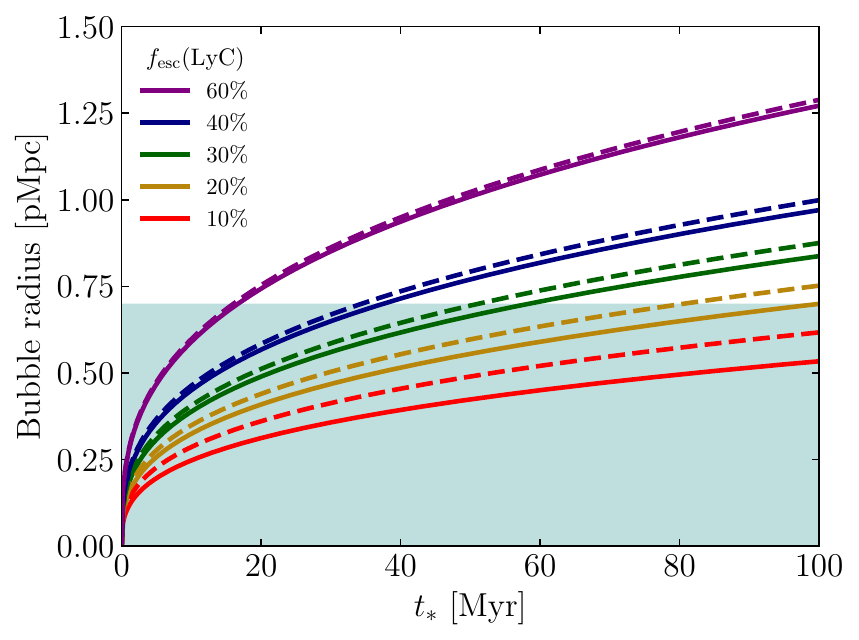}
    \caption{Radius of spherical bubbles hypothetically ionized by COLA1 as a function of the age parameter $t_*$, for different values of the LyC escape fraction. The dashed lines account for the contribution from the 4 galaxies at $z\approx 6.6$ with measured sizes, in addition to COLA1. The shaded area limits the required bubble size necessary to explain COLA1's Ly$\alpha$ profile, as predicted by \protect\cite{Mason&Gronke20}.}
    \label{fig:R_ion_bubble}
\end{figure}

Equation 6 in \cite{Mason&Gronke20} predicts the radius of an ionized bubble powered by a central source as a function of an age parameter $t_*$ for a simple spherical geometry around the ionizing source. Using this equation, and following \cite{Wistok23}, we compute the radius of an ionized bubble powered by COLA1 for different values of the LyC escape fraction (Fig.~\ref{fig:R_ion_bubble}). We have assumed an average hydrogen density $\bar{n}_{\rm H}=1.88\times 10^{-7}\cdot (1 + z)^3$ cm$^{-3}$ \citep{Mason&Gronke20} and an ionizing photon production of $\dot{N}_{\rm ion} = 2.1\times 10^{12} \cdot (1 - f_{\rm esc}{\rm (LyC)})^{-1} \cdot (L_{{\rm H}\beta}/{\rm erg\,s}^{-1})$ s$^{-1}$ \citep{Schaerer16}. For accounting the contribution from the other galaxies, in Fig.~\ref{fig:R_ion_bubble} we have conservatively assumed a high fiducial value of $f_{\rm esc}$(LyC) $=10\%$\footnote{For the computation of $\dot{N}_{\rm ion}$ for the galaxies other than COLA1 we adopted $\dot{N}_{\rm ion} = \xi_{\rm ion} \cdot L_{\rm UV}$ \citep[$\xi_{\rm ion} = 10^{25.59}$ Hz\,erg$^{-1}$;][]{Saxena23}, since a detection of \Hbeta is not available for some of the objects.}. According to this estimate, COLA1 alone could ionize a 0.7 pMpc bubble in a reasonable span of time \citep[$\lesssim 100$ Myr; see][]{Wistok23, Whitler23b} for escape fractions as low as $f_{\rm esc}\sim 20\%$, assuming it maintained the observed star-formation activity. The other four detected galaxies at $z\approx 6.6$ make only a minor contribution in ionizing the bubble when combined with COLA1, specially for high COLA1 escape fractions (dotted lines in Fig.~\ref{fig:R_ion_bubble}). For instance, for COLA1 LyC escape fraction of $40$\% (30\%, 20\%), the time required to ionize the bubble is $38$ Myr (58, $100$ Myr), with COLA1 contributing to $92\%$ (87\%, $80\%$) of the total ionizing photon budget of the $z\approx 6.6$ system detected in C1F.

Nevertheless, systems with extremely high $\Sigma_{\rm SFR}$ such as COLA1 are rare, and this might indicate that these are a product of short-lived star-formation bursts which many galaxies undergo during the EoR \citep[e.g.,][]{Endsley24}. In that case, COLA1 would have likely needed the ionizing photon input from other sources to carve the bubble, specially for $f_{\rm esc}{\rm (LyC)} \ll 50\%$. Depending on the star-formation burstiness, the contribution from numerous undetected faint sources ($M_{\rm UV}\gtrsim -17.5$) might be needed \citep[see e.g.,][]{Wistok23}. In turn, if bursty star-forming episodes are common, they could account for a significant part of the ionizing photon budget during reionization. It was found in simulations that the ionizing photon escape fraction of individual galaxies can rapidly fluctuate in Myr scales (e.g., \citealt{Rosdahl22}; see also \citealt{Matthee22}). In such scenarios, a relatively small fraction of relatively bright galaxies are responsible of the bulk of the ionizing emissivity.

\subsection{Is COLA1 powered by an AGN?}\label{sec:agn}

While the properties of COLA1 display particular similarities with some of the most extreme compact starbursts \citep[e.g.,][]{Vanzella22,MarquesChaves22,Ferrara23,Topping24}, the extreme ionizing conditions and the compact size could also indicate that COLA1s ionizing radiation originates from an AGN instead. The narrowness of the \lya line suggests that COLA1 is not powered by an AGN \citep[see][]{Matthee18}. We also look for hints of an AGN in the \Hbeta line measured in our grism data. First, we fit a Gaussian profile to the nebular \OIII$_{\lambda 5008}$ line. Next, we fit a double component Gaussian to the \Hbeta line in order to capture possible broad and narrow line components \citep[see][for a similar procedure]{Matthee23b}. The FWHM of one of the components is fixed to be the same as the fitted FWHM for \OIII$_{\lambda 5008}$, and the same value is used as lower limit for the other component. The best-fit for the broad component is FWHM $=750\pm 500$ km/s, with a contribution of $7^{+13}_{-7}$\% from the broad component to the total flux (S/N$_{\rm broad}\sim 1$). The best fitting parameters lead to an AGN bolometric luminosity of $\log_{10}(L_{\rm bol}/{\rm erg\,s}^{-1}) = 44.3$, which would be able to explain only $\sim 30\%$ of the UV luminosity for a typical AGN spectrum \citep{Shen20}. This result points towards a subdominant contribution to the UV luminosity.

The measured ratios \OIII$_{\lambda 4364} / {\rm H}\gamma = 0.36\pm 0.12$ and \OIII$_{\lambda 5008}$ / \OIII$_{\lambda 4364} = 34 \pm 11$ of COLA1 are typically seen in low-$z$ star-forming galaxies, whilst AGNs present higher \OIII$_{\lambda 4364}/{\rm H}\gamma$ for a given electron temperature \citep{Ubler23}. Hence, this is further evidence that an AGN is not the main driver of the ISM conditions in COLA1.

\subsection{Future directions}\label{sec:future_directions}

Further in-depth analysis of COLA1 and its surrounding environment is needed to determine the properties of the system and its ionized bubble with precision.
The particular shape of the rest-frame UV SED of COLA and its extreme UV slope are a challenge for the SED fitting models employed in this work (Sect.~\ref{sec:SED_fitting}). Deeper, extended UV spectroscopy would be useful, in particular to confirm the particular SED of COLA1 and to test other indirect indicators of $f_{\rm esc}$(LyC). These include coverage of [\ion{O}{II}] \citep[e.g.,][]{Mascia23a, Choustikov24}, \ion{Mg}{II} \citep[e.g.,][]{Chisholm20, Leclercq24} or the rest-frame UV \citep[e.g., \ion{C}{IV};][]{Naidu22, Schaerer22}. It would be particularly intruiging to identify possible strong emission-lines such as NIV or CIV \citep[e.g.,][]{Bunker23,Calabro24,Topping24} that could boost and impact the rest-frame UV photometry.

On the other hand, the study of the other detected galaxies in COLA1's over-density could yield valuable information about the size, morphology and optical depth of the ionized bubble. A natural future step is to follow-up these sources spectroscopically, aiming for the \lya line. If some of these galaxies are \lya emitters and they are located inside the bounds of the ionized bubble, their \lya line profiles and luminosities could be extremely valuable to determine the bubble properties. The enhanced \lya transmission of the ionized bubble would allow us to see the blue wings of the \lya lines and therefore estimate \lya and LyC escape fractions \citep[e.g.,][]{Bosman22}.

\section{Summary}\label{sec:summary}

The galaxy COLA1 is a luminous \lya emitter at redshift $z=6.5917$ with an unusual double-peaked profile discovered by \cite{Hu16} based on wide-field narrow-band imaging on the Subaru telescope. The detection of the blue \lya peak hints at the existence of an ionized bubble with a well determined size \citep[$R_{\rm ion}\sim 0.7$ pMpc;][]{Matthee18, Mason&Gronke20}, whereas the peak separation probes the detailed ISM conditions that are conducive to Lyman Continuum photon escape. This makes COLA1 a unique laboratory for studying the progress of reionization. However, the unknown systemic redshift made the interpretation of the \lya profile ambiguous. In this work, we presented results from a new JWST program that took WFSS and photometric data of the COLA1 field, in combination with VLT/X-shooter spectroscopy. The aim of this program was to describe the UV and optical properties of the galaxy COLA1, and to simultaneously characterise its over-density, traced by \OIII\ emitters that are part of the sample of 141 \OIII\ emitters at $5.35<z<6.9$ that were selected through \OIII\ doublet identification in the grism data. Below, we summarize our main findings.

\begin{itemize}
    \item We confirmed the systemic redshift of COLA1 to be $z=6.5917$, as measured using the \OIII\ doublet and \Hbeta line centroids. This redshift agrees with the systemic redshift that was previously estimated based on the double-peaked \lya line observed in the VLT/X-shooter spectrum: our rest-frame optical data therefore confirms that COLA1's \lya line shows the classical double-peaked profile shaped by resonant scattering in the ISM, as typically observed in low-redshift galaxies, and not strongly impacted by an IGM damping wing (Sect.~\ref{sec:z_and_f_measurement}, Figs.~\ref{fig:COLA1_line_fits} and \ref{fig:Lya_profile}).
    
    \item COLA1 has a fairly high UV luminosity $M_{\rm UV}=-21.35^{+0.07}_{-0.08}$, a very steep UV slope $\beta_{\rm UV}=-3.2\pm 0.4$ and we infer negligible dust attenuation from the Balmer decrement. The observed SED of COLA1 is unusual and the models used, despite being fairly flexible, are only marginally consistent with the measured UV properties. The metallicity ($12+\log_{10}\left(\text{O/H}\right)_{T_e} = 7.88^{+0.33}_{-0.30}$) and electron temperature ($T_{\rm e} =1.7^{+0.4}_{-0.3} \times 10^{4}$ K) can be directly estimated thanks to the H$\gamma$ (S/N $\approx 10$) and \OIII$_{\lambda 4363}$ (S/N $\approx 3$) measurements; both appearing as typical values for galaxies at similar redshifts in the literature (Sects.~\ref{sec:COLA1_ism_properties}, \ref{sec:uv_properties} and \ref{sec:SED_fitting}, Table~\ref{tab:COLA1_properties}).

    \item The detection of \Hbeta (S/N $\approx 25$) in combination with the measured $\log_{10}(L_{{\rm Ly}\alpha} / {\rm erg\,s}^{-1}) = 43.61 \pm 0.02$ in \cite{Matthee18}, lead us to measure a remarkably large \lya escape fraction, $f_{\rm esc}$(\lya ) $= 81\pm 5$\%. This result is in line with the expected little impact of the IGM on the \lya line, as an effect of an ionized bubble of radius $\sim 0.7$ pMpc (Sect.~\ref{sec:escape_frac}).

    \item Making use of different indirect methods, we estimate a considerable LyC escape fraction for COLA1, $f_{\rm esc}$(LyC) $= 20$--$50$\%. These estimates, along with the above described UV properties, reasonably point out COLA1 as a bona fide UV-luminous agent of reionization. Its star formation-rate is high ($\approx 10\ M_\sun\,{\rm yr}^{-1}$) in comparison with the typical values of the other C1F \OIII\ emitters ($3^{+5}_{-1}\ M_\sun\,{\rm yr}^{-1}$). COLA1 also presents a compact morphology ($R_{\rm UV}<0.26$ kpc) and eminently higher star-formation rate surface density than the average at similar redshifts, in line with known LyC leakers in the literature \citep[e.g.,][]{Vanzella16, Vanzella18} (Sects.~\ref{sec:escape_frac} and \ref{sec:star_formation}, Fig.~\ref{fig:Sigma_SFR}, Table~\ref{tab:fesc_LyC}).

    \item We detect several galaxy over-densities in the field at z=5.3--6.9 with $\delta + 1 \approx 3$--11. Five galaxies are detected to reside within the estimated size of the ionized region around COLA1 (3 with $L_{\OIII\lambda 5008}>10^{42}$ erg\,s$^{-1}$). A detailed comparison to mock observations from large N-body simulations, yields an over-density of $\delta + 1 = 1.96$. We show that this over-density is very common for a typical galaxy with COLA1's UV or \OIII\ luminosity (Sect.~\ref{sec:O3_sample_spatial_structure}, Fig.~\ref{fig:COLA1_overdensity}). 
    
    \item An analysis of the ionizing efficiency of all identified galaxies suggests that COLA1 could have significantly contributed to ionizing the $R_{\rm ion}\sim 0.7$ pMpc bubble needed to explain its peculiar \lya double-peaked profile, with minor contribution from the rest of detected galaxies at $z\approx 6.6$, the latter contributing to only $\lesssim20$\% of the ionizing photon budget. Further follow-up and analysis of these sources could bring more detailed insights about the extent and properties of the sources that ionized the bubble around COLA1 (Sect.~\ref{sec:COLA1_ionizing_power}, Fig.~\ref{fig:R_ion_bubble}).

\end{itemize}

The observations presented in this work simultaneously probed the properties of the unusual COLA1 galaxy (morphology, UV slope and optical lines) and those of galaxies within its environment. We have shown evidence of a luminous galaxy that is actively ionizing its surroundings by showcasing the main characteristics that are common in LyC leakers, in particular the strong Ly$\alpha$ emission with a narrow peak separation and a high SFR surface density. Simultaneously, we have shown that COLA1 resides in an over-density of galaxies and that these galaxies are located within the estimated size of the ionized bubble. However, the over-density is very common for normal galaxies that have similar UV or \OIII\ luminosities as COLA1. This suggests that the detection of the blue peak of the Ly$\alpha$ line in COLA1 is not due to an unusually large over-density of galaxies that could have led to an unusually large ionized bubble that facilitated the observability of these blue Ly$\alpha$ photons. An implication is that care should be taken when interpreting over-densities around galaxies with detected Ly$\alpha$ lines in connection with reionization models. Now, the key next step in determining the impact of similarly luminous galaxies for cosmic reionization is to evaluate how common galaxies with Ly$\alpha$ profiles like COLA1 are and what is the range in their over-densities.

\begin{acknowledgements}
The authors acknowledge the financial support from the MICIU with funding from the European Union NextGenerationEU and Generalitat Valenciana in the call Programa de Planes Complementarios de I+D+i (PRTR 2022) Project (VAL-JPAS), reference ASFAE/2022/025.

This work has been funded by project PID2019-109592GBI00/AEI/10.13039/501100011033 from the Spanish Ministerio de Ciencia e Innovación (MCIN)—Agencia Estatal de Investigación, by the Project of Excellence Prometeo/2020/085 from the Conselleria d’Innovació Universitats, Ciència i Societat Digital de la Generalitat Valenciana.

It has also be funded by the Project of Excellence Prometeo/2020/085 from the Conselleria d’Educació, Universitats, i Ocupació de la Generalitat Valenciana.

Funded by the European Union (ERC, AGENTS,  101076224). Views and opinions expressed are however those of the author(s) only and do not necessarily reflect those of the European Union or the European Research Council. Neither the European Union nor the granting authority can be held responsible for them.

ST acknowledges support by the Royal Society Research Grant G125142.

AH acknowledges support by the VILLUM FONDEN under grant 37459. The Cosmic Dawn Center (DAWN) is funded by the Danish National Research Foundation under grant DNRF140.

We acknowledge funding from JWST program GO-1933. Support for this work was provided by NASA through the NASA Hubble Fellowship grant HST-HF2-51515.001-A awarded by the Space Telescope Science Institute, which is operated by the Association of Universities for Research in Astronomy, Incorporated, under NASA contract NAS5-26555.

This work has received funding from the Swiss State Secretariat for Education, Research and Innovation (SERI) under contract number MB22.00072, as well as from the Swiss National Science Foundation (SNSF) through project grant 200020\_207349.

This work is based on observations made with the NASA/ESA/CSA James Webb Space Telescope. The data were obtained from the Mikulski Archive for Space Telescopes at the Space Telescope Science Institute, which is operated by the Association of Universities for Research in Astronomy, Inc., under NASA contract NAS 5-03127 for JWST. These observations are associated with program \# 1933. The specific observations analyzed can be accessed via https://doi.org/10.17909/s9ht-7n34.

\end{acknowledgements}

\bibliographystyle{aa}
\bibliography{my_bibliography}

\begin{thebibliography}{159}
\expandafter\ifx\csname natexlab\endcsname\relax\def\natexlab#1{#1}\fi

\bibitem[{{Atek} {et~al.}(2023){Atek}, {Chemerynska}, {Wang}, {Furtak}, {Weibel}, {Oesch}, {Weaver}, {Labb{\'e}}, {Bezanson}, {van Dokkum}, {Zitrin}, {Dayal}, {Williams}, {Nannayakkara}, {Price}, {Brammer}, {Goulding}, {Leja}, {Marchesini}, {Nelson}, {Pan}, \& {Whitaker}}]{Atek23}
{Atek}, H., {Chemerynska}, I., {Wang}, B., {et~al.} 2023, \mnras, 524, 5486

\bibitem[{{Begley} {et~al.}(2024){Begley}, {Cullen}, {McLure}, {Shapley}, {Dunlop}, {Carnall}, {McLeod}, {Donnan}, {Hamadouche}, \& {Stanton}}]{Begley24}
{Begley}, R., {Cullen}, F., {McLure}, R.~J., {et~al.} 2024, \mnras, 527, 4040

\bibitem[{{Behroozi} {et~al.}(2019){Behroozi}, {Wechsler}, {Hearin}, \& {Conroy}}]{Behroozi19}
{Behroozi}, P., {Wechsler}, R.~H., {Hearin}, A.~P., \& {Conroy}, C. 2019, \mnras, 488, 3143

\bibitem[{{Bertin} \& {Arnouts}(1996)}]{Bertin&Arnouts96}
{Bertin}, E. \& {Arnouts}, S. 1996, \aaps, 117, 393

\bibitem[{{Bosman} {et~al.}(2022){Bosman}, {Davies}, {Becker}, {Keating}, {Davies}, {Zhu}, {Eilers}, {D'Odorico}, {Bian}, {Bischetti}, {Cristiani}, {Fan}, {Farina}, {Haehnelt}, {Hennawi}, {Kulkarni}, {Mesinger}, {Meyer}, {Onoue}, {Pallottini}, {Qin}, {Ryan-Weber}, {Schindler}, {Walter}, {Wang}, \& {Yang}}]{Bosman22}
{Bosman}, S. E.~I., {Davies}, F.~B., {Becker}, G.~D., {et~al.} 2022, \mnras, 514, 55

\bibitem[{{Bosman} {et~al.}(2020){Bosman}, {Kakiichi}, {Meyer}, {Gronke}, {Laporte}, \& {Ellis}}]{Bosman20}
{Bosman}, S. E.~I., {Kakiichi}, K., {Meyer}, R.~A., {et~al.} 2020, \apj, 896, 49

\bibitem[{{Bouwens} {et~al.}(2015){Bouwens}, {Illingworth}, {Oesch}, {Caruana}, {Holwerda}, {Smit}, \& {Wilkins}}]{Bouwens15}
{Bouwens}, R.~J., {Illingworth}, G.~D., {Oesch}, P.~A., {et~al.} 2015, \apj, 811, 140

\bibitem[{{Bouwens} {et~al.}(2014){Bouwens}, {Illingworth}, {Oesch}, {Labb{\'e}}, {van Dokkum}, {Trenti}, {Franx}, {Smit}, {Gonzalez}, \& {Magee}}]{Bouwens14}
{Bouwens}, R.~J., {Illingworth}, G.~D., {Oesch}, P.~A., {et~al.} 2014, \apj, 793, 115

\bibitem[{{Bouwens} {et~al.}(2016){Bouwens}, {Smit}, {Labb{\'e}}, {Franx}, {Caruana}, {Oesch}, {Stefanon}, \& {Rasappu}}]{Bouwens16}
{Bouwens}, R.~J., {Smit}, R., {Labb{\'e}}, I., {et~al.} 2016, \apj, 831, 176

\bibitem[{{Bunker} {et~al.}(2023){Bunker}, {Saxena}, {Cameron}, {Willott}, {Curtis-Lake}, {Jakobsen}, {Carniani}, {Smit}, {Maiolino}, {Witstok}, {Curti}, {D'Eugenio}, {Jones}, {Ferruit}, {Arribas}, {Charlot}, {Chevallard}, {Giardino}, {de Graaff}, {Looser}, {L{\"u}tzgendorf}, {Maseda}, {Rawle}, {Rix}, {Del Pino}, {Alberts}, {Egami}, {Eisenstein}, {Endsley}, {Hainline}, {Hausen}, {Johnson}, {Rieke}, {Rieke}, {Robertson}, {Shivaei}, {Stark}, {Sun}, {Tacchella}, {Tang}, {Williams}, {Willmer}, {Baker}, {Baum}, {Bhatawdekar}, {Bowler}, {Boyett}, {Chen}, {Circosta}, {Helton}, {Ji}, {Kumari}, {Lyu}, {Nelson}, {Parlanti}, {Perna}, {Sandles}, {Scholtz}, {Suess}, {Topping}, {{\"U}bler}, {Wallace}, \& {Whitler}}]{Bunker23}
{Bunker}, A.~J., {Saxena}, A., {Cameron}, A.~J., {et~al.} 2023, \aap, 677, A88

\bibitem[{{Byler} {et~al.}(2017){Byler}, {Dalcanton}, {Conroy}, \& {Johnson}}]{Byler17}
{Byler}, N., {Dalcanton}, J.~J., {Conroy}, C., \& {Johnson}, B.~D. 2017, \apj, 840, 44

\bibitem[{{Calabro} {et~al.}(2024){Calabro}, {Castellano}, {Zavala}, {Pentericci}, {Arrabal Haro}, {Bakx}, {Burgarella}, {Casey}, {Dickinson}, {Finkelstein}, {Fontana}, {Llerena}, {Mascia}, {Merlin}, {Mitsuhashi}, {Napolitano}, {Paris}, {Perez-Gonzalez}, {Roberts-Borsani}, {Santini}, {Treu}, \& {Vanzella}}]{Calabro24}
{Calabro}, A., {Castellano}, M., {Zavala}, J.~A., {et~al.} 2024, ApJ, submitted, arXiv:2403.12683

\bibitem[{{Cameron} {et~al.}(2023{\natexlab{a}}){Cameron}, {Katz}, {Witten}, {Saxena}, {Laporte}, \& {Bunker}}]{Cameron23b}
{Cameron}, A.~J., {Katz}, H., {Witten}, C., {et~al.} 2023{\natexlab{a}}, arXiv e-prints, arXiv:2311.02051

\bibitem[{{Cameron} {et~al.}(2023{\natexlab{b}}){Cameron}, {Saxena}, {Bunker}, {D'Eugenio}, {Carniani}, {Maiolino}, {Curtis-Lake}, {Ferruit}, {Jakobsen}, {Arribas}, {Bonaventura}, {Charlot}, {Chevallard}, {Curti}, {Looser}, {Maseda}, {Rawle}, {Rodr{\'\i}guez Del Pino}, {Smit}, {{\"U}bler}, {Willott}, {Witstok}, {Egami}, {Eisenstein}, {Johnson}, {Hainline}, {Rieke}, {Robertson}, {Stark}, {Tacchella}, {Williams}, {Willmer}, {Bhatawdekar}, {Bowler}, {Boyett}, {Circosta}, {Helton}, {Jones}, {Kumari}, {Ji}, {Nelson}, {Parlanti}, {Sandles}, {Scholtz}, \& {Sun}}]{Cameron23}
{Cameron}, A.~J., {Saxena}, A., {Bunker}, A.~J., {et~al.} 2023{\natexlab{b}}, \aap, 677, A115

\bibitem[{{Cardelli} {et~al.}(1989){Cardelli}, {Clayton}, \& {Mathis}}]{Cardelli89}
{Cardelli}, J.~A., {Clayton}, G.~C., \& {Mathis}, J.~S. 1989, \apj, 345, 245

\bibitem[{{Chabrier}(2003)}]{Chabrier03}
{Chabrier}, G. 2003, Publications of the Astronomical Society of the Pacific, 115, 763

\bibitem[{{Chen} {et~al.}(2024){Chen}, {Stark}, {Mason}, {Topping}, {Whitler}, {Tang}, {Endsley}, \& {Charlot}}]{Chen23}
{Chen}, Z., {Stark}, D.~P., {Mason}, C., {et~al.} 2024, \mnras, 528, 7052

\bibitem[{{Chisholm} {et~al.}(2020){Chisholm}, {Prochaska}, {Schaerer}, {Gazagnes}, \& {Henry}}]{Chisholm20}
{Chisholm}, J., {Prochaska}, J.~X., {Schaerer}, D., {Gazagnes}, S., \& {Henry}, A. 2020, \mnras, 498, 2554

\bibitem[{{Chisholm} {et~al.}(2022){Chisholm}, {Saldana-Lopez}, {Flury}, {Schaerer}, {Jaskot}, {Amor{\'\i}n}, {Atek}, {Finkelstein}, {Fleming}, {Ferguson}, {Fern{\'a}ndez}, {Giavalisco}, {Hayes}, {Heckman}, {Henry}, {Ji}, {Marques-Chaves}, {Mauerhofer}, {McCandliss}, {Oey}, {{\"O}stlin}, {Rutkowski}, {Scarlata}, {Thuan}, {Trebitsch}, {Wang}, {Worseck}, \& {Xu}}]{Chisholm22}
{Chisholm}, J., {Saldana-Lopez}, A., {Flury}, S., {et~al.} 2022, \mnras, 517, 5104

\bibitem[{{Choi} {et~al.}(2017){Choi}, {Conroy}, \& {Byler}}]{Choi17}
{Choi}, J., {Conroy}, C., \& {Byler}, N. 2017, \apj, 838, 159

\bibitem[{{Choustikov} {et~al.}(2024){Choustikov}, {Katz}, {Saxena}, {Cameron}, {Devriendt}, {Slyz}, {Rosdahl}, {Blaizot}, \& {Michel-Dansac}}]{Choustikov24}
{Choustikov}, N., {Katz}, H., {Saxena}, A., {et~al.} 2024, \mnras, 529, 3751

\bibitem[{{Conroy} \& {Gunn}(2010)}]{FSPS3}
{Conroy}, C. \& {Gunn}, J.~E. 2010, \apj, 712, 833

\bibitem[{{Conroy} {et~al.}(2009){Conroy}, {Gunn}, \& {White}}]{FSPS1}
{Conroy}, C., {Gunn}, J.~E., \& {White}, M. 2009, \apj, 699, 486

\bibitem[{{Conroy} \& {Kratter}(2012)}]{Conroy12}
{Conroy}, C. \& {Kratter}, K.~M. 2012, \apj, 755, 123

\bibitem[{{Conroy} {et~al.}(2010){Conroy}, {White}, \& {Gunn}}]{FSPS2}
{Conroy}, C., {White}, M., \& {Gunn}, J.~E. 2010, \apj, 708, 58

\bibitem[{{Curti} {et~al.}(2023){Curti}, {D'Eugenio}, {Carniani}, {Maiolino}, {Sandles}, {Witstok}, {Baker}, {Bennett}, {Piotrowska}, {Tacchella}, {Charlot}, {Nakajima}, {Maheson}, {Mannucci}, {Amiri}, {Arribas}, {Belfiore}, {Bonaventura}, {Bunker}, {Chevallard}, {Cresci}, {Curtis-Lake}, {Hayden-Pawson}, {Jones}, {Kumari}, {Laseter}, {Looser}, {Marconi}, {Maseda}, {Scholtz}, {Smit}, {{\"U}bler}, \& {Wallace}}]{Curti23}
{Curti}, M., {D'Eugenio}, F., {Carniani}, S., {et~al.} 2023, \mnras, 518, 425

\bibitem[{{Dayal} \& {Giri}(2024)}]{Dayal23}
{Dayal}, P. \& {Giri}, S.~K. 2024, \mnras, 528, 2784

\bibitem[{{Dayal} {et~al.}(2020){Dayal}, {Volonteri}, {Choudhury}, {Schneider}, {Trebitsch}, {Gnedin}, {Atek}, {Hirschmann}, \& {Reines}}]{Dayal20}
{Dayal}, P., {Volonteri}, M., {Choudhury}, T.~R., {et~al.} 2020, \mnras, 495, 3065

\bibitem[{{de Barros} {et~al.}(2014){de Barros}, {Schaerer}, \& {Stark}}]{deBarros14}
{de Barros}, S., {Schaerer}, D., \& {Stark}, D.~P. 2014, \aap, 563, A81

\bibitem[{{Dickey} {et~al.}(2016){Dickey}, {van Dokkum}, {Oesch}, {Whitaker}, {Momcheva}, {Nelson}, {Leja}, {Brammer}, {Franx}, \& {Skelton}}]{Dickey16}
{Dickey}, C.~M., {van Dokkum}, P.~G., {Oesch}, P.~A., {et~al.} 2016, \apjl, 828, L11

\bibitem[{{Eilers} {et~al.}(2024){Eilers}, {Mackenzie}, {Pizzati}, {Matthee}, {Hennawi}, {Zhang}, {Bordoloi}, {Kashino}, {Lilly}, {Naidu}, {Simcoe}, {Yue}, {Frenk}, {Helly}, {Schaller}, \& {Schaye}}]{Eilers24}
{Eilers}, A.-C., {Mackenzie}, R., {Pizzati}, E., {et~al.} 2024, arXiv e-prints, arXiv:2403.07986

\bibitem[{{Endsley} \& {Stark}(2022)}]{Endsley22}
{Endsley}, R. \& {Stark}, D.~P. 2022, \mnras, 511, 6042

\bibitem[{{Endsley} {et~al.}(2023{\natexlab{a}}){Endsley}, {Stark}, {Whitler}, {Topping}, {Chen}, {Plat}, {Chisholm}, \& {Charlot}}]{Endsley23}
{Endsley}, R., {Stark}, D.~P., {Whitler}, L., {et~al.} 2023{\natexlab{a}}, \mnras, 524, 2312

\bibitem[{{Endsley} {et~al.}(2023{\natexlab{b}}){Endsley}, {Stark}, {Whitler}, {Topping}, {Johnson}, {Robertson}, {Tacchella}, {Alberts}, {Baker}, {Bhatawdekar}, {Boyett}, {Bunker}, {Cameron}, {Carniani}, {Charlot}, {Chen}, {Chevallard}, {Curtis-Lake}, {Danhaive}, {Egami}, {Eisenstein}, {Hainline}, {Helton}, {Ji}, {Looser}, {Maiolino}, {Nelson}, {Pusk{\'a}s}, {Rieke}, {Rieke}, {Rix}, {Sandles}, {Saxena}, {Simmonds}, {Smit}, {Sun}, {Williams}, {Willmer}, {Willott}, \& {Witstok}}]{Endsley24}
{Endsley}, R., {Stark}, D.~P., {Whitler}, L., {et~al.} 2023{\natexlab{b}}, MNRAS, submitted, arXiv:2306.05295

\bibitem[{{Ferland} {et~al.}(1998){Ferland}, {Korista}, {Verner}, {Ferguson}, {Kingdon}, \& {Verner}}]{Ferland98}
{Ferland}, G.~J., {Korista}, K.~T., {Verner}, D.~A., {et~al.} 1998, \pasp, 110, 761

\bibitem[{{Ferland} {et~al.}(2013){Ferland}, {Porter}, {van Hoof}, {Williams}, {Abel}, {Lykins}, {Shaw}, {Henney}, \& {Stancil}}]{Ferland13}
{Ferland}, G.~J., {Porter}, R.~L., {van Hoof}, P.~A.~M., {et~al.} 2013, \rmxaa, 49, 137

\bibitem[{{Ferrara}(2024)}]{Ferrara23}
{Ferrara}, A. 2024, \aap, 684, A207

\bibitem[{{Finkelstein} {et~al.}(2022){Finkelstein}, {Bagley}, {Song}, {Larson}, {Papovich}, {Dickinson}, {Finkelstein}, {Koekemoer}, {Pirzkal}, {Somerville}, {Yung}, {Behroozi}, {Ferguson}, {Giavalisco}, {Grogin}, {Hathi}, {Hutchison}, {Jung}, {Kocevski}, {Kawinwanichakij}, {Rojas-Ruiz}, {Ryan}, {Snyder}, \& {Tacchella}}]{Finkelstein22}
{Finkelstein}, S.~L., {Bagley}, M., {Song}, M., {et~al.} 2022, \apj, 928, 52

\bibitem[{{Finkelstein} {et~al.}(2019){Finkelstein}, {D'Aloisio}, {Paardekooper}, {Ryan}, {Behroozi}, {Finlator}, {Livermore}, {Upton Sanderbeck}, {Dalla Vecchia}, \& {Khochfar}}]{Finkelstein19}
{Finkelstein}, S.~L., {D'Aloisio}, A., {Paardekooper}, J.-P., {et~al.} 2019, \apj, 879, 36

\bibitem[{{Flury} {et~al.}(2022){Flury}, {Jaskot}, {Ferguson}, {Worseck}, {Makan}, {Chisholm}, {Saldana-Lopez}, {Schaerer}, {McCandliss}, {Xu}, {Wang}, {Oey}, {Ford}, {Heckman}, {Ji}, {Giavalisco}, {Amor{\'\i}n}, {Atek}, {Blaizot}, {Borthakur}, {Carr}, {Castellano}, {De Barros}, {Dickinson}, {Finkelstein}, {Fleming}, {Fontanot}, {Garel}, {Grazian}, {Hayes}, {Henry}, {Mauerhofer}, {Micheva}, {Ostlin}, {Papovich}, {Pentericci}, {Ravindranath}, {Rosdahl}, {Rutkowski}, {Santini}, {Scarlata}, {Teplitz}, {Thuan}, {Trebitsch}, {Vanzella}, \& {Verhamme}}]{Flury22}
{Flury}, S.~R., {Jaskot}, A.~E., {Ferguson}, H.~C., {et~al.} 2022, \apj, 930, 126

\bibitem[{{Fujimoto} {et~al.}(2023){Fujimoto}, {Arrabal Haro}, {Dickinson}, {Finkelstein}, {Kartaltepe}, {Larson}, {Burgarella}, {Bagley}, {Behroozi}, {Chworowsky}, {Hirschmann}, {Trump}, {Wilkins}, {Yung}, {Koekemoer}, {Papovich}, {Pirzkal}, {Ferguson}, {Fontana}, {Grogin}, {Grazian}, {Kewley}, {Kocevski}, {Lotz}, {Pentericci}, {Ravindranath}, {Somerville}, {Wilkins}, {Amor{\'\i}n}, {Backhaus}, {Calabr{\`o}}, {Casey}, {Cooper}, {Fern{\'a}ndez}, {Franco}, {Giavalisco}, {Hathi}, {Harish}, {Hutchison}, {Iyer}, {Jung}, {Lucas}, \& {Zavala}}]{Fujimoto23}
{Fujimoto}, S., {Arrabal Haro}, P., {Dickinson}, M., {et~al.} 2023, \apjl, 949, L25

\bibitem[{{Gnedin} \& {Madau}(2022)}]{Gnedin22}
{Gnedin}, N.~Y. \& {Madau}, P. 2022, Living Reviews in Computational Astrophysics, 8, 3

\bibitem[{{Greene} {et~al.}(2024){Greene}, {Labbe}, {Goulding}, {Furtak}, {Chemerynska}, {Kokorev}, {Dayal}, {Volonteri}, {Williams}, {Wang}, {Setton}, {Burgasser}, {Bezanson}, {Atek}, {Brammer}, {Cutler}, {Feldmann}, {Fujimoto}, {Glazebrook}, {de Graaff}, {Khullar}, {Leja}, {Marchesini}, {Maseda}, {Matthee}, {Miller}, {Naidu}, {Nanayakkara}, {Oesch}, {Pan}, {Papovich}, {Price}, {van Dokkum}, {Weaver}, {Whitaker}, \& {Zitrin}}]{Greene23}
{Greene}, J.~E., {Labbe}, I., {Goulding}, A.~D., {et~al.} 2024, \apj, 964, 39

\bibitem[{{Greig} \& {Mesinger}(2017)}]{Greig17}
{Greig}, B. \& {Mesinger}, A. 2017, \mnras, 465, 4838

\bibitem[{{Gronke} {et~al.}(2015){Gronke}, {Bull}, \& {Dijkstra}}]{Gronke15}
{Gronke}, M., {Bull}, P., \& {Dijkstra}, M. 2015, \apj, 812, 123

\bibitem[{{Gronke} {et~al.}(2021){Gronke}, {Ocvirk}, {Mason}, {Matthee}, {Bosman}, {Sorce}, {Lewis}, {Ahn}, {Aubert}, {Dawoodbhoy}, {Iliev}, {Shapiro}, \& {Yepes}}]{Gronke21}
{Gronke}, M., {Ocvirk}, P., {Mason}, C., {et~al.} 2021, \mnras, 508, 3697

\bibitem[{{Gurung-L{\'o}pez} {et~al.}(2020){Gurung-L{\'o}pez}, {Orsi}, {Bonoli}, {Padilla}, {Lacey}, \& {Baugh}}]{Gurung-Lopez20}
{Gurung-L{\'o}pez}, S., {Orsi}, {\'A}.~A., {Bonoli}, S., {et~al.} 2020, \mnras, 491, 3266

\bibitem[{{Haiman}(2002)}]{Haiman02}
{Haiman}, Z. 2002, \apjl, 576, L1

\bibitem[{{Hayes} \& {Scarlata}(2023)}]{Hayes23}
{Hayes}, M.~J. \& {Scarlata}, C. 2023, \apjl, 954, L14

\bibitem[{{Heckman} \& {Borthakur}(2016)}]{Heckman16}
{Heckman}, T.~M. \& {Borthakur}, S. 2016, \apj, 822, 9

\bibitem[{{Horne}(1986)}]{Horne86}
{Horne}, K. 1986, \pasp, 98, 609

\bibitem[{{Hu} {et~al.}(2016){Hu}, {Cowie}, {Songaila}, {Barger}, {Rosenwasser}, \& {Wold}}]{Hu16}
{Hu}, E.~M., {Cowie}, L.~L., {Songaila}, A., {et~al.} 2016, \apjl, 825, L7

\bibitem[{{Hu} {et~al.}(2024){Hu}, {Papovich}, {Dickinson}, {Kennicutt}, {Shen}, {Amor{\'\i}n}, {Arrabal Haro}, {Bagley}, {Bhatawdekar}, {Cleri}, {Cole}, {Dekel}, {de la Vega}, {Finkelstein}, {Grogin}, {Hathi}, {Hirschmann}, {Holwerda}, {Hutchison}, {Jung}, {Koekemoer}, {Kartaltepe}, {Lucas}, {Llerena}, {Mascia}, {Mobasher}, {Napolitano}, {Newman}, {Pentericci}, {P{\'e}rez-Gonz{\'a}lez}, {Trump}, {Wilkins}, \& {Yung}}]{Hu24}
{Hu}, W., {Papovich}, C., {Dickinson}, M., {et~al.} 2024, arXiv e-prints, arXiv:2401.12402

\bibitem[{{Hutter} {et~al.}(2021){Hutter}, {Dayal}, {Yepes}, {Gottl{\"o}ber}, {Legrand}, \& {Ucci}}]{Hutter21}
{Hutter}, A., {Dayal}, P., {Yepes}, G., {et~al.} 2021, \mnras, 503, 3698

\bibitem[{{Inoue} {et~al.}(2014){Inoue}, {Shimizu}, {Iwata}, \& {Tanaka}}]{Inoue14}
{Inoue}, A.~K., {Shimizu}, I., {Iwata}, I., \& {Tanaka}, M. 2014, \mnras, 442, 1805

\bibitem[{{Ishiyama} {et~al.}(2021){Ishiyama}, {Prada}, {Klypin}, {Sinha}, {Metcalf}, {Jullo}, {Altieri}, {Cora}, {Croton}, {de la Torre}, {Mill{\'a}n-Calero}, {Oogi}, {Ruedas}, \& {Vega-Mart{\'\i}nez}}]{Ishiyama21}
{Ishiyama}, T., {Prada}, F., {Klypin}, A.~A., {et~al.} 2021, \mnras, 506, 4210

\bibitem[{{Izotov} {et~al.}(2021){Izotov}, {Guseva}, {Fricke}, {Henkel}, {Schaerer}, \& {Thuan}}]{Izotov21}
{Izotov}, Y.~I., {Guseva}, N.~G., {Fricke}, K.~J., {et~al.} 2021, \aap, 646, A138

\bibitem[{{Izotov} {et~al.}(2016){Izotov}, {Schaerer}, {Thuan}, {Worseck}, {Guseva}, {Orlitov{\'a}}, \& {Verhamme}}]{Izotov16}
{Izotov}, Y.~I., {Schaerer}, D., {Thuan}, T.~X., {et~al.} 2016, \mnras, 461, 3683

\bibitem[{{Izotov} {et~al.}(2020){Izotov}, {Schaerer}, {Worseck}, {Verhamme}, {Guseva}, {Thuan}, {Orlitov{\'a}}, \& {Fricke}}]{Izotov20}
{Izotov}, Y.~I., {Schaerer}, D., {Worseck}, G., {et~al.} 2020, \mnras, 491, 468

\bibitem[{{Izotov} {et~al.}(2018){Izotov}, {Worseck}, {Schaerer}, {Guseva}, {Thuan}, {Fricke}, \& {Orlitov{\'a}}}]{Izotov18}
{Izotov}, Y.~I., {Worseck}, G., {Schaerer}, D., {et~al.} 2018, \mnras, 478, 4851

\bibitem[{{Johnson} {et~al.}(2021){Johnson}, {Leja}, {Conroy}, \& {Speagle}}]{Johnson21}
{Johnson}, B.~D., {Leja}, J., {Conroy}, C., \& {Speagle}, J.~S. 2021, \apjs, 254, 22

\bibitem[{{Jung} {et~al.}(2024){Jung}, {Finkelstein}, {Arrabal Haro}, {Dickinson}, {Ferguson}, {Hutchison}, {Kartaltepe}, {Larson}, {Simons}, {Papovich}, {Park}, {Pentericci}, {Trump}, {Amor{\'\i}n}, {Backhaus}, {Bagley}, {Casey}, {Cheng}, {Cleri}, {Cooper}, {Cooper}, {Gardner}, {Gawiser}, {Grazian}, {Hathi}, {Hirschmann}, {Koekemoer}, {Lucas}, {Mobasher}, {Pirzkal}, {Ravindranath}, {Straughn}, {Yung}, \& {de la Vega}}]{Jung23}
{Jung}, I., {Finkelstein}, S.~L., {Arrabal Haro}, P., {et~al.} 2024, \apj, 967, 73

\bibitem[{{Kakiichi} \& {Gronke}(2021)}]{Kakiichi&Gronke21}
{Kakiichi}, K. \& {Gronke}, M. 2021, \apj, 908, 30

\bibitem[{{Kashino} {et~al.}(2023){Kashino}, {Lilly}, {Matthee}, {Eilers}, {Mackenzie}, {Bordoloi}, \& {Simcoe}}]{Kashino23}
{Kashino}, D., {Lilly}, S.~J., {Matthee}, J., {et~al.} 2023, \apj, 950, 66

\bibitem[{{Katz} {et~al.}(2023){Katz}, {Saxena}, {Cameron}, {Carniani}, {Bunker}, {Arribas}, {Bhatawdekar}, {Bowler}, {Boyett}, {Cresci}, {Curtis-Lake}, {D'Eugenio}, {Kumari}, {Looser}, {Maiolino}, {{\"U}bler}, {Willott}, \& {Witstok}}]{Katz23}
{Katz}, H., {Saxena}, A., {Cameron}, A.~J., {et~al.} 2023, \mnras, 518, 592

\bibitem[{{Kerutt} {et~al.}(2024){Kerutt}, {Oesch}, {Wisotzki}, {Verhamme}, {Atek}, {Herenz}, {Illingworth}, {Kusakabe}, {Matthee}, {Mauerhofer}, {Montes}, {Naidu}, {Nelson}, {Reddy}, {Schaye}, {Simmonds}, {Urrutia}, \& {Vitte}}]{Kerutt23}
{Kerutt}, J., {Oesch}, P.~A., {Wisotzki}, L., {et~al.} 2024, \aap, 684, A42

\bibitem[{{Kim} {et~al.}(2023){Kim}, {Bayliss}, {Rigby}, {Gladders}, {Chisholm}, {Sharon}, {Dahle}, {Rivera-Thorsen}, {Florian}, {Khullar}, {Mahler}, {Mainali}, {Napier}, {Navarre}, {Owens}, \& {Roberson}}]{Kim23}
{Kim}, K.~J., {Bayliss}, M.~B., {Rigby}, J.~R., {et~al.} 2023, \apjl, 955, L17

\bibitem[{{Kimm} {et~al.}(2022){Kimm}, {Bieri}, {Geen}, {Rosdahl}, {Blaizot}, {Michel-Dansac}, \& {Garel}}]{Kimm22}
{Kimm}, T., {Bieri}, R., {Geen}, S., {et~al.} 2022, \apjs, 259, 21

\bibitem[{{Kocevski} {et~al.}(2023){Kocevski}, {Onoue}, {Inayoshi}, {Trump}, {Arrabal Haro}, {Grazian}, {Dickinson}, {Finkelstein}, {Kartaltepe}, {Hirschmann}, {Aird}, {Holwerda}, {Fujimoto}, {Juneau}, {Amor{\'\i}n}, {Backhaus}, {Bagley}, {Barro}, {Bell}, {Bisigello}, {Calabr{\`o}}, {Cleri}, {Cooper}, {Ding}, {Grogin}, {Ho}, {Hutchison}, {Inoue}, {Jiang}, {Jones}, {Koekemoer}, {Li}, {Li}, {McGrath}, {Molina}, {Papovich}, {P{\'e}rez-Gonz{\'a}lez}, {Pirzkal}, {Wilkins}, {Yang}, \& {Yung}}]{Kocevski23}
{Kocevski}, D.~D., {Onoue}, M., {Inayoshi}, K., {et~al.} 2023, \apjl, 954, L4

\bibitem[{{Kokorev} {et~al.}(2024){Kokorev}, {Caputi}, {Greene}, {Dayal}, {Trebitsch}, {Cutler}, {Fujimoto}, {Labb{\'e}}, {Miller}, {Iani}, {Navarro-Carrera}, \& {Rinaldi}}]{Kokorev24}
{Kokorev}, V., {Caputi}, K.~I., {Greene}, J.~E., {et~al.} 2024, \apj, 968, 38

\bibitem[{{Kramarenko} {et~al.}(2024){Kramarenko}, {Kerutt}, {Verhamme}, {Oesch}, {Barrufet}, {Matthee}, {Kusakabe}, {Goovaerts}, \& {Thai}}]{Kramarenko24}
{Kramarenko}, I.~G., {Kerutt}, J., {Verhamme}, A., {et~al.} 2024, \mnras, 527, 9853

\bibitem[{{Kulkarni} {et~al.}(2019){Kulkarni}, {Worseck}, \& {Hennawi}}]{Kulkarni19}
{Kulkarni}, G., {Worseck}, G., \& {Hennawi}, J.~F. 2019, \mnras, 488, 1035

\bibitem[{{Labb{\'e}} {et~al.}(2013){Labb{\'e}}, {Oesch}, {Bouwens}, {Illingworth}, {Magee}, {Gonz{\'a}lez}, {Carollo}, {Franx}, {Trenti}, {van Dokkum}, \& {Stiavelli}}]{Labbe13}
{Labb{\'e}}, I., {Oesch}, P.~A., {Bouwens}, R.~J., {et~al.} 2013, \apjl, 777, L19

\bibitem[{{Leclercq} {et~al.}(2017){Leclercq}, {Bacon}, {Wisotzki}, {Mitchell}, {Garel}, {Verhamme}, {Blaizot}, {Hashimoto}, {Herenz}, {Conseil}, {Cantalupo}, {Inami}, {Contini}, {Richard}, {Maseda}, {Schaye}, {Marino}, {Akhlaghi}, {Brinchmann}, \& {Carollo}}]{Leclercq17}
{Leclercq}, F., {Bacon}, R., {Wisotzki}, L., {et~al.} 2017, \aap, 608, A8

\bibitem[{{Leclercq} {et~al.}(2024){Leclercq}, {Chisholm}, {King}, {Zeimann}, {Jaskot}, {Henry}, {Hayes}, {Flury}, {Izotov}, {Prochaska}, {Verhamme}, {Amor{\'\i}n}, {Atek}, {Bait}, {Blaizot}, {Carr}, {Ji}, {Le Reste}, {Ferguson}, {Gazagnes}, {Heckman}, {Komarova}, {Marques-Chaves}, {{\"O}stlin}, {Saldana-Lopez}, {Scarlata}, {Schaerer}, {Thuan}, {Trebitsch}, {Worseck}, {Wang}, \& {Xu}}]{Leclercq24}
{Leclercq}, F., {Chisholm}, J., {King}, W., {et~al.} 2024, A\&A, submitted, arXiv:2401.14981

\bibitem[{{Lecroq} {et~al.}(2024){Lecroq}, {Charlot}, {Bressan}, {Bruzual}, {Costa}, {Iorio}, {Spera}, {Mapelli}, {Chen}, {Chevallard}, \& {Dall'Amico}}]{Lecroq24}
{Lecroq}, M., {Charlot}, S., {Bressan}, A., {et~al.} 2024, \mnras, 527, 9480

\bibitem[{{Leonova} {et~al.}(2022){Leonova}, {Oesch}, {Qin}, {Naidu}, {Wyithe}, {de Barros}, {Bouwens}, {Ellis}, {Endsley}, {Hutter}, {Illingworth}, {Kerutt}, {Labb{\'e}}, {Laporte}, {Magee}, {Mutch}, {Roberts-Borsani}, {Smit}, {Stark}, {Stefanon}, {Tacchella}, \& {Zitrin}}]{Leonova22}
{Leonova}, E., {Oesch}, P.~A., {Qin}, Y., {et~al.} 2022, \mnras, 515, 5790

\bibitem[{{Lu} {et~al.}(2023){Lu}, {Mason}, {Hutter}, {Mesinger}, {Qin}, {Stark}, \& {Endsley}}]{Lu23}
{Lu}, T.-Y., {Mason}, C., {Hutter}, A., {et~al.} 2023, MNRAS, submitted, arXiv:2304.11192

\bibitem[{{Luridiana} {et~al.}(2015){Luridiana}, {Morisset}, \& {Shaw}}]{Luridiana15}
{Luridiana}, V., {Morisset}, C., \& {Shaw}, R.~A. 2015, \aap, 573, A42

\bibitem[{{Ma} {et~al.}(2018){Ma}, {Hopkins}, {Garrison-Kimmel}, {Faucher-Gigu{\`e}re}, {Quataert}, {Boylan-Kolchin}, {Hayward}, {Feldmann}, \& {Kere{\v{s}}}}]{Ma18}
{Ma}, X., {Hopkins}, P.~F., {Garrison-Kimmel}, S., {et~al.} 2018, \mnras, 478, 1694

\bibitem[{{Mainali} {et~al.}(2022){Mainali}, {Rigby}, {Chisholm}, {Bayliss}, {Bordoloi}, {Gladders}, {Rivera-Thorsen}, {Dahle}, {Sharon}, {Florian}, {Berg}, {Sharma}, {Owens}, {Kjellgren}, {Kim}, \& {Wayne}}]{Mainali22}
{Mainali}, R., {Rigby}, J.~R., {Chisholm}, J., {et~al.} 2022, \apj, 940, 160

\bibitem[{{Maji} {et~al.}(2022){Maji}, {Verhamme}, {Rosdahl}, {Garel}, {Blaizot}, {Mauerhofer}, {Pittavino}, {Victoria Feser}, {Chuniaud}, {Kimm}, {Katz}, \& {Haehnelt}}]{Maji22}
{Maji}, M., {Verhamme}, A., {Rosdahl}, J., {et~al.} 2022, \aap, 663, A66

\bibitem[{{Marques-Chaves} {et~al.}(2022){Marques-Chaves}, {Schaerer}, {{\'A}lvarez-M{\'a}rquez}, {Verhamme}, {Ceverino}, {Chisholm}, {Colina}, {Dessauges-Zavadsky}, {P{\'e}rez-Fournon}, {Saldana-Lopez}, {Upadhyaya}, \& {Vanzella}}]{MarquesChaves22}
{Marques-Chaves}, R., {Schaerer}, D., {{\'A}lvarez-M{\'a}rquez}, J., {et~al.} 2022, \mnras, 517, 2972

\bibitem[{{Mascia} {et~al.}(2024){Mascia}, {Pentericci}, {Calabr{\`o}}, {Santini}, {Napolitano}, {Arrabal Haro}, {Castellano}, {Dickinson}, {Ocvirk}, {Lewis}, {Amor{\'\i}n}, {Bagley}, {Bhatawdekar}, {Cleri}, {Costantin}, {Dekel}, {Finkelstein}, {Fontana}, {Giavalisco}, {Grogin}, {Hathi}, {Hirschmann}, {Holwerda}, {Jung}, {Kartaltepe}, {Koekemoer}, {Lucas}, {Papovich}, {P{\'e}rez-Gonz{\'a}lez}, {Pirzkal}, {Trump}, {Wilkins}, \& {Yung}}]{Mascia23b}
{Mascia}, S., {Pentericci}, L., {Calabr{\`o}}, A., {et~al.} 2024, \aap, 685, A3

\bibitem[{{Mascia} {et~al.}(2023{\natexlab{a}}){Mascia}, {Pentericci}, {Calabr{\`o}}, {Treu}, {Santini}, {Yang}, {Napolitano}, {Roberts-Borsani}, {Bergamini}, {Grillo}, {Rosati}, {Vulcani}, {Castellano}, {Boyett}, {Fontana}, {Glazebrook}, {Henry}, {Mason}, {Merlin}, {Morishita}, {Nanayakkara}, {Paris}, {Roy}, {Williams}, {Wang}, {Brammer}, {Brada{\v{c}}}, {Chen}, {Kelly}, {Koekemoer}, {Trenti}, \& {Windhorst}}]{Mascia23a}
{Mascia}, S., {Pentericci}, L., {Calabr{\`o}}, A., {et~al.} 2023{\natexlab{a}}, \aap, 672, A155

\bibitem[{{Mascia} {et~al.}(2023{\natexlab{b}}){Mascia}, {Pentericci}, {Saxena}, {Belfiori}, {Calabr{\`o}}, {Castellano}, {Saldana-Lopez}, {Talia}, {Amor{\'\i}n}, {Cullen}, {Garilli}, {Guaita}, {LLerena}, {McLure}, {Moresco}, {Santini}, \& {Schaerer}}]{Mascia22}
{Mascia}, S., {Pentericci}, L., {Saxena}, A., {et~al.} 2023{\natexlab{b}}, \aap, 674, A221

\bibitem[{{Mason} \& {Gronke}(2020)}]{Mason&Gronke20}
{Mason}, C.~A. \& {Gronke}, M. 2020, \mnras, 499, 1395

\bibitem[{{Mason} {et~al.}(2018){Mason}, {Treu}, {Dijkstra}, {Mesinger}, {Trenti}, {Pentericci}, {de Barros}, \& {Vanzella}}]{Mason18}
{Mason}, C.~A., {Treu}, T., {Dijkstra}, M., {et~al.} 2018, \apj, 856, 2

\bibitem[{{Matthee} {et~al.}(2023){Matthee}, {Mackenzie}, {Simcoe}, {Kashino}, {Lilly}, {Bordoloi}, \& {Eilers}}]{Matthee23a}
{Matthee}, J., {Mackenzie}, R., {Simcoe}, R.~A., {et~al.} 2023, \apj, 950, 67

\bibitem[{{Matthee} {et~al.}(2024){Matthee}, {Naidu}, {Brammer}, {Chisholm}, {Eilers}, {Goulding}, {Greene}, {Kashino}, {Labbe}, {Lilly}, {Mackenzie}, {Oesch}, {Weibel}, {Wuyts}, {Xiao}, {Bordoloi}, {Bouwens}, {van Dokkum}, {Illingworth}, {Kramarenko}, {Maseda}, {Mason}, {Meyer}, {Nelson}, {Reddy}, {Shivaei}, {Simcoe}, \& {Yue}}]{Matthee23b}
{Matthee}, J., {Naidu}, R.~P., {Brammer}, G., {et~al.} 2024, \apj, 963, 129

\bibitem[{{Matthee} {et~al.}(2022){Matthee}, {Naidu}, {Pezzulli}, {Gronke}, {Sobral}, {Oesch}, {Hayes}, {Erb}, {Schaerer}, {Amor{\'\i}n}, {Tacchella}, {Paulino-Afonso}, {Llerena}, {Calhau}, \& {R{\"o}ttgering}}]{Matthee22}
{Matthee}, J., {Naidu}, R.~P., {Pezzulli}, G., {et~al.} 2022, \mnras, 512, 5960

\bibitem[{{Matthee} {et~al.}(2018){Matthee}, {Sobral}, {Gronke}, {Paulino-Afonso}, {Stefanon}, \& {R{\"o}ttgering}}]{Matthee18}
{Matthee}, J., {Sobral}, D., {Gronke}, M., {et~al.} 2018, \aap, 619, A136

\bibitem[{{Matthee} {et~al.}(2021){Matthee}, {Sobral}, {Hayes}, {Pezzulli}, {Gronke}, {Schaerer}, {Naidu}, {R{\"o}ttgering}, {Calhau}, {Paulino-Afonso}, {Santos}, \& {Amor{\'\i}n}}]{Matthee21}
{Matthee}, J., {Sobral}, D., {Hayes}, M., {et~al.} 2021, \mnras, 505, 1382

\bibitem[{{Merlin} {et~al.}(2022){Merlin}, {Bonchi}, {Paris}, {Belfiori}, {Fontana}, {Castellano}, {Nonino}, {Polenta}, {Santini}, {Yang}, {Glazebrook}, {Treu}, {Roberts-Borsani}, {Trenti}, {Birrer}, {Brammer}, {Grillo}, {Calabr{\`o}}, {Marchesini}, {Mason}, {Mercurio}, {Morishita}, {Strait}, {Boyett}, {Leethochawalit}, {Nanayakkara}, {Vulcani}, {Bradac}, \& {Wang}}]{Merlin22}
{Merlin}, E., {Bonchi}, A., {Paris}, D., {et~al.} 2022, \apjl, 938, L14

\bibitem[{{Meyer} {et~al.}(2021){Meyer}, {Laporte}, {Ellis}, {Verhamme}, \& {Garel}}]{Meyer21}
{Meyer}, R.~A., {Laporte}, N., {Ellis}, R.~S., {Verhamme}, A., \& {Garel}, T. 2021, \mnras, 500, 558

\bibitem[{{Miller}(1974)}]{Miller74}
{Miller}, J.~S. 1974, \araa, 12, 331

\bibitem[{{Naidu} {et~al.}(2018){Naidu}, {Forrest}, {Oesch}, {Tran}, \& {Holden}}]{Naidu18}
{Naidu}, R.~P., {Forrest}, B., {Oesch}, P.~A., {Tran}, K.-V.~H., \& {Holden}, B.~P. 2018, \mnras, 478, 791

\bibitem[{{Naidu} {et~al.}(2022{\natexlab{a}}){Naidu}, {Matthee}, {Oesch}, {Conroy}, {Sobral}, {Pezzulli}, {Hayes}, {Erb}, {Amor{\'\i}n}, {Gronke}, {Schaerer}, {Tacchella}, {Kerutt}, {Paulino-Afonso}, {Calhau}, {Llerena}, \& {R{\"o}ttgering}}]{Naidu22}
{Naidu}, R.~P., {Matthee}, J., {Oesch}, P.~A., {et~al.} 2022{\natexlab{a}}, \mnras, 510, 4582

\bibitem[{{Naidu} {et~al.}(2022{\natexlab{b}}){Naidu}, {Oesch}, {Setton}, {Matthee}, {Conroy}, {Johnson}, {Weaver}, {Bouwens}, {Brammer}, {Dayal}, {Illingworth}, {Barrufet}, {Belli}, {Bezanson}, {Bose}, {Heintz}, {Leja}, {Leonova}, {Marques-Chaves}, {Stefanon}, {Toft}, {van der Wel}, {van Dokkum}, {Weibel}, \& {Whitaker}}]{Naidu22schrodinger}
{Naidu}, R.~P., {Oesch}, P.~A., {Setton}, D.~J., {et~al.} 2022{\natexlab{b}}, ApJL, submitted, arXiv:2208.02794

\bibitem[{{Naidu} {et~al.}(2020){Naidu}, {Tacchella}, {Mason}, {Bose}, {Oesch}, \& {Conroy}}]{Naidu20}
{Naidu}, R.~P., {Tacchella}, S., {Mason}, C.~A., {et~al.} 2020, \apj, 892, 109

\bibitem[{{Nakajima} {et~al.}(2018){Nakajima}, {Fletcher}, {Ellis}, {Robertson}, \& {Iwata}}]{Nakajima18}
{Nakajima}, K., {Fletcher}, T., {Ellis}, R.~S., {Robertson}, B.~E., \& {Iwata}, I. 2018, \mnras, 477, 2098

\bibitem[{{Nakajima} {et~al.}(2023){Nakajima}, {Ouchi}, {Isobe}, {Harikane}, {Zhang}, {Ono}, {Umeda}, \& {Oguri}}]{Nakajima23}
{Nakajima}, K., {Ouchi}, M., {Isobe}, Y., {et~al.} 2023, \apjs, 269, 33

\bibitem[{{Nakane} {et~al.}(2024){Nakane}, {Ouchi}, {Nakajima}, {Harikane}, {Ono}, {Umeda}, {Isobe}, {Zhang}, \& {Xu}}]{Nakane23}
{Nakane}, M., {Ouchi}, M., {Nakajima}, K., {et~al.} 2024, \apj, 967, 28

\bibitem[{{Ocvirk} {et~al.}(2020){Ocvirk}, {Aubert}, {Sorce}, {Shapiro}, {Deparis}, {Dawoodbhoy}, {Lewis}, {Teyssier}, {Yepes}, {Gottl{\"o}ber}, {Ahn}, {Iliev}, \& {Hoffman}}]{Ocvirk20}
{Ocvirk}, P., {Aubert}, D., {Sorce}, J.~G., {et~al.} 2020, \mnras, 496, 4087

\bibitem[{{Ocvirk} {et~al.}(2021){Ocvirk}, {Lewis}, {Gillet}, {Chardin}, {Aubert}, {Deparis}, \& {Th{\'e}lie}}]{Ocvirk21}
{Ocvirk}, P., {Lewis}, J. S.~W., {Gillet}, N., {et~al.} 2021, \mnras, 507, 6108

\bibitem[{{Oesch} {et~al.}(2023){Oesch}, {Brammer}, {Naidu}, {Bouwens}, {Chisholm}, {Illingworth}, {Matthee}, {Nelson}, {Qin}, {Reddy}, {Shapley}, {Shivaei}, {van Dokkum}, {Weibel}, {Whitaker}, {Wuyts}, {Covelo-Paz}, {Endsley}, {Fudamoto}, {Giovinazzo}, {Herard-Demanche}, {Kerutt}, {Kramarenko}, {Labbe}, {Leonova}, {Lin}, {Magee}, {Marchesini}, {Maseda}, {Mason}, {Matharu}, {Meyer}, {Neufeld}, {Prieto Lyon}, {Schaerer}, {Sharma}, {Shuntov}, {Smit}, {Stefanon}, {Wyithe}, \& {Xiao}}]{Oesch23}
{Oesch}, P.~A., {Brammer}, G., {Naidu}, R.~P., {et~al.} 2023, \mnras, 525, 2864

\bibitem[{{Oke} \& {Gunn}(1983)}]{Oke83}
{Oke}, J.~B. \& {Gunn}, J.~E. 1983, \apj, 266, 713

\bibitem[{{Osterbrock}(1989)}]{Osterbrock89}
{Osterbrock}, D.~E. 1989, {Astrophysics of gaseous nebulae and active galactic nuclei}

\bibitem[{{Pahl} {et~al.}(2023){Pahl}, {Shapley}, {Steidel}, {Reddy}, {Chen}, {Rudie}, \& {Strom}}]{Pahl23}
{Pahl}, A.~J., {Shapley}, A., {Steidel}, C.~C., {et~al.} 2023, \mnras, 521, 3247

\bibitem[{{Pentericci} {et~al.}(2014){Pentericci}, {Vanzella}, {Fontana}, {Castellano}, {Treu}, {Mesinger}, {Dijkstra}, {Grazian}, {Brada{\v{c}}}, {Conselice}, {Cristiani}, {Dunlop}, {Galametz}, {Giavalisco}, {Giallongo}, {Koekemoer}, {McLure}, {Maiolino}, {Paris}, \& {Santini}}]{Pentericci14}
{Pentericci}, L., {Vanzella}, E., {Fontana}, A., {et~al.} 2014, \apj, 793, 113

\bibitem[{{Pilyugin} {et~al.}(2006){Pilyugin}, {V{\'\i}lchez}, \& {Thuan}}]{Pilyugin06}
{Pilyugin}, L.~S., {V{\'\i}lchez}, J.~M., \& {Thuan}, T.~X. 2006, \mnras, 370, 1928

\bibitem[{{Planck Collaboration} {et~al.}(2016){Planck Collaboration}, {Adam}, {Aghanim}, {Ashdown}, {Aumont}, {Baccigalupi}, {Ballardini}, {Banday}, {Barreiro}, {Bartolo}, {Basak}, {Battye}, {Benabed}, {Bernard}, {Bersanelli}, {Bielewicz}, {Bock}, {Bonaldi}, {Bonavera}, {Bond}, {Borrill}, {Bouchet}, {Boulanger}, {Bucher}, {Burigana}, {Calabrese}, {Cardoso}, {Carron}, {Chiang}, {Colombo}, {Combet}, {Comis}, {Couchot}, {Coulais}, {Crill}, {Curto}, {Cuttaia}, {Davis}, {de Bernardis}, {de Rosa}, {de Zotti}, {Delabrouille}, {Di Valentino}, {Dickinson}, {Diego}, {Dor{\'e}}, {Douspis}, {Ducout}, {Dupac}, {Elsner}, {En{\ss}lin}, {Eriksen}, {Falgarone}, {Fantaye}, {Finelli}, {Forastieri}, {Frailis}, {Fraisse}, {Franceschi}, {Frolov}, {Galeotta}, {Galli}, {Ganga}, {G{\'e}nova-Santos}, {Gerbino}, {Ghosh}, {Gonz{\'a}lez-Nuevo}, {G{\'o}rski}, {Gruppuso}, {Gudmundsson}, {Hansen}, {Helou}, {Henrot-Versill{\'e}}, {Herranz}, {Hivon}, {Huang}, {Ili{\'c}}, {Jaffe}, {Jones}, {Keih{\"a}nen}, {Keskitalo}, {Kisner}, {Knox},
  {Krachmalnicoff}, {Kunz}, {Kurki-Suonio}, {Lagache}, {L{\"a}hteenm{\"a}ki}, {Lamarre}, {Langer}, {Lasenby}, {Lattanzi}, {Lawrence}, {Le Jeune}, {Levrier}, {Lewis}, {Liguori}, {Lilje}, {L{\'o}pez-Caniego}, {Ma}, {Mac{\'\i}as-P{\'e}rez}, {Maggio}, {Mangilli}, {Maris}, {Martin}, {Mart{\'\i}nez-Gonz{\'a}lez}, {Matarrese}, {Mauri}, {McEwen}, {Meinhold}, {Melchiorri}, {Mennella}, {Migliaccio}, {Miville-Desch{\^e}nes}, {Molinari}, {Moneti}, {Montier}, {Morgante}, {Moss}, {Naselsky}, {Natoli}, {Oxborrow}, {Pagano}, {Paoletti}, {Partridge}, {Patanchon}, {Patrizii}, {Perdereau}, {Perotto}, {Pettorino}, {Piacentini}, {Plaszczynski}, {Polastri}, {Polenta}, {Puget}, {Rachen}, {Racine}, {Reinecke}, {Remazeilles}, {Renzi}, {Rocha}, {Rossetti}, {Roudier}, {Rubi{\~n}o-Mart{\'\i}n}, {Ruiz-Granados}, {Salvati}, {Sandri}, {Savelainen}, {Scott}, {Sirri}, {Sunyaev}, {Suur-Uski}, {Tauber}, {Tenti}, {Toffolatti}, {Tomasi}, {Tristram}, {Trombetti}, {Valiviita}, {Van Tent}, {Vielva}, {Villa}, {Vittorio}, {Wandelt}, {Wehus}, {White},
  {Zacchei}, \& {Zonca}}]{Planck16}
{Planck Collaboration}, {Adam}, R., {Aghanim}, N., {et~al.} 2016, \aap, 596, A108

\bibitem[{{Planck Collaboration} {et~al.}(2020){Planck Collaboration}, {Aghanim}, {Akrami}, {Ashdown}, {Aumont}, {Baccigalupi}, {Ballardini}, {Banday}, {Barreiro}, {Bartolo}, {Basak}, {Battye}, {Benabed}, {Bernard}, {Bersanelli}, {Bielewicz}, {Bock}, {Bond}, {Borrill}, {Bouchet}, {Boulanger}, {Bucher}, {Burigana}, {Butler}, {Calabrese}, {Cardoso}, {Carron}, {Challinor}, {Chiang}, {Chluba}, {Colombo}, {Combet}, {Contreras}, {Crill}, {Cuttaia}, {de Bernardis}, {de Zotti}, {Delabrouille}, {Delouis}, {Di Valentino}, {Diego}, {Dor{\'e}}, {Douspis}, {Ducout}, {Dupac}, {Dusini}, {Efstathiou}, {Elsner}, {En{\ss}lin}, {Eriksen}, {Fantaye}, {Farhang}, {Fergusson}, {Fernandez-Cobos}, {Finelli}, {Forastieri}, {Frailis}, {Fraisse}, {Franceschi}, {Frolov}, {Galeotta}, {Galli}, {Ganga}, {G{\'e}nova-Santos}, {Gerbino}, {Ghosh}, {Gonz{\'a}lez-Nuevo}, {G{\'o}rski}, {Gratton}, {Gruppuso}, {Gudmundsson}, {Hamann}, {Handley}, {Hansen}, {Herranz}, {Hildebrandt}, {Hivon}, {Huang}, {Jaffe}, {Jones}, {Karakci}, {Keih{\"a}nen},
  {Keskitalo}, {Kiiveri}, {Kim}, {Kisner}, {Knox}, {Krachmalnicoff}, {Kunz}, {Kurki-Suonio}, {Lagache}, {Lamarre}, {Lasenby}, {Lattanzi}, {Lawrence}, {Le Jeune}, {Lemos}, {Lesgourgues}, {Levrier}, {Lewis}, {Liguori}, {Lilje}, {Lilley}, {Lindholm}, {L{\'o}pez-Caniego}, {Lubin}, {Ma}, {Mac{\'\i}as-P{\'e}rez}, {Maggio}, {Maino}, {Mandolesi}, {Mangilli}, {Marcos-Caballero}, {Maris}, {Martin}, {Martinelli}, {Mart{\'\i}nez-Gonz{\'a}lez}, {Matarrese}, {Mauri}, {McEwen}, {Meinhold}, {Melchiorri}, {Mennella}, {Migliaccio}, {Millea}, {Mitra}, {Miville-Desch{\^e}nes}, {Molinari}, {Montier}, {Morgante}, {Moss}, {Natoli}, {N{\o}rgaard-Nielsen}, {Pagano}, {Paoletti}, {Partridge}, {Patanchon}, {Peiris}, {Perrotta}, {Pettorino}, {Piacentini}, {Polastri}, {Polenta}, {Puget}, {Rachen}, {Reinecke}, {Remazeilles}, {Renzi}, {Rocha}, {Rosset}, {Roudier}, {Rubi{\~n}o-Mart{\'\i}n}, {Ruiz-Granados}, {Salvati}, {Sandri}, {Savelainen}, {Scott}, {Shellard}, {Sirignano}, {Sirri}, {Spencer}, {Sunyaev}, {Suur-Uski}, {Tauber}, {Tavagnacco},
  {Tenti}, {Toffolatti}, {Tomasi}, {Trombetti}, {Valenziano}, {Valiviita}, {Van Tent}, {Vibert}, {Vielva}, {Villa}, {Vittorio}, {Wandelt}, {Wehus}, {White}, {White}, {Zacchei}, \& {Zonca}}]{Planck18}
{Planck Collaboration}, {Aghanim}, N., {Akrami}, Y., {et~al.} 2020, \aap, 641, A6

\bibitem[{{Qin} {et~al.}(2022){Qin}, {Wyithe}, {Oesch}, {Illingworth}, {Leonova}, {Mutch}, \& {Naidu}}]{Qin22}
{Qin}, Y., {Wyithe}, J. S.~B., {Oesch}, P.~A., {et~al.} 2022, \mnras, 510, 3858

\bibitem[{{Raiter} {et~al.}(2010){Raiter}, {Fosbury}, \& {Teimoorinia}}]{Raiter10}
{Raiter}, A., {Fosbury}, R.~A.~E., \& {Teimoorinia}, H. 2010, \aap, 510, A109

\bibitem[{{Reddy} {et~al.}(2020){Reddy}, {Shapley}, {Kriek}, {Steidel}, {Shivaei}, {Sanders}, {Mobasher}, {Coil}, {Siana}, {Freeman}, {Azadi}, {Fetherolf}, {Leung}, {Price}, \& {Zick}}]{Reddy20}
{Reddy}, N.~A., {Shapley}, A.~E., {Kriek}, M., {et~al.} 2020, \apj, 902, 123

\bibitem[{{Reddy} {et~al.}(2018){Reddy}, {Shapley}, {Sanders}, {Kriek}, {Coil}, {Shivaei}, {Freeman}, {Mobasher}, {Siana}, {Azadi}, {Fetherolf}, {Fornasini}, {Leung}, {Price}, {Zick}, \& {Barro}}]{Reddy18}
{Reddy}, N.~A., {Shapley}, A.~E., {Sanders}, R.~L., {et~al.} 2018, \apj, 869, 92

\bibitem[{{Robertson}(2022)}]{Robertson22}
{Robertson}, B.~E. 2022, \araa, 60, 121

\bibitem[{{Rosdahl} {et~al.}(2022){Rosdahl}, {Blaizot}, {Katz}, {Kimm}, {Garel}, {Haehnelt}, {Keating}, {Martin-Alvarez}, {Michel-Dansac}, \& {Ocvirk}}]{Rosdahl22}
{Rosdahl}, J., {Blaizot}, J., {Katz}, H., {et~al.} 2022, \mnras, 515, 2386

\bibitem[{{Roy} {et~al.}(2023){Roy}, {Henry}, {Treu}, {Jones}, {Prieto-Lyon}, {Mason}, {Heckman}, {Nanayakkara}, {Pentericci}, {Mascia}, {Brada{\v{c}}}, {Vanzella}, {Scarlata}, {Boyett}, {Trenti}, \& {Wang}}]{Roy23}
{Roy}, N., {Henry}, A., {Treu}, T., {et~al.} 2023, \apjl, 952, L14

\bibitem[{{Saxena} {et~al.}(2024){Saxena}, {Bunker}, {Jones}, {Stark}, {Cameron}, {Witstok}, {Arribas}, {Baker}, {Baum}, {Bhatawdekar}, {Bowler}, {Boyett}, {Carniani}, {Charlot}, {Chevallard}, {Curti}, {Curtis-Lake}, {Eisenstein}, {Endsley}, {Hainline}, {Helton}, {Johnson}, {Kumari}, {Looser}, {Maiolino}, {Rieke}, {Rix}, {Robertson}, {Sandles}, {Simmonds}, {Smit}, {Tacchella}, {Williams}, {Willmer}, \& {Willott}}]{Saxena23}
{Saxena}, A., {Bunker}, A.~J., {Jones}, G.~C., {et~al.} 2024, \aap, 684, A84

\bibitem[{{Saxena} {et~al.}(2022){Saxena}, {Cryer}, {Ellis}, {Pentericci}, {Calabr{\`o}}, {Mascia}, {Saldana-Lopez}, {Schaerer}, {Katz}, {Llerena}, \& {Amor{\'\i}n}}]{Saxena22}
{Saxena}, A., {Cryer}, E., {Ellis}, R.~S., {et~al.} 2022, \mnras, 517, 1098

\bibitem[{{Saxena} {et~al.}(2023){Saxena}, {Robertson}, {Bunker}, {Endsley}, {Cameron}, {Charlot}, {Simmonds}, {Tacchella}, {Witstok}, {Willott}, {Carniani}, {Curtis-Lake}, {Ferruit}, {Jakobsen}, {Arribas}, {Chevallard}, {Curti}, {D'Eugenio}, {De Graaff}, {Jones}, {Looser}, {Maseda}, {Rawle}, {Rix}, {Del Pino}, {Smit}, {{\"U}bler}, {Eisenstein}, {Hainline}, {Hausen}, {Johnson}, {Rieke}, {Williams}, {Willmer}, {Baker}, {Bhatawdekar}, {Bowler}, {Boyett}, {Chen}, {Egami}, {Ji}, {Kumari}, {Nelson}, {Perna}, {Sandles}, {Scholtz}, \& {Shivaei}}]{Saxena23b}
{Saxena}, A., {Robertson}, B.~E., {Bunker}, A.~J., {et~al.} 2023, \aap, 678, A68

\bibitem[{{Schaerer}(2003)}]{Schaerer03}
{Schaerer}, D. 2003, \aap, 397, 527

\bibitem[{{Schaerer} {et~al.}(2016){Schaerer}, {Izotov}, {Verhamme}, {Orlitov{\'a}}, {Thuan}, {Worseck}, \& {Guseva}}]{Schaerer16}
{Schaerer}, D., {Izotov}, Y.~I., {Verhamme}, A., {et~al.} 2016, \aap, 591, L8

\bibitem[{{Schaerer} {et~al.}(2022){Schaerer}, {Izotov}, {Worseck}, {Berg}, {Chisholm}, {Jaskot}, {Nakajima}, {Ravindranath}, {Thuan}, \& {Verhamme}}]{Schaerer22}
{Schaerer}, D., {Izotov}, Y.~I., {Worseck}, G., {et~al.} 2022, \aap, 658, L11

\bibitem[{{Sharma} {et~al.}(2018){Sharma}, {Theuns}, \& {Frenk}}]{Sharma18}
{Sharma}, M., {Theuns}, T., \& {Frenk}, C. 2018, \mnras, 477, L111

\bibitem[{{Shen} {et~al.}(2020){Shen}, {Hopkins}, {Faucher-Gigu{\`e}re}, {Alexander}, {Richards}, {Ross}, \& {Hickox}}]{Shen20}
{Shen}, X., {Hopkins}, P.~F., {Faucher-Gigu{\`e}re}, C.-A., {et~al.} 2020, \mnras, 495, 3252

\bibitem[{{Shibuya} {et~al.}(2015){Shibuya}, {Ouchi}, \& {Harikane}}]{Shibuya15}
{Shibuya}, T., {Ouchi}, M., \& {Harikane}, Y. 2015, \apjs, 219, 15

\bibitem[{{Simmonds} {et~al.}(2024){Simmonds}, {Tacchella}, {Hainline}, {Johnson}, {McClymont}, {Robertson}, {Saxena}, {Sun}, {Witten}, {Baker}, {Bhatawdekar}, {Boyett}, {Bunker}, {Charlot}, {Curtis-Lake}, {Egami}, {Eisenstein}, {Hausen}, {Maiolino}, {Maseda}, {Scholtz}, {Williams}, {Willott}, \& {Witstok}}]{Simmonds24}
{Simmonds}, C., {Tacchella}, S., {Hainline}, K., {et~al.} 2024, \mnras, 527, 6139

\bibitem[{{Simmonds} {et~al.}(2023){Simmonds}, {Tacchella}, {Maseda}, {Williams}, {Baker}, {Witten}, {Johnson}, {Robertson}, {Saxena}, {Sun}, {Witstok}, {Bhatawdekar}, {Boyett}, {Bunker}, {Charlot}, {Curtis-Lake}, {Egami}, {Eisenstein}, {Ji}, {Maiolino}, {Sandles}, {Smit}, {{\"U}bler}, \& {Willott}}]{Simmonds23}
{Simmonds}, C., {Tacchella}, S., {Maseda}, M., {et~al.} 2023, \mnras, 523, 5468

\bibitem[{{Sobral} \& {Matthee}(2019)}]{sobralmatthee19}
{Sobral}, D. \& {Matthee}, J. 2019, \aap, 623, A157

\bibitem[{{Songaila} {et~al.}(2018){Songaila}, {Hu}, {Barger}, {Cowie}, {Hasinger}, {Rosenwasser}, \& {Waters}}]{Songaila18}
{Songaila}, A., {Hu}, E.~M., {Barger}, A.~J., {et~al.} 2018, \apj, 859, 91

\bibitem[{{Steidel} {et~al.}(2018){Steidel}, {Bogosavljevi{\'c}}, {Shapley}, {Reddy}, {Rudie}, {Pettini}, {Trainor}, \& {Strom}}]{Steidel18}
{Steidel}, C.~C., {Bogosavljevi{\'c}}, M., {Shapley}, A.~E., {et~al.} 2018, \apj, 869, 123

\bibitem[{Storey \& Zeippen(2000)}]{Storey00}
Storey, P.~J. \& Zeippen, C.~J. 2000, Monthly Notices of the Royal Astronomical Society, 312, 813

\bibitem[{{Tacchella} {et~al.}(2022){Tacchella}, {Finkelstein}, {Bagley}, {Dickinson}, {Ferguson}, {Giavalisco}, {Graziani}, {Grogin}, {Hathi}, {Hutchison}, {Jung}, {Koekemoer}, {Larson}, {Papovich}, {Pirzkal}, {Rojas-Ruiz}, {Song}, {Schneider}, {Somerville}, {Wilkins}, \& {Yung}}]{Tacchella22}
{Tacchella}, S., {Finkelstein}, S.~L., {Bagley}, M., {et~al.} 2022, \apj, 927, 170

\bibitem[{{Tacchella} {et~al.}(2023){Tacchella}, {Johnson}, {Robertson}, {Carniani}, {D'Eugenio}, {Kumari}, {Maiolino}, {Nelson}, {Suess}, {{\"U}bler}, {Williams}, {Adebusola}, {Alberts}, {Arribas}, {Bhatawdekar}, {Bonaventura}, {Bowler}, {Bunker}, {Cameron}, {Curti}, {Egami}, {Eisenstein}, {Frye}, {Hainline}, {Helton}, {Ji}, {Looser}, {Lyu}, {Perna}, {Rawle}, {Rieke}, {Rieke}, {Saxena}, {Sandles}, {Shivaei}, {Simmonds}, {Sun}, {Willmer}, {Willott}, \& {Witstok}}]{Tacchella23}
{Tacchella}, S., {Johnson}, B.~D., {Robertson}, B.~E., {et~al.} 2023, \mnras, 522, 6236

\bibitem[{{Tang} {et~al.}(2023){Tang}, {Stark}, {Chen}, {Mason}, {Topping}, {Endsley}, {Senchyna}, {Plat}, {Lu}, {Whitler}, {Robertson}, \& {Charlot}}]{Tang23}
{Tang}, M., {Stark}, D.~P., {Chen}, Z., {et~al.} 2023, \mnras, 526, 1657

\bibitem[{{Theios} {et~al.}(2019){Theios}, {Steidel}, {Strom}, {Rudie}, {Trainor}, \& {Reddy}}]{Theios19}
{Theios}, R.~L., {Steidel}, C.~C., {Strom}, A.~L., {et~al.} 2019, \apj, 871, 128

\bibitem[{{Topping} {et~al.}(2022){Topping}, {Stark}, {Endsley}, {Plat}, {Whitler}, {Chen}, \& {Charlot}}]{Topping22}
{Topping}, M.~W., {Stark}, D.~P., {Endsley}, R., {et~al.} 2022, \apj, 941, 153

\bibitem[{{Topping} {et~al.}(2024{\natexlab{a}}){Topping}, {Stark}, {Endsley}, {Whitler}, {Hainline}, {Johnson}, {Robertson}, {Tacchella}, {Chen}, {Alberts}, {Baker}, {Bunker}, {Carniani}, {Charlot}, {Chevallard}, {Curtis-Lake}, {DeCoursey}, {Egami}, {Eisenstein}, {Ji}, {Maiolino}, {Williams}, {Willmer}, {Willott}, \& {Witstok}}]{Topping23}
{Topping}, M.~W., {Stark}, D.~P., {Endsley}, R., {et~al.} 2024{\natexlab{a}}, \mnras, 529, 4087

\bibitem[{{Topping} {et~al.}(2024{\natexlab{b}}){Topping}, {Stark}, {Senchyna}, {Plat}, {Zitrin}, {Endsley}, {Charlot}, {Furtak}, {Maseda}, {Smit}, {Mainali}, {Chevallard}, {Molyneux}, \& {Rigby}}]{Topping24}
{Topping}, M.~W., {Stark}, D.~P., {Senchyna}, P., {et~al.} 2024{\natexlab{b}}, \mnras, 529, 3301

\bibitem[{{Trebitsch} {et~al.}(2021){Trebitsch}, {Dubois}, {Volonteri}, {Pfister}, {Cadiou}, {Katz}, {Rosdahl}, {Kimm}, {Pichon}, {Beckmann}, {Devriendt}, \& {Slyz}}]{Trebitsch21}
{Trebitsch}, M., {Dubois}, Y., {Volonteri}, M., {et~al.} 2021, \aap, 653, A154

\bibitem[{{{\"U}bler} {et~al.}(2024){{\"U}bler}, {Maiolino}, {P{\'e}rez-Gonz{\'a}lez}, {D'Eugenio}, {Perna}, {Curti}, {Arribas}, {Bunker}, {Carniani}, {Charlot}, {Rodr{\'\i}guez Del Pino}, {Baker}, {B{\"o}ker}, {Cresci}, {Dunlop}, {Grogin}, {Jones}, {Kumari}, {Lamperti}, {Laporte}, {Marshall}, {Mazzolari}, {Parlanti}, {Rawle}, {Scholtz}, {Venturi}, \& {Witstok}}]{Ubler23}
{{\"U}bler}, H., {Maiolino}, R., {P{\'e}rez-Gonz{\'a}lez}, P.~G., {et~al.} 2024, \mnras, 531, 355

\bibitem[{{Vanzella} {et~al.}(2022){Vanzella}, {Castellano}, {Bergamini}, {Meneghetti}, {Zanella}, {Calura}, {Caminha}, {Rosati}, {Cupani}, {Me{\v{s}}tri{\'c}}, {Brammer}, {Tozzi}, {Mercurio}, {Grillo}, {Sani}, {Cristiani}, {Nonino}, {Merlin}, \& {Pignataro}}]{Vanzella22}
{Vanzella}, E., {Castellano}, M., {Bergamini}, P., {et~al.} 2022, \aap, 659, A2

\bibitem[{{Vanzella} {et~al.}(2016){Vanzella}, {de Barros}, {Vasei}, {Alavi}, {Giavalisco}, {Siana}, {Grazian}, {Hasinger}, {Suh}, {Cappelluti}, {Vito}, {Amorin}, {Balestra}, {Brusa}, {Calura}, {Castellano}, {Comastri}, {Fontana}, {Gilli}, {Mignoli}, {Pentericci}, {Vignali}, \& {Zamorani}}]{Vanzella16}
{Vanzella}, E., {de Barros}, S., {Vasei}, K., {et~al.} 2016, \apj, 825, 41

\bibitem[{{Vanzella} {et~al.}(2018){Vanzella}, {Nonino}, {Cupani}, {Castellano}, {Sani}, {Mignoli}, {Calura}, {Meneghetti}, {Gilli}, {Comastri}, {Mercurio}, {Caminha}, {Caputi}, {Rosati}, {Grillo}, {Cristiani}, {Balestra}, {Fontana}, \& {Giavalisco}}]{Vanzella18}
{Vanzella}, E., {Nonino}, M., {Cupani}, G., {et~al.} 2018, \mnras, 476, L15

\bibitem[{{Verhamme} {et~al.}(2015){Verhamme}, {Orlitov{\'a}}, {Schaerer}, \& {Hayes}}]{Verhamme15}
{Verhamme}, A., {Orlitov{\'a}}, I., {Schaerer}, D., \& {Hayes}, M. 2015, \aap, 578, A7

\bibitem[{{Verhamme} {et~al.}(2017){Verhamme}, {Orlitov{\'a}}, {Schaerer}, {Izotov}, {Worseck}, {Thuan}, \& {Guseva}}]{Verhamme17}
{Verhamme}, A., {Orlitov{\'a}}, I., {Schaerer}, D., {et~al.} 2017, \aap, 597, A13

\bibitem[{{Verhamme} {et~al.}(2006){Verhamme}, {Schaerer}, \& {Maselli}}]{Verhamme06}
{Verhamme}, A., {Schaerer}, D., \& {Maselli}, A. 2006, \aap, 460, 397

\bibitem[{{Villa-V{\'e}lez} {et~al.}(2021){Villa-V{\'e}lez}, {Buat}, {Theul{\'e}}, {Boquien}, \& {Burgarella}}]{Villa-Velez21}
{Villa-V{\'e}lez}, J.~A., {Buat}, V., {Theul{\'e}}, P., {Boquien}, M., \& {Burgarella}, D. 2021, \aap, 654, A153

\bibitem[{{Weaver} {et~al.}(2022){Weaver}, {Kauffmann}, {Ilbert}, {McCracken}, {Moneti}, {Toft}, {Brammer}, {Shuntov}, {Davidzon}, {Hsieh}, {Laigle}, {Anastasiou}, {Jespersen}, {Vinther}, {Capak}, {Casey}, {McPartland}, {Milvang-Jensen}, {Mobasher}, {Sanders}, {Zalesky}, {Arnouts}, {Aussel}, {Dunlop}, {Faisst}, {Franx}, {Furtak}, {Fynbo}, {Gould}, {Greve}, {Gwyn}, {Kartaltepe}, {Kashino}, {Koekemoer}, {Kokorev}, {Le F{\`e}vre}, {Lilly}, {Masters}, {Magdis}, {Mehta}, {Peng}, {Riechers}, {Salvato}, {Sawicki}, {Scarlata}, {Scoville}, {Shirley}, {Silverman}, {Sneppen}, {Smolc̆i{\'c}}, {Steinhardt}, {Stern}, {Tanaka}, {Taniguchi}, {Teplitz}, {Vaccari}, {Wang}, \& {Zamorani}}]{Weaver22}
{Weaver}, J.~R., {Kauffmann}, O.~B., {Ilbert}, O., {et~al.} 2022, \apjs, 258, 11

\bibitem[{{Weinberger} {et~al.}(2018){Weinberger}, {Kulkarni}, {Haehnelt}, {Choudhury}, \& {Puchwein}}]{Weinberger18}
{Weinberger}, L.~H., {Kulkarni}, G., {Haehnelt}, M.~G., {Choudhury}, T.~R., \& {Puchwein}, E. 2018, \mnras, 479, 2564

\bibitem[{{Whitler} {et~al.}(2024){Whitler}, {Stark}, {Endsley}, {Chen}, {Mason}, {Topping}, \& {Charlot}}]{Whitler23b}
{Whitler}, L., {Stark}, D.~P., {Endsley}, R., {et~al.} 2024, \mnras, 529, 855

\bibitem[{{Wilkins} {et~al.}(2013){Wilkins}, {Bunker}, {Coulton}, {Croft}, {di Matteo}, {Khandai}, \& {Feng}}]{Wilkins13}
{Wilkins}, S.~M., {Bunker}, A., {Coulton}, W., {et~al.} 2013, \mnras, 430, 2885

\bibitem[{{Wilkins} {et~al.}(2012){Wilkins}, {Gonzalez-Perez}, {Lacey}, \& {Baugh}}]{Wilkins12}
{Wilkins}, S.~M., {Gonzalez-Perez}, V., {Lacey}, C.~G., \& {Baugh}, C.~M. 2012, \mnras, 424, 1522

\bibitem[{{Windhorst} {et~al.}(2023){Windhorst}, {Cohen}, {Jansen}, {Summers}, {Tompkins}, {Conselice}, {Driver}, {Yan}, {Coe}, {Frye}, {Grogin}, {Koekemoer}, {Marshall}, {O'Brien}, {Pirzkal}, {Robotham}, {Ryan}, {Willmer}, {Carleton}, {Diego}, {Keel}, {Porto}, {Redshaw}, {Scheller}, {Wilkins}, {Willner}, {Zitrin}, {Adams}, {Austin}, {Arendt}, {Beacom}, {Bhatawdekar}, {Bradley}, {Broadhurst}, {Cheng}, {Civano}, {Dai}, {Dole}, {D'Silva}, {Duncan}, {Fazio}, {Ferrami}, {Ferreira}, {Finkelstein}, {Furtak}, {Gim}, {Griffiths}, {Hammel}, {Harrington}, {Hathi}, {Holwerda}, {Honor}, {Huang}, {Hyun}, {Im}, {Joshi}, {Kamieneski}, {Kelly}, {Larson}, {Li}, {Lim}, {Ma}, {Maksym}, {Manzoni}, {Meena}, {Milam}, {Nonino}, {Pascale}, {Petric}, {Pierel}, {del Carmen Polletta}, {R{\"o}ttgering}, {Rutkowski}, {Smail}, {Straughn}, {Strolger}, {Swirbul}, {Trussler}, {Wang}, {Welch}, {B. Wyithe}, {Yun}, {Zackrisson}, {Zhang}, \& {Zhao}}]{Windhorst23}
{Windhorst}, R.~A., {Cohen}, S.~H., {Jansen}, R.~A., {et~al.} 2023, \aj, 165, 13

\bibitem[{{Witstok} {et~al.}(2024){Witstok}, {Smit}, {Saxena}, {Jones}, {Helton}, {Sun}, {Maiolino}, {Kumari}, {Stark}, {Bunker}, {Arribas}, {Baker}, {Bhatawdekar}, {Boyett}, {Cameron}, {Carniani}, {Charlot}, {Chevallard}, {Curti}, {Curtis-Lake}, {Eisenstein}, {Endsley}, {Hainline}, {Ji}, {Johnson}, {Looser}, {Nelson}, {Perna}, {Rix}, {Robertson}, {Sandles}, {Scholtz}, {Simmonds}, {Tacchella}, {{\"U}bler}, {Williams}, {Willmer}, \& {Willott}}]{Wistok23}
{Witstok}, J., {Smit}, R., {Saxena}, A., {et~al.} 2024, \aap, 682, A40

\bibitem[{{Zackrisson} {et~al.}(2013){Zackrisson}, {Inoue}, \& {Jensen}}]{Zackrisson13}
{Zackrisson}, E., {Inoue}, A.~K., \& {Jensen}, H. 2013, \apj, 777, 39

\end{thebibliography}

\appendix

\section{C1F \OIII\ emitter sample catalog}\label{sec:O3_cat}

In Table~\ref{tab:O3_cat} we list the C1F sample, composed of 141 \OIII\ emitters, selected as is described in Sect.~\ref{sec:doublet_identification}. We tabulate the IDs, sky coordinates of the F356W detection in the J2000 system, \OIII\ doublet redshifts (see Sect.~\ref{sec:z_and_f_measurement}), \texttt{Prospector} UV magnitudes (see Sect.~\ref{sec:SED_fitting}) and \OIII$\lambda$5008 luminosities measured from the grism data.

\begin{table*}
    \centering
    \caption{Catalog of the C1F \OIII\ emitter sample.}
    \label{tab:O3_cat}
    \resizebox{0.47\linewidth}{!}{
    \begin{tabular}{rccccc}
        \toprule
        ID & RA & DEC & $z$ & $M_{\rm UV}$ & \makecell{$\log_{10}(L_{\rm \OIII\lambda 5008}$\\$/{\rm erg\,s}^{-1})$}\\
        \midrule
        1628 & 10:02:40.92s & 02$^\circ$12\arcmin27.17\arcsec & 5.353 & $-19.47\pm 0.07$ & $42.05\pm 0.06$ \\
9199 & 10:02:37.01s & 02$^\circ$13\arcmin10.64\arcsec & 5.362 & $-17.52\pm 0.28$ & $42.01\pm 0.07$ \\
2756 & 10:02:37.43s & 02$^\circ$10\arcmin55.89\arcsec & 5.363 & $-18.34\pm 0.29$ & $42.22\pm 0.03$ \\
14494 & 10:02:32.98s & 02$^\circ$13\arcmin17.66\arcsec & 5.364 & $-18.53\pm 0.39$ & $41.92\pm 0.07$ \\
8985 & 10:02:37.47s & 02$^\circ$13\arcmin21.03\arcsec & 5.365 & $-18.29\pm 0.28$ & $42.11\pm 0.05$ \\
5587 & 10:02:39.37s & 02$^\circ$13\arcmin00.41\arcsec & 5.368 & $-18.42\pm 0.12$ & $42.10\pm 0.07$ \\
9060 & 10:02:36.30s & 02$^\circ$12\arcmin41.41\arcsec & 5.369 & $-20.42\pm 0.07$ & $42.91\pm 0.01$ \\
12554 & 10:02:31.21s & 02$^\circ$12\arcmin47.95\arcsec & 5.369 & $-19.91\pm 0.10$ & $42.47\pm 0.02$ \\
12017 & 10:02:30.38s & 02$^\circ$12\arcmin29.70\arcsec & 5.370 & $-19.86\pm 0.13$ & $42.80\pm 0.02$ \\
18147 & 10:02:35.15s & 02$^\circ$12\arcmin53.17\arcsec & 5.371 & $-17.71\pm 0.38$ & $41.96\pm 0.10$ \\
9165 & 10:02:37.41s & 02$^\circ$13\arcmin21.45\arcsec & 5.372 & $-19.91\pm 0.10$ & $42.30\pm 0.03$ \\
17602 & 10:02:34.32s & 02$^\circ$12\arcmin46.06\arcsec & 5.373 & $-17.95\pm 0.24$ & $41.91\pm 0.06$ \\
9161 & 10:02:36.20s & 02$^\circ$12\arcmin42.01\arcsec & 5.373 & $-19.20\pm 0.11$ & $42.39\pm 0.03$ \\
17961 & 10:02:35.30s & 02$^\circ$13\arcmin09.72\arcsec & 5.375 & $-19.10\pm 0.11$ & $42.30\pm 0.02$ \\
7921 & 10:02:37.64s & 02$^\circ$13\arcmin01.13\arcsec & 5.375 & $-17.44\pm 0.46$ & $41.84\pm 0.05$ \\
12966 & 10:02:30.38s & 02$^\circ$12\arcmin11.41\arcsec & 5.376 & $-19.23\pm 0.18$ & $42.05\pm 0.08$ \\
5888 & 10:02:33.48s & 02$^\circ$09\arcmin53.88\arcsec & 5.376 & $-18.66\pm 0.16$ & $42.01\pm 0.03$ \\
6212 & 10:02:35.36s & 02$^\circ$11\arcmin03.78\arcsec & 5.376 & $-18.65\pm 0.16$ & $42.21\pm 0.03$ \\
19319 & 10:02:36.30s & 02$^\circ$13\arcmin03.88\arcsec & 5.377 & $-20.46\pm 0.09$ & $42.20\pm 0.06$ \\
15579 & 10:02:31.24s & 02$^\circ$12\arcmin01.95\arcsec & 5.378 & $-19.32\pm 0.07$ & $42.50\pm 0.01$ \\
11580 & 10:02:29.23s & 02$^\circ$12\arcmin06.52\arcsec & 5.379 & $-19.34\pm 0.11$ & $42.31\pm 0.03$ \\
7152 & 10:02:39.81s & 02$^\circ$13\arcmin54.07\arcsec & 5.380 & $-19.87\pm 0.06$ & $42.19\pm 0.04$ \\
7375 & 10:02:40.17s & 02$^\circ$14\arcmin11.42\arcsec & 5.414 & $-19.53\pm 0.11$ & $42.25\pm 0.03$ \\
6611 & 10:02:41.50s & 02$^\circ$14\arcmin34.35\arcsec & 5.415 & $-19.72\pm 0.11$ & $42.51\pm 0.02$ \\
14699 & 10:02:31.21s & 02$^\circ$12\arcmin08.06\arcsec & 5.439 & $-17.68\pm 0.41$ & $41.93\pm 0.06$ \\
2140 & 10:02:38.11s & 02$^\circ$11\arcmin06.21\arcsec & 5.440 & $-18.21\pm 0.23$ & $42.07\pm 0.04$ \\
16304 & 10:02:34.43s & 02$^\circ$13\arcmin17.27\arcsec & 5.447 & $-19.84\pm 0.09$ & $42.56\pm 0.02$ \\
3563 & 10:02:37.19s & 02$^\circ$11\arcmin02.24\arcsec & 5.448 & $-19.56\pm 0.12$ & $42.47\pm 0.04$ \\
4355 & 10:02:42.18s & 02$^\circ$14\arcmin01.98\arcsec & 5.449 & $-20.57\pm 0.06$ & $42.75\pm 0.02$ \\
5466 & 10:02:41.72s & 02$^\circ$14\arcmin15.21\arcsec & 5.452 & $-18.55\pm 0.07$ & $41.88\pm 0.06$ \\
1257 & 10:02:44.06s & 02$^\circ$14\arcmin04.98\arcsec & 5.452 & $-18.99\pm 0.27$ & $42.34\pm 0.07$ \\
485 & 10:02:44.52s & 02$^\circ$13\arcmin59.56\arcsec & 5.452 & $-19.51\pm 0.12$ & $42.19\pm 0.05$ \\
803 & 10:02:44.28s & 02$^\circ$14\arcmin00.56\arcsec & 5.453 & $-20.26\pm 0.08$ & $42.66\pm 0.01$ \\
1272 & 10:02:44.29s & 02$^\circ$14\arcmin10.56\arcsec & 5.453 & $-19.17\pm 0.22$ & $42.46\pm 0.04$ \\
2294 & 10:02:42.92s & 02$^\circ$13\arcmin47.68\arcsec & 5.470 & $-17.07\pm 0.29$ & $41.81\pm 0.12$ \\
3988 & 10:02:33.83s & 02$^\circ$09\arcmin21.29\arcsec & 5.487 & $-19.52\pm 0.09$ & $42.22\pm 0.03$ \\
13704 & 10:02:36.47s & 02$^\circ$13\arcmin08.53\arcsec & 5.500 & $-19.02\pm 0.05$ & $42.10\pm 0.05$ \\
1180 & 10:02:43.97s & 02$^\circ$13\arcmin59.09\arcsec & 5.500 & $-19.09\pm 0.14$ & $42.53\pm 0.03$ \\
1450 & 10:02:35.75s & 02$^\circ$09\arcmin35.84\arcsec & 5.524 & $-17.73\pm 0.19$ & $41.80\pm 0.07$ \\
6489 & 10:02:33.24s & 02$^\circ$10\arcmin01.89\arcsec & 5.541 & $-17.20\pm 0.34$ & $41.84\pm 0.07$ \\
5864 & 10:02:41.67s & 02$^\circ$14\arcmin21.91\arcsec & 5.548 & $-17.67\pm 0.15$ & $41.92\pm 0.03$ \\
5267 & 10:02:41.44s & 02$^\circ$14\arcmin01.36\arcsec & 5.549 & $-17.49\pm 0.24$ & $41.61\pm 0.07$ \\
9276 & 10:02:30.67s & 02$^\circ$09\arcmin42.83\arcsec & 5.559 & $-18.58\pm 0.38$ & $41.96\pm 0.04$ \\
4631 & 10:02:41.91s & 02$^\circ$14\arcmin00.10\arcsec & 5.575 & $-18.61\pm 0.14$ & $42.04\pm 0.05$ \\
6510 & 10:02:41.89s & 02$^\circ$14\arcmin46.99\arcsec & 5.629 & $-17.82\pm 0.17$ & $41.83\pm 0.10$ \\
16499 & 10:02:36.27s & 02$^\circ$14\arcmin16.58\arcsec & 5.697 & $-18.55\pm 0.21$ & $42.11\pm 0.03$ \\
16205 & 10:02:36.31s & 02$^\circ$14\arcmin21.48\arcsec & 5.698 & $-20.62\pm 0.07$ & $43.10\pm 0.01$ \\
7826 & 10:02:39.98s & 02$^\circ$14\arcmin15.56\arcsec & 5.703 & $-18.49\pm 0.09$ & $42.14\pm 0.03$ \\
16463 & 10:02:37.72s & 02$^\circ$15\arcmin04.38\arcsec & 5.724 & $-18.31\pm 0.33$ & $42.04\pm 0.05$ \\
3570 & 10:02:36.99s & 02$^\circ$10\arcmin58.08\arcsec & 5.731 & $-18.00\pm 0.28$ & $42.04\pm 0.07$ \\
3932 & 10:02:40.23s & 02$^\circ$12\arcmin50.47\arcsec & 5.737 & $-19.67\pm 0.10$ & $42.39\pm 0.03$ \\
6408 & 10:02:38.87s & 02$^\circ$13\arcmin04.30\arcsec & 5.765 & $-18.36\pm 0.17$ & $41.83\pm 0.06$ \\
6135 & 10:02:38.99s & 02$^\circ$13\arcmin01.70\arcsec & 5.768 & $-17.04\pm 0.42$ & $41.79\pm 0.11$ \\
12951 & 10:02:35.13s & 02$^\circ$14\arcmin49.33\arcsec & 5.772 & $-20.05\pm 0.06$ & $42.73\pm 0.01$ \\
18558 & 10:02:35.62s & 02$^\circ$12\arcmin59.77\arcsec & 5.773 & $-17.82\pm 0.20$ & $42.13\pm 0.04$ \\
17085 & 10:02:29.67s & 02$^\circ$10\arcmin22.02\arcsec & 5.773 & $-17.68\pm 0.20$ & $41.75\pm 0.09$ \\
7037 & 10:02:36.32s & 02$^\circ$11\arcmin55.88\arcsec & 5.775 & $-18.79\pm 0.17$ & $42.06\pm 0.05$ \\
16724 & 10:02:29.27s & 02$^\circ$10\arcmin18.72\arcsec & 5.775 & $-19.73\pm 0.06$ & $42.43\pm 0.02$ \\
6384 & 10:02:35.74s & 02$^\circ$11\arcmin20.44\arcsec & 5.777 & $-18.30\pm 0.14$ & $42.12\pm 0.03$ \\
18752 & 10:02:29.34s & 02$^\circ$09\arcmin15.90\arcsec & 5.792 & $-21.10\pm 0.12$ & $43.26\pm 0.01$ \\
9709 & 10:02:29.61s & 02$^\circ$09\arcmin15.12\arcsec & 5.795 & $-21.68\pm 0.06$ & $42.78\pm 0.03$ \\
17097 & 10:02:28.40s & 02$^\circ$09\arcmin39.26\arcsec & 5.800 & $-19.94\pm 0.09$ & $42.24\pm 0.06$ \\
19268 & 10:02:35.33s & 02$^\circ$12\arcmin34.03\arcsec & 5.802 & $-18.79\pm 0.11$ & $41.94\pm 0.04$ \\
9876 & 10:02:35.60s & 02$^\circ$12\arcmin29.33\arcsec & 5.803 & $-17.43\pm 0.18$ & $41.89\pm 0.04$ \\
9741 & 10:02:36.27s & 02$^\circ$12\arcmin54.03\arcsec & 5.847 & $-18.85\pm 0.12$ & $41.69\pm 0.09$ \\
9289 & 10:02:30.07s & 02$^\circ$09\arcmin24.14\arcsec & 5.848 & $-18.89\pm 0.17$ & $42.81\pm 0.02$ \\
16805 & 10:02:33.23s & 02$^\circ$12\arcmin32.95\arcsec & 5.850 & $-18.78\pm 0.27$ & $41.99\pm 0.04$ \\
17335 & 10:02:38.44s & 02$^\circ$15\arcmin03.18\arcsec & 5.854 & $-17.88\pm 0.23$ & $41.74\pm 0.10$ \\
8125 & 10:02:39.92s & 02$^\circ$14\arcmin20.68\arcsec & 5.856 & $-17.40\pm 0.45$ & $41.74\pm 0.13$ \\
9640 & 10:02:40.52s & 02$^\circ$15\arcmin15.93\arcsec & 5.856 & $-18.44\pm 0.25$ & $42.03\pm 0.04$ \\
        \bottomrule
        \end{tabular}
        }
        \hspace{0px}
        \resizebox{0.47\linewidth}{!}{
        \begin{tabular}{rccccc}
        \toprule
        ID & RA & DEC & $z$ & $M_{\rm UV}$ & \makecell{$\log_{10}(L_{\rm \OIII\lambda 5008}$\\$/{\rm erg\,s}^{-1})$}\\
        \midrule
        5196 & 10:02:41.44s & 02$^\circ$13\arcmin58.97\arcsec & 5.866 & $-19.65\pm 0.05$ & $42.00\pm 0.05$ \\
6556 & 10:02:33.77s & 02$^\circ$10\arcmin20.68\arcsec & 5.873 & $-17.73\pm 0.63$ & $42.07\pm 0.07$ \\
7699 & 10:02:34.83s & 02$^\circ$11\arcmin23.24\arcsec & 5.874 & $-17.99\pm 0.34$ & $41.68\pm 0.04$ \\
6613 & 10:02:36.02s & 02$^\circ$11\arcmin34.89\arcsec & 5.876 & $-19.68\pm 0.04$ & $42.23\pm 0.04$ \\
9275 & 10:02:34.71s & 02$^\circ$11\arcmin54.54\arcsec & 5.876 & $-20.12\pm 0.04$ & $42.70\pm 0.01$ \\
12710 & 10:02:26.94s & 02$^\circ$10\arcmin23.96\arcsec & 5.876 & $-20.59\pm 0.11$ & $42.38\pm 0.03$ \\
9085 & 10:02:34.37s & 02$^\circ$11\arcmin39.81\arcsec & 5.877 & $-19.69\pm 0.05$ & $42.32\pm 0.02$ \\
7754 & 10:02:34.86s & 02$^\circ$11\arcmin25.97\arcsec & 5.879 & $-18.35\pm 0.10$ & $41.92\pm 0.07$ \\
6123 & 10:02:32.70s & 02$^\circ$09\arcmin30.20\arcsec & 5.881 & $-21.31\pm 0.04$ & $43.14\pm 0.01$ \\
7198 & 10:02:32.53s & 02$^\circ$09\arcmin55.64\arcsec & 5.884 & -- & $42.44\pm 0.03$ \\
1195 & 10:02:37.82s & 02$^\circ$10\arcmin36.99\arcsec & 5.893 & $-20.33\pm 0.27$ & $42.26\pm 0.06$ \\
4382 & 10:02:33.95s & 02$^\circ$09\arcmin33.06\arcsec & 5.910 & $-18.87\pm 0.19$ & $42.15\pm 0.08$ \\
3590 & 10:02:38.43s & 02$^\circ$11\arcmin43.33\arcsec & 5.929 & $-20.39\pm 0.06$ & $42.46\pm 0.02$ \\
18345 & 10:02:38.17s & 02$^\circ$14\arcmin28.14\arcsec & 5.946 & $-18.99\pm 0.18$ & $41.75\pm 0.08$ \\
3136 & 10:02:35.37s & 02$^\circ$09\arcmin56.64\arcsec & 5.984 & $-18.42\pm 0.25$ & $42.21\pm 0.07$ \\
7059 & 10:02:33.37s & 02$^\circ$10\arcmin20.43\arcsec & 6.021 & $-17.47\pm 0.28$ & $42.40\pm 0.09$ \\
17816 & 10:02:38.09s & 02$^\circ$14\arcmin38.43\arcsec & 6.025 & $-18.44\pm 0.28$ & $41.90\pm 0.11$ \\
602 & 10:02:44.55s & 02$^\circ$14\arcmin04.16\arcsec & 6.074 & $-18.23\pm 0.18$ & $42.18\pm 0.03$ \\
4935 & 10:02:42.47s & 02$^\circ$14\arcmin27.46\arcsec & 6.084 & $-17.99\pm 0.20$ & $41.90\pm 0.10$ \\
6763 & 10:02:31.91s & 02$^\circ$09\arcmin25.24\arcsec & 6.104 & $-18.21\pm 0.18$ & $41.70\pm 0.09$ \\
11833 & 10:02:30.61s & 02$^\circ$12\arcmin42.76\arcsec & 6.110 & $-18.71\pm 0.27$ & $41.80\pm 0.08$ \\
11538 & 10:02:27.27s & 02$^\circ$10\arcmin59.74\arcsec & 6.134 & $-19.38\pm 0.17$ & $42.38\pm 0.06$ \\
11651 & 10:02:29.60s & 02$^\circ$12\arcmin13.48\arcsec & 6.134 & $-18.72\pm 0.12$ & $42.08\pm 0.03$ \\
8082 & 10:02:35.78s & 02$^\circ$12\arcmin03.24\arcsec & 6.136 & $-19.13\pm 0.07$ & $42.35\pm 0.02$ \\
6033 & 10:02:39.50s & 02$^\circ$13\arcmin15.23\arcsec & 6.138 & $-18.92\pm 0.16$ & $42.18\pm 0.05$ \\
7245 & 10:02:39.70s & 02$^\circ$13\arcmin52.82\arcsec & 6.140 & $-19.91\pm 0.09$ & $42.37\pm 0.02$ \\
6028 & 10:02:37.79s & 02$^\circ$12\arcmin18.86\arcsec & 6.144 & $-18.07\pm 0.13$ & $41.85\pm 0.05$ \\
492 & 10:02:42.16s & 02$^\circ$12\arcmin42.18\arcsec & 6.159 & $-18.69\pm 0.23$ & $42.20\pm 0.04$ \\
1551 & 10:02:40.11s & 02$^\circ$11\arcmin59.52\arcsec & 6.165 & $-19.44\pm 0.11$ & $42.09\pm 0.03$ \\
13523 & 10:02:34.99s & 02$^\circ$14\arcmin31.95\arcsec & 6.167 & $-19.17\pm 0.12$ & $42.21\pm 0.03$ \\
12026 & 10:02:30.54s & 02$^\circ$12\arcmin36.04\arcsec & 6.177 & $-18.17\pm 0.15$ & $42.41\pm 0.02$ \\
8942 & 10:02:31.13s & 02$^\circ$09\arcmin50.46\arcsec & 6.259 & $-19.77\pm 0.12$ & $42.58\pm 0.03$ \\
6888 & 10:02:33.34s & 02$^\circ$10\arcmin14.30\arcsec & 6.265 & $-19.63\pm 0.08$ & $42.45\pm 0.03$ \\
6012 & 10:02:41.50s & 02$^\circ$14\arcmin20.38\arcsec & 6.305 & $-18.63\pm 0.09$ & $41.64\pm 0.11$ \\
1979 & 10:02:34.24s & 02$^\circ$08\arcmin56.31\arcsec & 6.351 & $-18.77\pm 0.13$ & $42.21\pm 0.06$ \\
14368 & 10:02:37.94s & 02$^\circ$15\arcmin58.13\arcsec & 6.473 & $-19.36\pm 0.13$ & $42.01\pm 0.05$ \\
4077 & 10:02:42.08s & 02$^\circ$13\arcmin54.90\arcsec & 6.506 & $-17.79\pm 0.21$ & $41.91\pm 0.09$ \\
7206 & 10:02:42.00s & 02$^\circ$15\arcmin07.62\arcsec & 6.509 & -- & $42.19\pm 0.04$ \\
17093 & 10:02:28.47s & 02$^\circ$09\arcmin41.99\arcsec & 6.518 & $-19.62\pm 0.15$ & $42.28\pm 0.04$ \\
5338 & 10:02:36.31s & 02$^\circ$11\arcmin14.26\arcsec & 6.522 & $-18.71\pm 0.25$ & $41.84\pm 0.07$ \\
8364 & 10:02:32.84s & 02$^\circ$10\arcmin33.20\arcsec & 6.523 & $-18.81\pm 0.18$ & $42.00\pm 0.08$ \\
6985 & 10:02:35.75s & 02$^\circ$11\arcmin35.30\arcsec & 6.526 & $-19.64\pm 0.08$ & $42.14\pm 0.03$ \\
9607 & 10:02:33.09s & 02$^\circ$11\arcmin12.51\arcsec & 6.527 & $-18.74\pm 0.38$ & $42.31\pm 0.03$ \\
5630 & 10:02:36.10s & 02$^\circ$11\arcmin13.50\arcsec & 6.574 & $-19.06\pm 0.08$ & $41.92\pm 0.05$ \\
17640 & 10:02:32.27s & 02$^\circ$11\arcmin32.18\arcsec & 6.582 & $-19.30\pm 0.08$ & $42.28\pm 0.02$ \\
9269 & 10:02:35.38s & 02$^\circ$12\arcmin13.93\arcsec & 6.591 & $-21.21\pm 0.04$ & $43.12\pm 0.01$ \\
18837 & 10:02:37.51s & 02$^\circ$13\arcmin56.81\arcsec & 6.594 & $-19.10\pm 0.14$ & $42.06\pm 0.05$ \\
9336 & 10:02:36.01s & 02$^\circ$12\arcmin40.39\arcsec & 6.596 & $-17.56\pm 0.18$ & $41.76\pm 0.08$ \\
4669 & 10:02:40.21s & 02$^\circ$13\arcmin06.56\arcsec & 6.604 & $-17.42\pm 0.34$ & $42.06\pm 0.10$ \\
16172 & 10:02:27.76s & 02$^\circ$09\arcmin42.85\arcsec & 6.701 & $-19.12\pm 0.11$ & $42.11\pm 0.05$ \\
4301 & 10:02:41.66s & 02$^\circ$13\arcmin45.09\arcsec & 6.709 & $-19.52\pm 0.10$ & $41.63\pm 0.09$ \\
5901 & 10:02:37.28s & 02$^\circ$11\arcmin59.23\arcsec & 6.721 & $-18.90\pm 0.14$ & $42.12\pm 0.05$ \\
8502 & 10:02:34.26s & 02$^\circ$11\arcmin23.06\arcsec & 6.728 & $-19.06\pm 0.13$ & $42.00\pm 0.07$ \\
17133 & 10:02:31.74s & 02$^\circ$11\arcmin28.47\arcsec & 6.737 & $-18.44\pm 0.16$ & $42.08\pm 0.07$ \\
16747 & 10:02:29.73s & 02$^\circ$10\arcmin32.94\arcsec & 6.738 & $-20.16\pm 0.05$ & $42.35\pm 0.04$ \\
7217 & 10:02:33.69s & 02$^\circ$10\arcmin34.72\arcsec & 6.742 & -- & $42.46\pm 0.05$ \\
16154 & 10:02:31.81s & 02$^\circ$11\arcmin59.51\arcsec & 6.750 & $-18.79\pm 0.15$ & $42.09\pm 0.03$ \\
17228 & 10:02:34.56s & 02$^\circ$13\arcmin00.52\arcsec & 6.758 & $-18.85\pm 0.16$ & $42.45\pm 0.03$ \\
12655 & 10:02:29.36s & 02$^\circ$11\arcmin43.82\arcsec & 6.758 & $-19.30\pm 0.15$ & $42.27\pm 0.06$ \\
16065 & 10:02:34.20s & 02$^\circ$13\arcmin17.65\arcsec & 6.760 & $-17.37\pm 0.24$ & $42.19\pm 0.03$ \\
13825 & 10:02:26.37s & 02$^\circ$09\arcmin45.11\arcsec & 6.767 & $-19.34\pm 0.10$ & $42.28\pm 0.03$ \\
3018 & 10:02:33.47s & 02$^\circ$08\arcmin51.47\arcsec & 6.771 & $-18.81\pm 0.12$ & $42.17\pm 0.09$ \\
3079 & 10:02:35.35s & 02$^\circ$09\arcmin54.19\arcsec & 6.795 & $-19.78\pm 0.05$ & $42.24\pm 0.06$ \\
2167 & 10:02:33.12s & 02$^\circ$08\arcmin23.16\arcsec & 6.816 & $-17.64\pm 0.25$ & $41.91\pm 0.10$ \\
8365 & 10:02:31.35s & 02$^\circ$09\arcmin44.38\arcsec & 6.840 & $-18.80\pm 0.35$ & $42.19\pm 0.08$ \\
126 & 10:02:45.22s & 02$^\circ$14\arcmin12.33\arcsec & 6.873 & $-18.78\pm 0.17$ & $42.19\pm 0.11$ \\
17174 & 10:02:38.71s & 02$^\circ$15\arcmin20.92\arcsec & 6.897 & $-18.42\pm 0.18$ & $41.87\pm 0.12$ \\
6133 & 10:02:36.59s & 02$^\circ$11\arcmin42.18\arcsec & 6.911 & $-18.36\pm 0.12$ & $42.07\pm 0.06$ \\
9912 & 10:02:39.80s & 02$^\circ$14\arcmin46.64\arcsec & 6.916 & $-17.71\pm 0.12$ & $42.00\pm 0.08$ \\
13835 & 10:02:32.35s & 02$^\circ$13\arcmin00.89\arcsec & 6.933 & $-19.37\pm 0.09$ & $42.29\pm 0.04$ \\
5843 & 10:02:36.85s & 02$^\circ$11\arcmin43.19\arcsec & 6.937 & $-19.08\pm 0.05$ & $42.31\pm 0.03$ \\

        \bottomrule
\end{tabular}
}
\end{table*}

\section{Foreground galaxy removal}\label{sec:tail_removal}

\begin{figure}
    \centering
    \includegraphics[width=\linewidth]{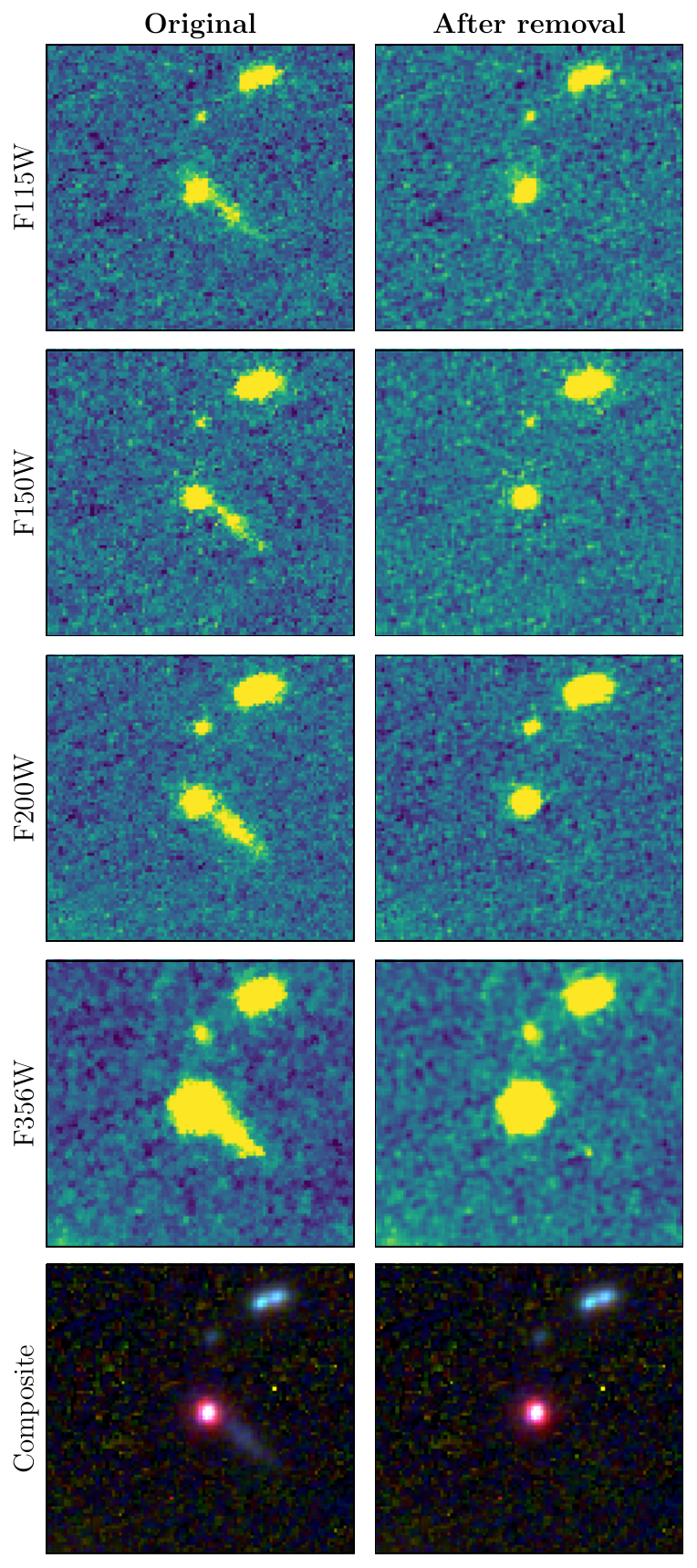}
    \caption{Broad band images in the four photometric filters of the COLA1 survey, and the composite color image of COLA1. We show the images before (left column) and after (right column) the foreground galaxy contaminant removal.}
    \label{fig:tail_removal}
\end{figure}

A blue `tail' can be seen in the composite image of COLA1 (Fig.~\ref{fig:COLA1_2D_spec}). No emission lines are detected in our grism data from this source. Given the extended shape and colors of this source, and particularly a detection in the $B$ band in the ground based data it is very likely a foreground source. The contamination from this source must be removed from COLA1's photometric fluxes. We model this object with a four component 2D Gaussian model, independently in each filter. Three Gaussians with fixed centroid are used to model COLA1 and the wings of the PSF. Two of these three Gaussians are allowed to be asymmetric, whilst one of them is fixed to be symmetric. The fourth Gaussian is used to fit the 'tail'. Once the Gaussians are fitted to our data, we subtract the object component and re-do the photometry. We conservatively increase the errors of the photometry by 20\% of the change in the fluxes due to this process, in quadrature. Figure \ref{fig:tail_removal} shows broad-band images before and after applying this procedure.

\section{Examples of spectra}\label{sec:example_spec}

In Fig.~\ref{fig:example_spec} we show examples of inspection images of selected \OIII\ emitters in this work. The first five objects shown in Fig.~\ref{fig:example_spec} correspond to those in the environment of COLA1 ($z=z_{\rm COLA1}\pm 0.02$). The object with ID NUMBER 1272 illustrates how the morphology of sources is captured in the emission lines of 2D grism spectra, in many cases allowing a secure match with the imaging. On the other hand, the object with ID NUMBER 18752 has a prominent extended morphology as well, and its \OIII\ doublet is only seen in module B. Since modules A and B disperse light in opposite directions, the spatial information is mirrored in the spectral direction for module B.

\begin{figure*}
    \centering
    \includegraphics[width=0.49\linewidth]{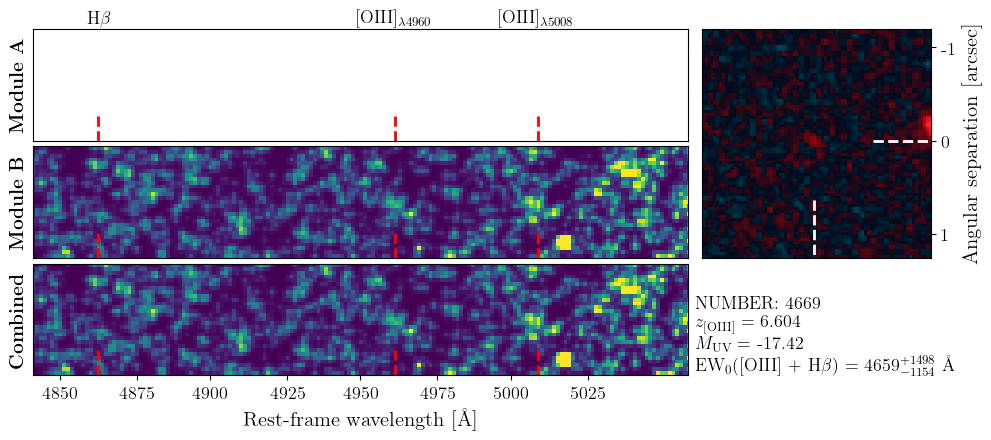}
    \includegraphics[width=0.49\linewidth]{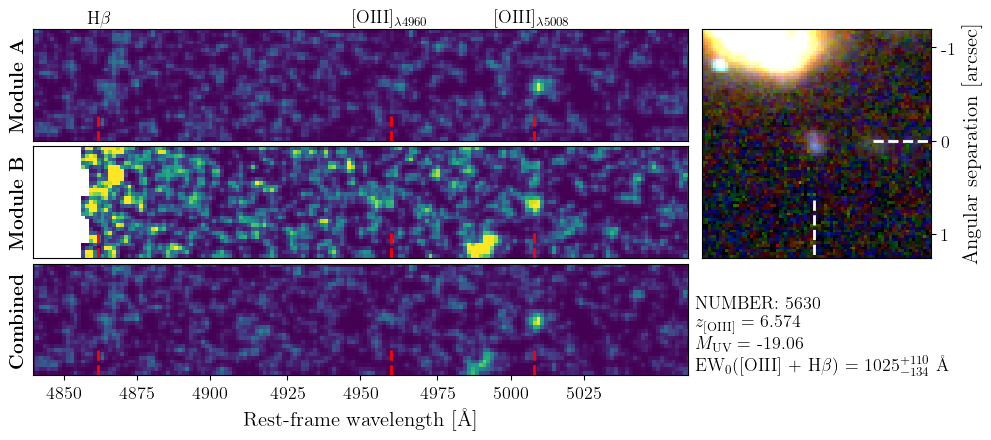}
    \includegraphics[width=0.49\linewidth]{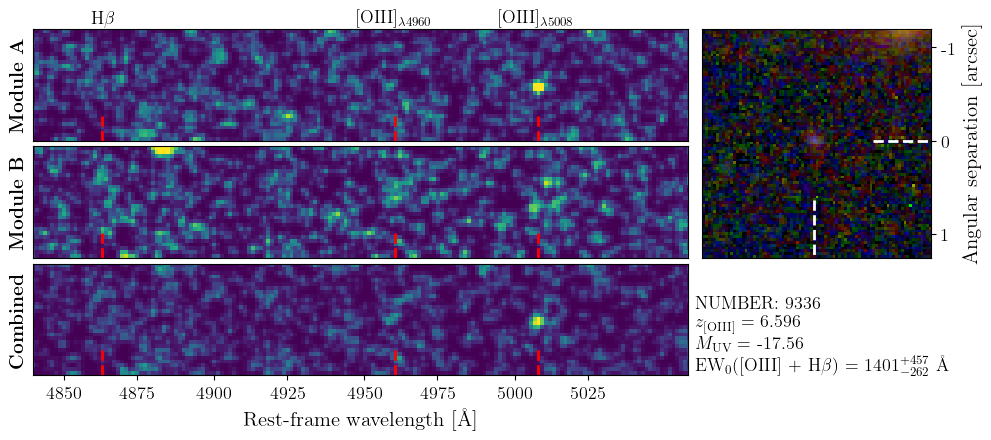}
    \includegraphics[width=0.49\linewidth]{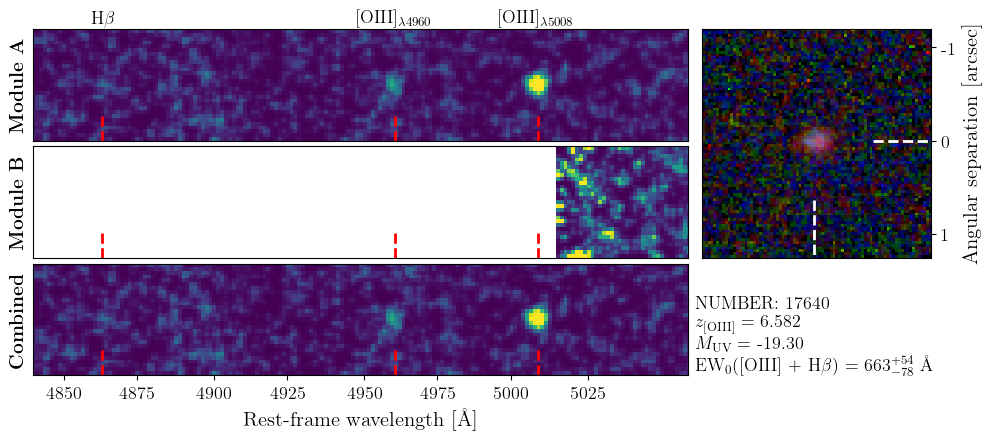}
    \includegraphics[width=0.49\linewidth]{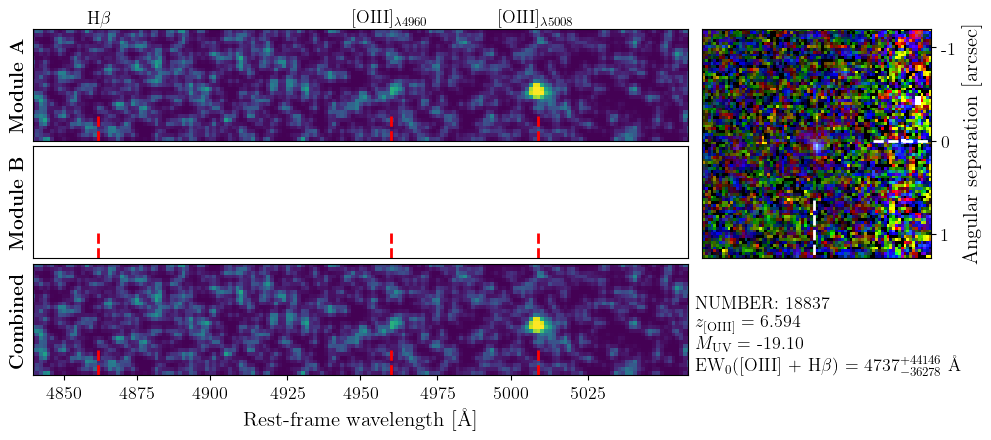}
    \includegraphics[width=0.49\linewidth]{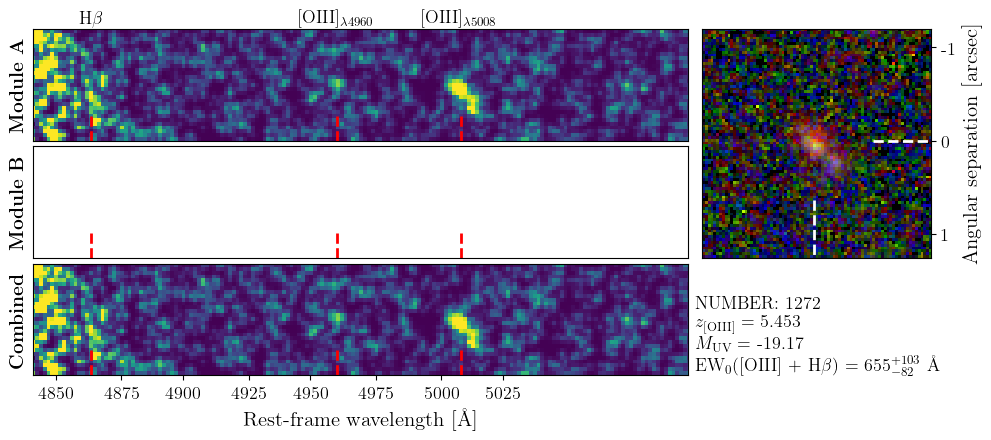}
    \includegraphics[width=0.49\linewidth]{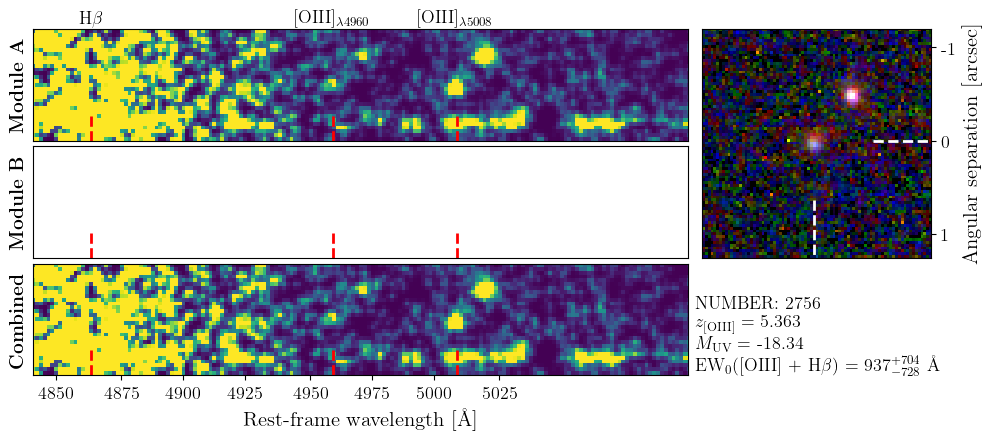}
    \includegraphics[width=0.49\linewidth]{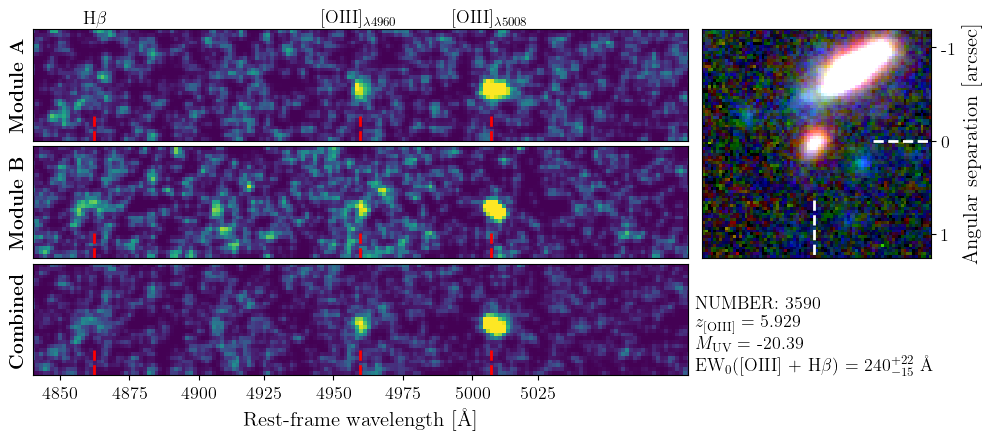}
    \includegraphics[width=0.49\linewidth]{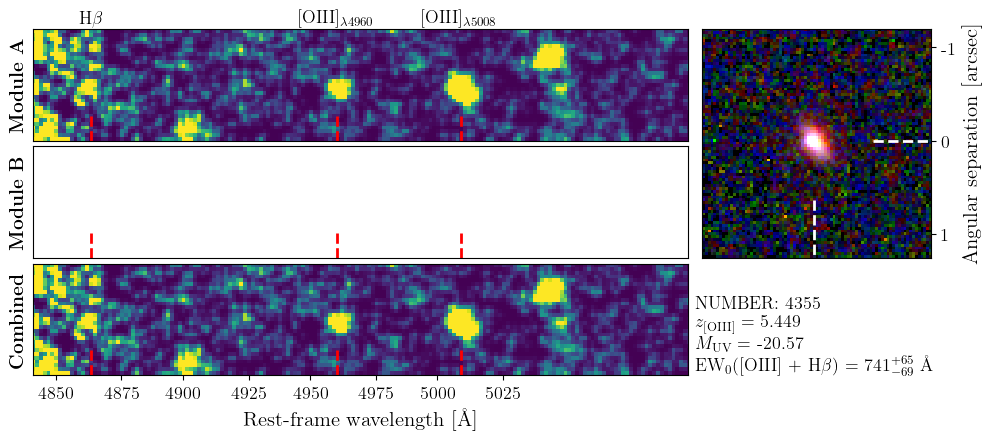}
    \includegraphics[width=0.49\linewidth]{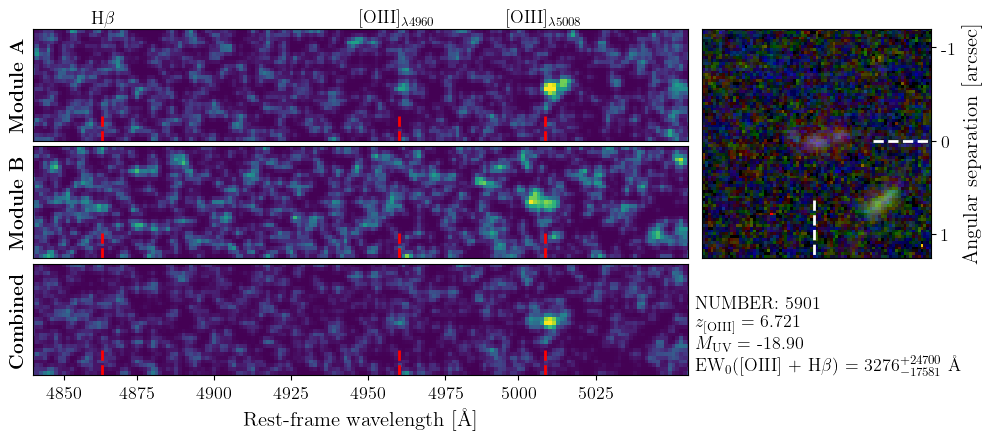}

    \caption{Examples of \OIII\ emitter grism 2D spectra.}
    \label{fig:example_spec}
\end{figure*}

\begin{figure*}\ContinuedFloat
    \includegraphics[width=0.49\linewidth]{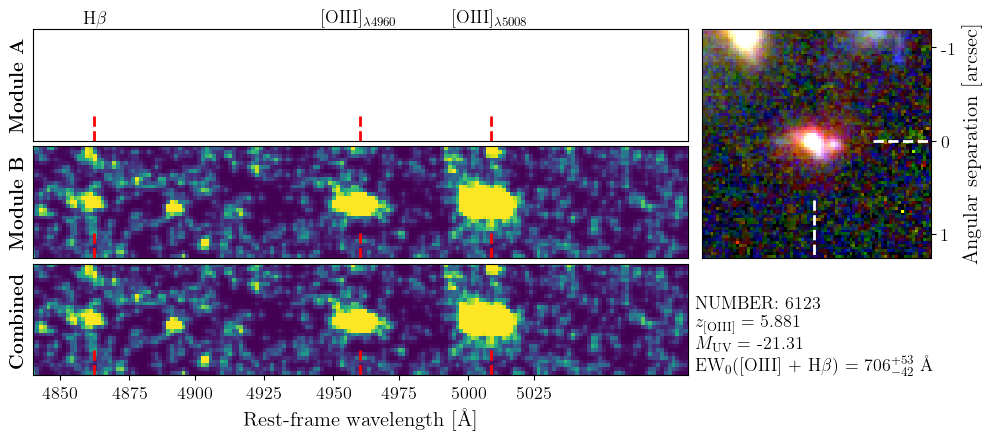}
    \includegraphics[width=0.49\linewidth]{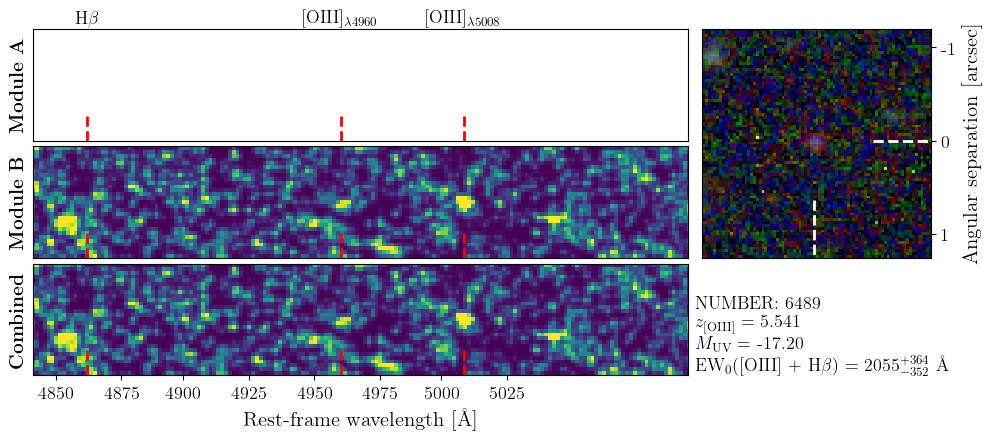}
    \includegraphics[width=0.49\linewidth]{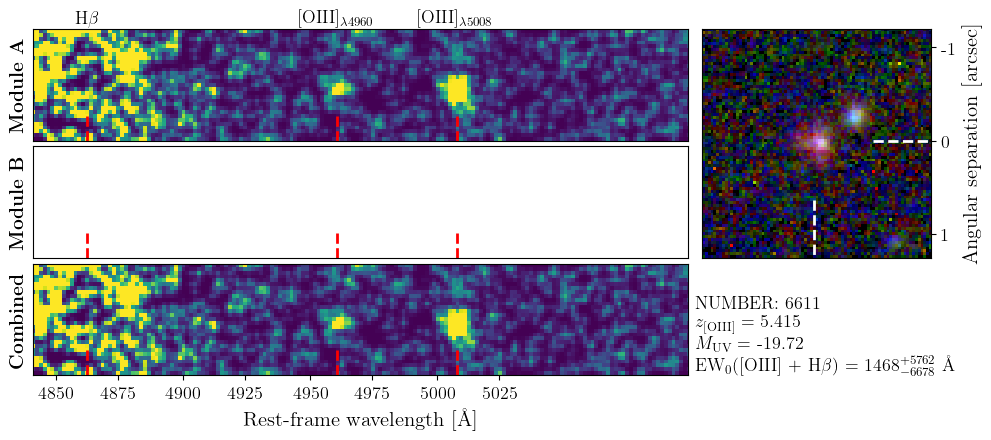}
    \includegraphics[width=0.49\linewidth]{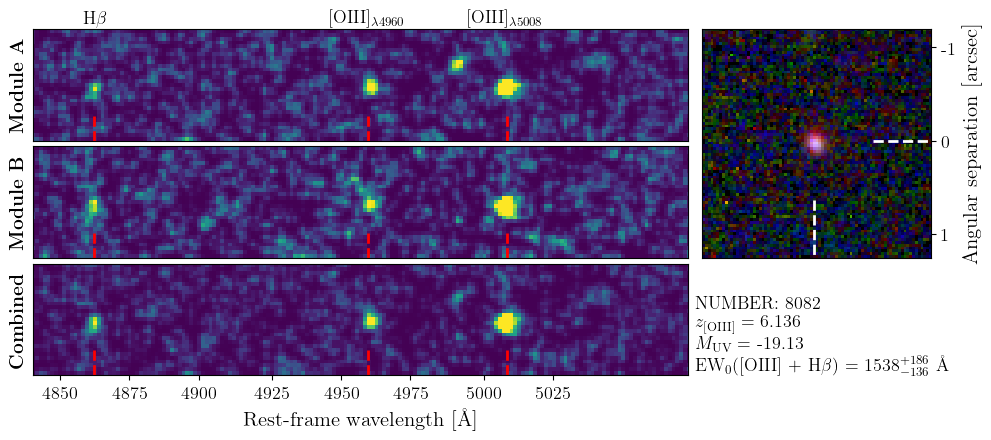}
    \includegraphics[width=0.49\linewidth]{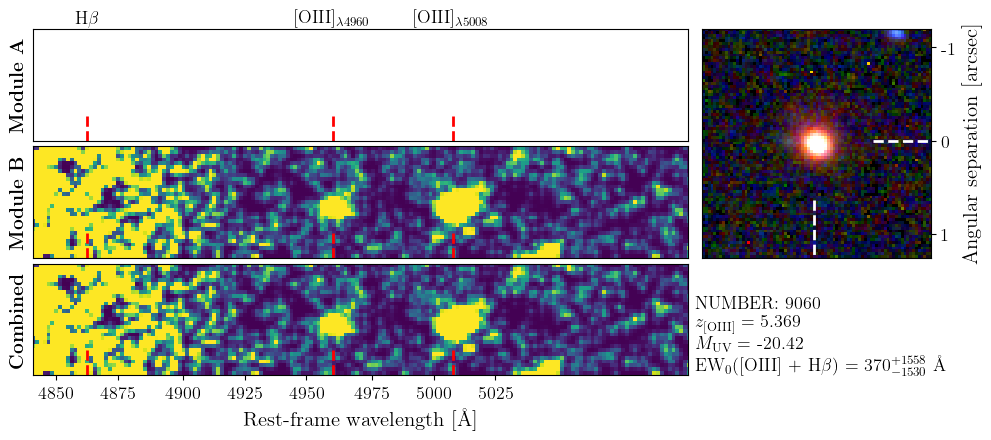}
    \includegraphics[width=0.49\linewidth]{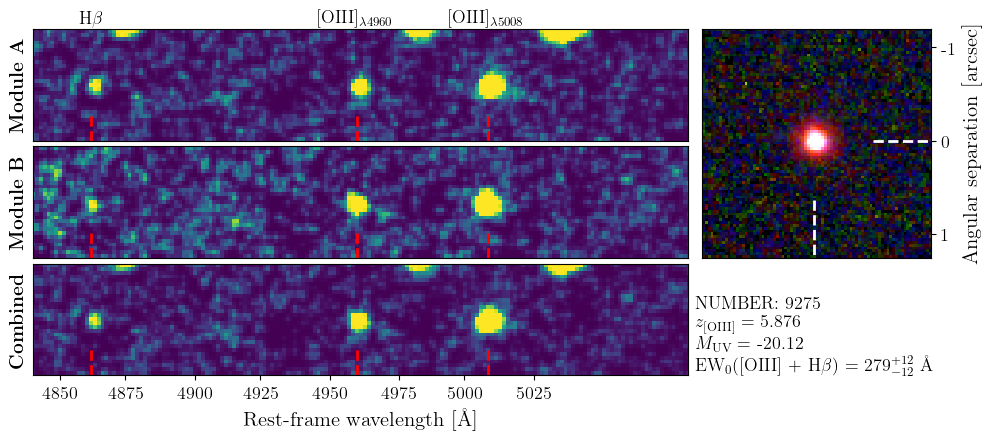}
    \includegraphics[width=0.49\linewidth]{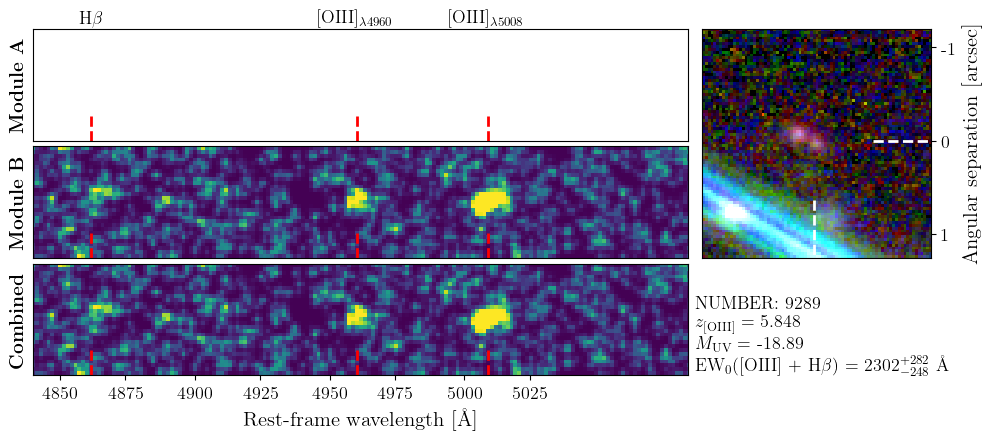}
    \includegraphics[width=0.49\linewidth]{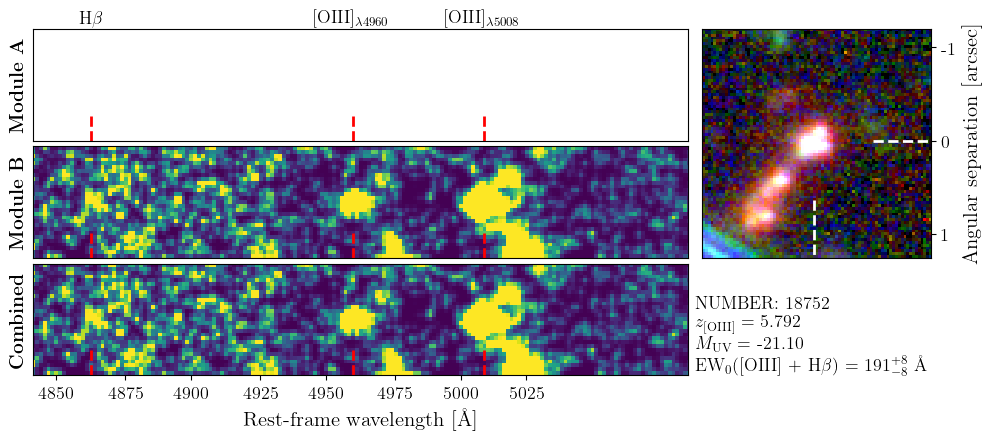}

    \caption{continued}
        \label{fig:example_spec2}

\end{figure*}

\end{document}